\documentclass[11pt]{article}

\usepackage[letterpaper,margin=1in]{geometry}
\usepackage{setspace}
\usepackage{times}
\usepackage{fix-cm}
\usepackage{titling}
\usepackage{amsmath,amssymb,amsfonts,amsthm}
\usepackage{bbm}
\usepackage{chngcntr}
\usepackage{xparse}
\allowdisplaybreaks[2]

\usepackage{graphicx}
\graphicspath{{Figures/}{figures/}}
\usepackage{booktabs}
\usepackage{multicol,multirow}
\usepackage{makecell}
\usepackage{array}
\usepackage{hhline}
\usepackage{threeparttable}
\usepackage{rotating}
\usepackage{float}
\usepackage{caption}
\usepackage[caption=false]{subfig}
\let\subfigure\subfloat
\usepackage{algorithm}
\floatname{algorithm}{Framework}
\usepackage{algorithmic}
\usepackage{fancyvrb}
\usepackage{adjustbox}
\usepackage{placeins}


\usepackage{enumitem}
\usepackage{natbib}
\bibpunct[, ]{(}{)}{,}{a}{}{,}%
\usepackage{bibunits}
\defaultbibliography{mybib}
\defaultbibliographystyle{apalike}
\usepackage{fewerfloatpages}
\usepackage{xcolor}
\usepackage{hyperref}
\hypersetup{
    colorlinks=true,
    linkcolor=blue,
    citecolor=blue,
    urlcolor=blue,
    hypertexnames=false
}
\usepackage[capitalize,nameinlink]{cleveref}

\usepackage{endnotes}
\let\footnote=\endnote

\newcolumntype{C}[1]{>{\centering\let\newline\\\arraybackslash\hspace{0pt}}m{#1}}

\IfFileExists{input_metadata.tex}{%

\newcommand{\PaperTitle}{Tradable It\^o Signatures: A Model-Free, Interpretable Framework for Dynamic Hedging}

\newcommand{\PaperKeywords}{Dynamic hedging, Derivative replication, It\^o signature, Path-dependent derivatives, Interpretable machine learning.}

\newcommand{\PaperJEL}{G11; G13; C32; C45; C63}%
}{%
    \newcommand{\PaperTitle}{Paper Title for INFORMS Finance 2026}%
    \newcommand{\PaperKeywords}{financial engineering; fintech; risk management; machine learning; optimization}%
}
\providecommand{\PaperKeywords}{financial engineering; fintech; risk management; machine learning; optimization}
\providecommand{\PaperJEL}{G11; G13; G17; C45; C61; C63}

\theoremstyle{plain}
\newtheorem{theorem}{Theorem}
\newtheorem{corollary}{Corollary}

\newtheorem{lemma}{Lemma}

\theoremstyle{definition}
\newtheorem{definition}{Definition}
\newtheorem{assumption}{Assumption}
\newtheorem{example}{Example}

\theoremstyle{remark}
\newtheorem{remark}{Remark}

\crefname{theorem}{Theorem}{Theorems}
\crefname{corollary}{Corollary}{Corollaries}
\crefname{proposition}{Proposition}{Propositions}
\crefname{lemma}{Lemma}{Lemmas}
\crefname{definition}{Definition}{Definitions}
\crefname{assumption}{Assumption}{Assumptions}
\crefname{example}{Example}{Examples}
\crefname{remark}{Remark}{Remarks}
\crefname{algorithm}{Framework}{Frameworks}
\crefname{section}{Section}{Sections}
\crefname{appendix}{Appendix}{Appendices}
\crefname{figure}{Figure}{Figures}
\crefname{table}{Table}{Tables}
\crefname{equation}{Equation}{Equations}

\let\amsthmproof\proof
\let\amsthmendproof\endproof
\RenewDocumentCommand{\proof}{o g}{%
  \IfNoValueTF{#1}{%
    \IfNoValueTF{#2}{\amsthmproof}{\amsthmproof[#2]}%
  }{\amsthmproof[#1]}%
}
\RenewDocumentCommand{\endproof}{}{\amsthmendproof}
\NewDocumentCommand{\Halmos}{}{}

\newenvironment{APPENDICES}{\appendix}{}
\usepackage[final]{microtype}

\AtBeginDocument{%
  \emergencystretch=4em
  \tolerance=3000
  \hbadness=10000
  \binoppenalty=0
  \relpenalty=0
  \allowdisplaybreaks[4]
}

\makeatletter

\newcommand{\Rmnum}[1]{\expandafter\@slowromancap\romannumeral #1@}
\makeatother

\title{\bfseries \PaperTitle}

\author{%
\begin{minipage}{0.96\textwidth}
\centering
Xin Guo\\
\small UC Berkeley Department of Industrial Engineering and Operations Research\\
\small \texttt{xinguo@berkeley.edu}\\[0.75em]
Binnan Wang\\
\small Peking University School of Mathematical Sciences and Laboratory for Mathematical Economics and Quantitative Finance\\
\small \texttt{wangbinnan@stu.pku.edu.cn}\\[0.75em]
Ruixun Zhang\thanks{Corresponding author}\\
\small Peking University School of Mathematical Sciences, Center for Statistical Science, National Engineering Laboratory for Big Data Analysis and Applications, and Laboratory for Mathematical Economics and Quantitative Finance\\
\small \texttt{zhangruixun@pku.edu.cn}
\end{minipage}%
}

\begin{document}

\maketitle

\begin{abstract}
We propose an interpretable machine-learning framework for dynamic hedging using the It\^o signature transform, which turns asset-price paths into a set of linear features that universally represent nonlinear functions on time-series. We show that each discretized It\^o signature component can be perfectly replicated by a simple self-financing strategy using only the underlying assets and cash, which turns It\^o signature components into tradable and transparent hedging bases. This allows nonlinear derivative payoffs to be approximated by linear combinations of signature terms and hedged through the corresponding combination of trading strategies. We further establish a new approximation result for the It\^o signature and derive theoretical bounds for both in-sample and out-of-sample hedging errors. Our method is computationally efficient, easy to implement, and avoids the estimation of future conditional expectations, which makes it attractive for real-world applications. In simulations, our method delivers strong sample efficiency at substantially lower computational cost than neural-network benchmarks. In an empirical study of S\&P 500 index options, it performs robustly across vanilla and path-dependent contracts, with the signature-kernel weighted version providing further gains by localizing estimation to similar historical market paths. Overall, the paper identifies the It\^o signature as a practical, transparent, and model-agnostic implementation framework for dynamic hedging.
\end{abstract}

\vspace{0.5em}
\noindent\textbf{Keywords:} \PaperKeywords

\vspace{0.5em}
\noindent\textbf{JEL Classification:} \PaperJEL

\vspace{1em}

\clearpage
\begingroup
\begin{spacing}{0.90}
\setlength{\parskip}{0pt}
\tableofcontents
\end{spacing}
\endgroup
\clearpage
\begin{bibunit}
\section{Introduction}

\paragraph{Hedging Problems.} 
Dynamic hedging is a central problem in quantitative finance. In the classical Black--Scholes framework, delta hedging provides an exact continuous-time replication strategy for European options under geometric Brownian motion \citep{black1973pricing}.  Classical model-based approaches, while theoretically elegant and economically important, typically rely on restrictive assumptions on the underlying price dynamics, market structures, or payoff forms, which makes them less suitable for general path-dependent and exotic derivatives.

Recent advances in machine learning provide a model-free alternative to classical hedging methods. In particular, deep hedging methods parameterize trading strategies by neural networks and estimate hedge positions directly from historical path information under a chosen loss criterion \citep{buehler2019deep, zhang2021option, ruf2022hedging, lutkebohmert2022robust, almeida2023can}. These methods substantially enlarge the class of hedging rules that can be learned from data. However, they face a different set of challenges. Because the hedge is produced by a black-box nonlinear model, it is often difficult to understand how the position is formed and why it changes across market states. This lack of transparency can make the strategy difficult to evaluate in practice \citep{jaeger2021interpretable,nazemi2024interpretable}. In addition, neural-network-based hedging usually depends on nontrivial model designs and intensive training. Its performance may also be sensitive to sample size and to shifts in the market environment. These issues are particularly relevant in financial applications, where historical data are often limited and hedge implementation typically requires a transparent rule that can be monitored and interpreted.

\paragraph{Signature-Based Hedging.} 
One of the latest tools developed for hedging is the signature transform. 
For a price path, the signature transform consists of iterated integrals of its increments and provides a structured hierarchy of path-dependent features \citep{morrill2020generalised,lyons2022signature,bayer2025signature}. Low-order terms capture cumulative changes, co-movements, and interactions with time, whereas higher-order terms encode richer nonlinear dependence on the realized path. This representation is naturally relevant for hedging because many derivative payoffs are nonlinear functionals of the underlying price path, while signature components provide a feature space for representing such payoffs \citep{horvath2023optimal}.

A key reason for the success of signature methods is their universal nonlinearity property: under appropriate conditions, any continuous functional of a path can be approximated arbitrarily well by a linear functional of its signature \citep{lyons2022signature,cuchiero2023signature}. This property is attractive for statistical learning because it converts a nonlinear approximation problem into a linear one. For dynamic hedging, this suggests a natural strategy: approximate a derivative payoff by a linear combination of signature terms and then use the fitted linear representation to construct the hedge.

A major obstacle, however, is that the standard universal nonlinearity result is naturally tied to the Stratonovich signature or to rough-path signatures, rather than to the It\^o signature that is more directly aligned with non-anticipative trading. \endnote{The reason is structural: the Stratonovich integral obeys the classical chain rule, which yields the algebraic closure properties needed for Stone--Weierstrass-type arguments \citep{cuchiero2023signature}.} Accordingly, existing signature-based hedging methods typically build on Stratonovich or rough-path objects \citep{lyons2020non, cirone2025roughkernelhedging, cuchiero2025universal, abijaber2025a, abijaber2025b}. This creates an implementation difficulty. A dynamic hedge must be adapted to the information available at each trading time, whereas Stratonovich-based representations generally require an additional translation into a non-anticipative trading rule. In practice, this translation is often implemented through the Hoff lead--lag transform \citep{flint2016discretely} together with the estimation of expected future signatures \citep{lyons2020non, cirone2025roughkernelhedging}. It increases the state dimension, causes the number of signature terms to grow exponentially with the truncation order, and introduces an additional source of estimation error. 

\paragraph{The It\^o Signature Gap.}
From a financial perspective, the It\^o signature is the more natural object for dynamic hedging because it is non-anticipative and directly compatible with self-financing trading strategies \citep{allan2024cadlag}. The difficulty is that the It\^o signature does not, in general, enjoy the same convenient universal nonlinearity property as the Stratonovich signature. This creates the central motivation that this paper addresses: the mathematically convenient signature for approximation is not the most natural object for trading, while the financially natural object for trading lacks the standard approximation theory.

\paragraph{Our Work.}
In this work, we develop a new hedging framework based on the It\^o signature that is both flexible and transparent. Our key idea is to treat discretized It\^o signature components not merely as predictive features, but as tradable bases for hedge construction. We show that each discretized It\^o signature component can be exactly replicated by a simple self-financing trading strategy using only the underlying assets and cash. This converts the hedging problem into a linear allocation problem over transparent trading bases. Building on this representation, we develop an It\^o-signature-based approach to approximate nonlinear derivative payoffs, analyze its theoretical hedging error, and evaluate its performance in both simulations and market option data. In the empirical analysis, we also use the signature kernel as a path-similarity weighting scheme to deal with non-stationary data. The framework is flexible enough to handle complex path-dependent payoffs, while remaining transparent, computationally tractable, and directly linked to implementable trading strategies.

\paragraph{Main Results.} 
Our contributions are as follows. 
\begin{enumerate} 
\item We establish the tradability of discretized It\^o signatures. Each nonconstant discretized It\^o-signature component is the terminal gain of an adapted self-financing strategy in the underlying assets and cash. This result gives every It\^o-signature component a direct financial interpretation and forms the basis of the hedging framework (Theorem~\ref{thm:hedging strategy}). 
\item We develop the theoretical foundation for It\^o-signature hedging. We prove the convergence of discretized It\^o signatures to their continuous-time counterparts, establish approximation results for It\^o-signature representations, and translate these results into self-financing hedging guarantees. We then derive finite-sample statistical bounds for estimating the truncated signature hedge and combine discretization, approximation, and estimation errors into a total out-of-sample hedging error bound (Theorems~\ref{thm:lpconvergence}--\ref{thm:total_hedging_error} and Corollaries~\ref{cor:ito_un_constant}--\ref{cor:gen_hedge_error}).
\item In simulations, the It\^o-signature hedge is shown to be accurate, stable, and computationally efficient, especially when training data are limited. The method achieves strong hedging performance with substantially less computation than neural-network and Stratonovich-signature benchmarks. Its linear structure also allows us to interpret which tradable basis strategies are used for different payoff structures (Section~\ref{sec:simulation}). 
\item In an empirical study of S\&P 500 index options, the It\^o-signature hedge performs well for both vanilla and path-dependent contracts. The It\^o signature captures path dependence, while the signature-kernel weights help with limited and non-stationary financial data. The combined method is especially useful for short-maturity, out-of-the-money, and path-dependent options (Section~\ref{sec:emp}).
\end{enumerate} 

Overall, the paper identifies the discretized It\^o signature as a practical basis for transparent dynamic hedging.
Our framework combines implementability, interpretability, and data efficiency, and it offers a model-free alternative to black-box hedging methods.

\paragraph{Related Literature.}
For classic dynamic hedging, the literature extends the Black--Scholes benchmark approach to more general settings, including stochastic volatility models \citep{di1995mean, renault1996option,ait2013leverage}, jump-diffusion models \citep{cont2007hedging,bates2022empirical}, discrete-time trading \citep{bertsimas2001hedging}, and minimum-variance hedging objectives \citep{hull2017optimal}. A related strand uses machine learning to learn hedge ratios directly from data \citep{buehler2019deep, zhang2021option, ruf2022hedging, lutkebohmert2022robust}. Our It\^o-signature hedging framework differs from these approaches by avoiding a parametric specification of the price dynamics in implementation and by producing a hedge that is linear in tradable basis strategies rather than generated by a black-box nonlinear model \citep{fan2009option,xiu2014hermite}.

Path signature originates in algebraic topology \citep{Chen1954, Chen1957} and receives a modern formulation in rough path theory \citep{lyons2007differential, friz2010multidimensional}. It has since become a useful tool in machine learning and mathematical finance \citep{reizensteinIisignatureLibraryEfficient2018,kiraly2019kernels,sugiura2020machine,morrill2021neural,salvi2021higher,chevyrev2025primer,guo2025consistency}. Empirical work further suggests that signature-based linear models can compete effectively with more complex nonlinear learners while retaining interpretability and requiring substantially less architecture tuning \citep{levin2016learning,pan2023path,bleistein2023learning,guTransportationMarketRate2024}.

Existing signature-based hedging methods typically build on Stratonovich or rough-path objects \citep{cuchiero2025universal, abijaber2025a, abijaber2025b}. These methods provide powerful approximation tools, but they generally require an additional step to translate the fitted representation into an adapted trading rule, often through the Hoff lead--lag transform and the estimation of expected future signatures \citep{flint2016discretely,lyons2020non,cirone2025roughkernelhedging}. Our main distinction is that discretized It\^o-signature coordinates serve directly as tradable self-financing bases, so the coefficients fitted in the payoff approximation are the same coefficients used to construct the hedge.

Recent work has begun to narrow the It\^o signature approximation gap. For example, \citet{harang2024universalapproximationnongeometricrough} establish approximation results for polynomial functionals built from the It\^o signature, while \citet{kwossek2025approximation} prove a universal approximation result for the It\^o signature of time-augmented paths with quadratic variation. These contributions clarify the approximation power of the It\^o signature but do not directly yield an implementable dynamic hedging framework. Our paper bridges this gap and evaluates the performance of new hedging framework both theoretically and empirically.

Signature kernels measure the similarity between two paths by taking an inner product of their signature features. Theoretical work develops this idea as a kernel method for ordered paths, with tractable computation and path-distribution extensions \citep{kiraly2019kernels,salvi2021signature,chevyrev2022signature,lee2025signature}. This path-similarity view is useful for non-stationary time series, because it lets the estimator rely more on histories that are close to the current regime. However, its empirical use in financial settings with limited data is still underexplored. A closely related example is \citet{guTransportationMarketRate2024}, who use signature-kernel similarity weights to forecast transportation marketplace rates under seasonality and regime shifts. We adopt this  idea for hedging, using the signature kernel to give more weight to historical market paths that are more similar to the current one.

\paragraph{Outline.}
\Cref{sec:framework} introduces the setup and our It\^o-signature-based hedging strategy. \Cref{sec: theoretical analysis} and \Cref{sec:statistical_learning} develop the theoretical results. \Cref{sec:simulation} reports the simulation evidence. \Cref{sec:emp} presents the empirical results. \Cref{sec:conclusion} concludes. All proofs can be found in Appendix~\ref{appendix:lemmaandproofs}.

\section{Tradable It\^o-Signature Hedging Framework\label{sec:framework}}
This section introduces the tradable signature basis used throughout the paper. We first define the signature components of a time-augmented price path. We then show that, in the It\^o sense, the self-financing hedging strategy can be linearly represented by signature features.

\subsection{Definition of the Signature Transform}\label{subsec:signatureandUNproperty}

For the purpose of dynamic hedging for nonlinear and path-dependent payoffs, we begin with the signature representation of a path. 
Let \( (\Omega,\mathcal F,\{\mathcal F_t\}_{t\in[0,T]},\mathbb P) \) be a filtered probability space satisfying the usual conditions. Let \(X=\{X_t\}_{t\in[0,T]}\) be a \(d\)-dimensional continuous \(\{\mathcal F_t\}\)-adapted semimartingale. We define its time-augmented path by
\begin{equation}\label{eq:timeaug}
    \tilde X_t:=(X_t,t)\in\mathbb R^{d+1},\qquad t\in[0,T].
\end{equation}

We will also use the following unified notation, where
\(
\Gamma=(i_1,\dots,i_m)\in\{1,\dots,d+1\}^m
\)
denotes a multi-index and \(|\Gamma|=m\) denotes its length. We write \(\emptyset\) for the empty multi-index and \(\Gamma^-=(i_1,\dots,i_{m-1})\) for the truncated multi-index obtained by removing the last entry. We write \(\pi_n=\{0=t_0<t_1<\cdots<t_n=T\}\) as a partition of \([0,T]\) with mesh
\(
\|\pi_n\|:=\max_{0\le r\le n-1 }(t_{r+1}-t_r)\to 0\) as \( n\to \infty\).
\begin{definition}[Continuous and discrete signature components]
We denote by
$
S(\tilde X)_t^{\Gamma,\mathrm{I}}
$ and $
S(\tilde X)_t^{\Gamma,\mathrm{S}}
$
the continuous It\^o and Stratonovich signature components at time \(t\), respectively, defined recursively by
\(
S(\tilde X)_t^{\emptyset,\mathrm{I}}
=
S(\tilde X)_t^{\emptyset,\mathrm{S}}
\equiv 1,
\)
and for \(\Gamma=(i_1,\dots,i_m)=(\Gamma^-,i_m)\), they admit the iterated-integral representations
\[
S(\tilde X)_t^{\Gamma,\mathrm{I}}
=
\int_{0<t_1<\cdots<t_m<t}
d\tilde X_{t_1}^{i_1}\cdots d\tilde X_{t_m}^{i_m}
=\int_0^t S(\tilde X)_u^{(i_1,\dots,i_{m-1}),\mathrm{I}}\,d\tilde X_u^{i_m},
\]
and
\[
S(\tilde X)_t^{\Gamma,\mathrm{S}}
=
\int_{0<t_1<\cdots<t_m<t}
d\tilde X_{t_1}^{i_1}\circ\cdots\circ d\tilde X_{t_m}^{i_m}
=\int_0^t S(\tilde X)_u^{(i_1,\dots,i_{m-1}),\mathrm{S}}\circ d\tilde X_u^{i_m}.
\]

For \(k=0,\dots,n\), we denote by $
S(\tilde X)_{t_k}^{\Gamma,\mathrm{I},\pi_n}
$
the corresponding discrete It\^o-signature coordinate with partition $\pi_n$, defined recursively by
\(
S(\tilde X)_{t_k}^{\emptyset,\mathrm{I},\pi_n}:=1,
\)
and for \(\Gamma=(i_1,\dots,i_m)=(\Gamma^-,i_m)\),
\[
S(\tilde X)_{t_k}^{\Gamma,\mathrm{I},\pi_n}
:=
\sum_{r=0}^{k-1}
S(\tilde X)_{t_r}^{\Gamma^-,\mathrm{I},\pi_n}
\bigl(\tilde X_{t_{r+1}}^{i_m}-\tilde X_{t_r}^{i_m}\bigr).
\]
\end{definition}
For a fixed order \(m\), there are \((d+1)^m\) signature coordinates of level \(m\). The number of terms therefore grows exponentially with the order, which makes truncation essential in applications.

Throughout the paper,  \(\mathrm I\) and \(\mathrm S\) always stand for It\^o and Stratonovich, respectively. The It\^o signature aligns naturally with non-anticipative trading and self-financing strategies, and will be our  focus in this section. Later in
Section~\ref{subsec:ito_universality},  we will exploit the universal
nonlinearity of the Stratonovich signature to develop analogous  It\^o-signature representations.

\subsection{The Tradability Theorem: Discretized Signature as PnL of Tradable Strategy}

The main result of this subsection is that every discretized It\^o-signature coordinate is either the cumulative profit and loss (PnL) of an adapted self-financing strategy or a cash position. In other words, signature coordinates are more than nonlinear path features, they form tradable basis elements for hedge construction.

\begin{theorem}[Discretized It\^o signature as tradable PnL]\label{thm:hedging strategy}
For any multi-index
\(
\Gamma=(i_1,\dots,i_m)\in\{1,\dots,d+1\}^m
\)
and any \(k\in\{1,\dots,n\}\), define
\(
r(\Gamma):=\max\{q\in\{1,\dots,m\}: i_q\le d\}
\)
whenever \(\Gamma\) contains at least one risky-asset coordinate. Then there exist \(\mathcal F_{t_j}\)-measurable trading positions \(\{\theta_{j,k}^{\Gamma,\pi_n}\}_{j=0}^{k-1}\) such that
\[
S(\tilde X)_{t_k}^{\Gamma,\mathrm I,\pi_n}
=
\sum_{j=0}^{k-1}\theta_{j,k}^{\Gamma,\pi_n}
\bigl(X_{t_{j+1}}^{\,i_{r(\Gamma)}}-X_{t_j}^{\,i_{r(\Gamma)}}\bigr).
\]
Moreover, if \(\Gamma\) contains no risky-asset coordinates, that is, if
\(
i_1=\cdots=i_m=d+1,
\)
then \(S(\tilde X)_{t_k}^{\Gamma,\mathrm I,\pi_n}\) is deterministic for a fixed partition and represents a cash position.
\end{theorem}

Theorem~\ref{thm:hedging strategy} gives each discretized It\^o-signature component a direct financial interpretation. If the multi-index contains at least one risky-asset coordinate, it represents the cumulative PnL of an adapted self-financing strategy on the underlying assets and cash; if it contains only time coordinates, then it represents a cash position. The construction is non-anticipative, uses no auxiliary derivatives or latent state variables, and is implemented directly on the observed trading grid. This last feature is important in practice because dynamic hedging takes place in discrete time, and the resulting strategy fits naturally into discrete-time hedging frameworks \citep{schweizer1995variance,bertsimas2001hedging,remillard2013optimal}.

Another distinctive feature of the construction is its recursive structure. Higher-order signature strategies are built from lower-order trading gains, the PnL of a lower-order signature strategy is the position for a higher-order signature strategy. In this sense, the hedge does not arise from a black-box mapping from the path to a trading position. Instead, it arises from a linear combination of basis strategies, each of which has a direct dynamic interpretation.

The trading representation in Theorem~\ref{thm:hedging strategy} is the main distinction between our framework and existing signature-based hedging approaches, in which signature terms are only used as nonlinear regressors \citep{lyons2020non,abijaber2025b,cirone2025roughkernelhedging}. 

We now give a simple example to illustrate the mechanism behind Theorem~\ref{thm:hedging strategy}.
In this example, the regressors themselves carry trading content: once the payoff is fitted by a linear combination of discretized It\^o-signature coordinates, the same linear combination can be implemented through the associated self-financing strategies.

\begin{example}\label{ex3}
Consider two risky assets \(S\) and \(V\). Let \(T=1\), and use the uniform
partition
\(
\pi_n=\left\{t_i=\frac{i}{n}: i=0,1,\dots,n\right\}
\)
of \([0,1]\). Consider the time-augmented path
\(
\tilde X_t=(S_t,V_t,t)\in\mathbb R^3.
\)
Assume that the derivative payoff is approximated by
\[
f(S,V)
\approx
l_0
+
l_1\,S(\tilde X)_{t_n}^{(1),\mathrm I,\pi_n}
+
l_2\,S(\tilde X)_{t_n}^{(1,2),\mathrm I,\pi_n}
+
l_3\,S(\tilde X)_{t_n}^{(1,3),\mathrm I,\pi_n}.
\]
Here the It\^o signature components correspond to simple trading rules on the
uniform grid. The term \(
S(\tilde X)_{t_n}^{(1),\mathrm I,\pi_n}=S_T-S_0\)
is generated by holding one share of asset \(S\). The term
\(
S(\tilde X)_{t_n}^{(1,2),\mathrm I,\pi_n}
=
\sum_{i=0}^{n-1}
S(\tilde X)_{t_i}^{(1),\mathrm I,\pi_n}
\bigl(V_{t_{i+1}}-V_{t_i}\bigr)
\)
is generated by holding \(S(\tilde X)_{t_i}^{(1),\mathrm I,\pi_n}\) shares of
asset \(V\) over \([t_i,t_{i+1}]\). The term
\(S(\tilde X)_{t_n}^{(1,3),\mathrm I,\pi_n}
=
\sum_{i=0}^{n-1}
\frac{n-i-1}{n}
\bigl(S_{t_{i+1}}-S_{t_i}\bigr)
\)
is generated by holding \(\frac{n-i-1}{n}\) shares of asset \(S\) over
\([t_i,t_{i+1}]\).
Combining the three components, the hedging portfolio starts from initial cash \(c=l_0\) and over
each interval \([t_i,t_{i+1}]\) holds
\(
l_1+l_3\frac{n-i-1}{n}
\)
shares of asset \(S\) and
\(
l_2\,S(\tilde X)_{t_i}^{(1),\mathrm I,\pi_n}
\)
shares of asset \(V\). At maturity, the terminal wealth equals
\(l_0
+
l_1\,S(\tilde X)_{t_n}^{(1),\mathrm I,\pi_n}
+
l_2\,S(\tilde X)_{t_n}^{(1,2),\mathrm I,\pi_n}
+
l_3\,S(\tilde X)_{t_n}^{(1,3),\mathrm I,\pi_n},
\)
which approximates the target payoff \(f(S,V)\).
\end{example}

In fact, the trading positions can be explicitly given.   To illustrate, consider for ease of exposition a simple  setting: one risky asset, unit time horizon, and a uniform
trading grid.

\begin{example}
    [Explicit strategy under a uniform partition]\label{ex:specialstrategy}
Under the uniform partition \(
\pi_n=\left\{t_l=\frac{l}{n}: l=0,1,\dots,n\right\}
\), assume \(T=1\) and \(d=1\), and the market contains a single risky asset \(X\) with time-augmented path
\(
\tilde X_t=(X_t,t)\in\mathbb R^2.
\)
Let
\(
\Gamma=(i_1,\dots,i_m)\in\{1,2\}^m
\)
satisfy
\[
i_m=\cdots=i_{q+1}=2,\qquad i_q=1,\qquad 1\le q\le m-1.
\]
Define
\(
\Gamma_0=(i_1,\dots,i_{q-1}).
\)
Then for every \(k=1,\dots,n\),
\(
S(\tilde X)_{t_k}^{\Gamma,\mathrm I,\pi_n}
=
\sum_{l=0}^{k-1} c_{k,l}^{\Gamma,\pi_n}\bigl(X_{t_{l+1}}-X_{t_l}\bigr),
\)
where
\[
c_{k,l}^{\Gamma,\pi_n}
=
\frac{\binom{k-1-l}{m-q}}{n^{m-q}}\,
S(\tilde X)_{t_l}^{\Gamma_0,\mathrm I,\pi_n},
\qquad 0\le l\le k-1,
\]
with the convention \(\binom{a}{b}=0\) whenever \(a<b\).
\end{example}
Note that  the binomial coefficient in
the trading weight is specific to the uniform partition, more general
deterministic partitions lead to analogous recursive strategies but not to this
simple binomial expression.

\subsection{Interpretable Hedging Framework using It\^o Signature}

\begin{figure}[htbp]
    \centering
    \includegraphics[width=\textwidth]{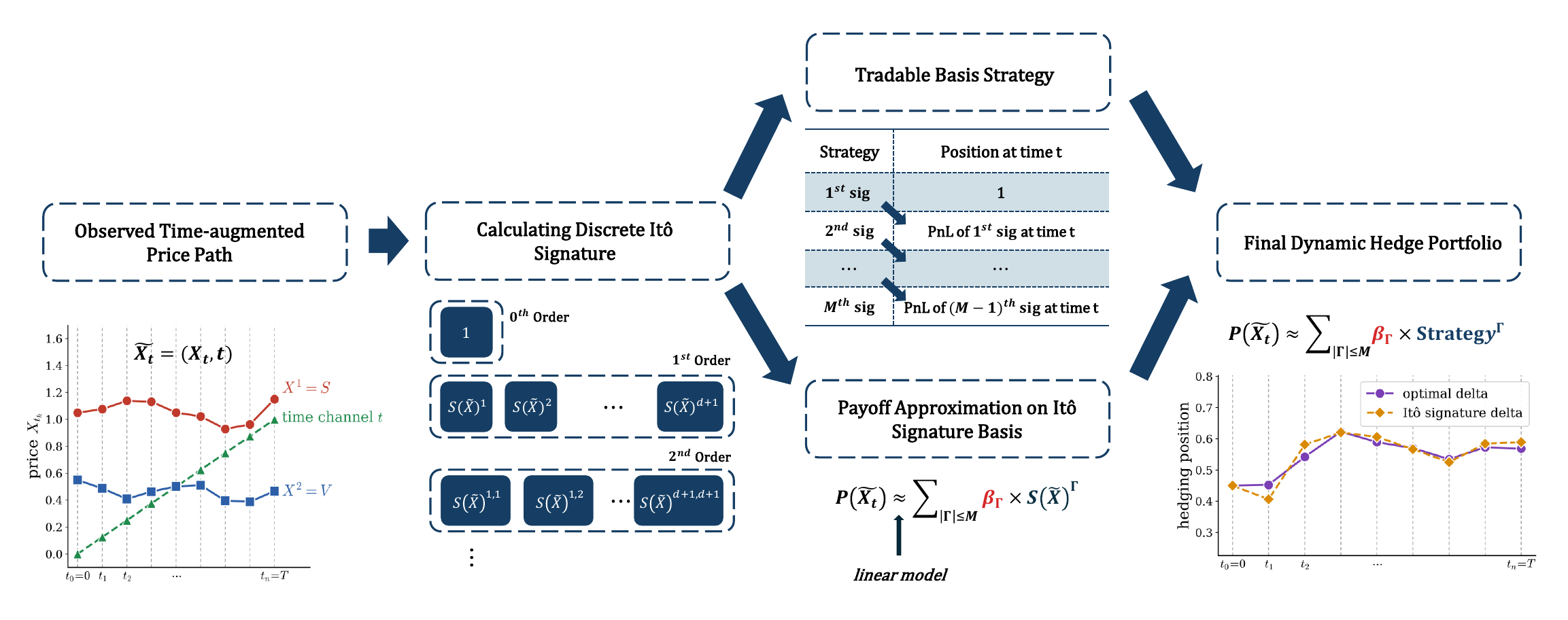}
    \caption{\textbf{It\^o-signature hedging framework.}
    Starting from the time-augmented price path \(\tilde{X}=(X,t)\), the framework first approximates the derivative payoff by a linear combination of discretized It\^o signature terms. Theorem~\ref{thm:hedging strategy} then maps each term into a self-financing basis strategy. The final hedge takes the same linear combination of these basis strategies with coefficients \(\beta\).}
    \label{fig:framework_flowchart}
\end{figure}

Theorem~\ref{thm:hedging strategy} leads to a natural hedging framework where the same set of discretized It\^o-signature coordinates is used for both payoff approximation and hedge implementation. This dynamic hedging as a single linear system is summarized in Framework~\ref{fram:0}. This framework also works on the observed trading grid and uses only the underlying assets and cash, which makes it directly implementable in discrete time. Figure~\ref{fig:framework_flowchart} summarizes this logic.

\begin{algorithm}[htbp]
\caption{It\^o Signature Hedging Framework}
\label{fram:0}
\begin{algorithmic}[0]
\STATE \textbf{Step 1: Payoff Approximation}
\STATE Approximate the derivative payoff \(P(\tilde{X})\) by a linear combination of discretized It\^o-signature coordinates:
\begin{equation}\label{eq:discretelinear}
    P(\tilde{X})
    \approx
    \beta_0
    +
    \sum_{r=1}^{m}\;
    \sum_{|\Gamma|=r}
    \beta_{\Gamma}\,
    S(\tilde X)_{t_n}^{\Gamma,\mathrm{I},\pi_n}.
\end{equation}
Estimate the coefficients \(\beta\) with a linear method such as ordinary least squares (OLS) or Lasso.

\STATE \textbf{Step 2: Basis-Strategy Construction}
\STATE For each nonconstant discrete It\^o-signature coordinate \(S(\tilde X)_{t_n}^{\Gamma,\mathrm{I},\pi_n}\) appearing in \eqref{eq:discretelinear}, construct the associated self-financing trading strategy using Theorem~\ref{thm:hedging strategy}.

\STATE \textbf{Step 3: Hedge Assembly}
\STATE Form the final hedging portfolio by taking the same linear combination of the basis strategies as in Step 1, with weights given by the estimated coefficients \(\beta\).
\end{algorithmic}
\end{algorithm}


\section{Theoretical Analysis of Population Hedging Error}\label{sec: theoretical analysis}

The population-level performance of the implementable hedge in Equation~\eqref{eq:discretelinear} is governed by two sources of error. The first is a discretization error, which arises from replacing continuous-time It\^o-signature coordinates by their tradable discrete-time counterparts. The second is a population approximation error, which measures how well the target payoff can be represented by a finite-order continuous-time It\^o-signature expansion.

This section studies these two errors in the It\^o-signature hedging framework,
and combines them into an implementable \(\varepsilon\)-hedging statement. The main result (Theorem~\ref{thm:approx_hedging}) shows that discrete It\^o-signature portfolios can approximate a target payoff arbitrarily close in the \(L^2\) sense.
We emphasize that our hedging procedure is model-free in implementation, as  the hedge is fitted
directly from discretized It\^o-signature features of observed price paths.
The diffusion setup introduced below is used only to derive the theoretical error bounds, useful for subsequent analysis in  Sections \ref{sec: theoretical analysis}--\ref{sec:statistical_learning}.

\subsection{Asset Dynamics}\label{subsec:assetdynamic}

Given a filtered probability space
\(
(\Omega,\mathcal F,\mathbb P,\{\mathcal F_t\}_{t\in[0,T]})
\)
and a \(d\)-dimensional Brownian motion \(B=\{B_t\}_{t\in[0,T]}\) adapted to \(\{\mathcal F_t\}\), consider a market with \(d\) tradable assets, whose price process \(X=\{X_t\}_{t\in[0,T]}\) takes values in \(\mathbb R^d\) and evolves according to the diffusion
\begin{equation}\label{eq:asset_dynamics_section1}
    \mathrm{d}X_t=\mu(X_t,t)\,\mathrm{d}t+\sigma(X_t,t)\,\mathrm{d}B_t,
    \qquad
    X_0=x_0\in\mathbb R^d,
\end{equation}
where \(\mu:\mathbb R^d\times[0,T]\to\mathbb R^d\) and \(\sigma:\mathbb R^d\times[0,T]\to\mathbb R^{d\times d}\) satisfy appropriate technical conditions to ensure the well-posedness, as below.

\begin{assumption}[\textbf{Global Lipschitz continuity}]\label{ass:1}
There exists a constant \(L>0\) such that for all \(x,y\in\mathbb R^d\) and all \(t\in[0,T]\),
$
|\mu(x,t)-\mu(y,t)|+\|\sigma(x,t)-\sigma(y,t)\|
\le L|x-y|.$
\end{assumption}

\begin{assumption}[\textbf{Linear growth}]\label{ass:2}
There exists a constant \(K>0\) such that for all \(x\in\mathbb R^d\) and all \(t\in[0,T]\),
$
|\mu(x,t)|+\|\sigma(x,t)\|
\le K(1+|x|).
$
\end{assumption}

These assumptions serve  two purposes. First, they guarantee that the continuous-time trading environment is well posed, so that the self-financing interpretation of signature-based portfolios is meaningful. Second, they provide the moment and continuity properties needed for the discrete-to-continuous approximation results established below. To capture time-inhomogeneous trading effects, we consider the time-augmented process
\(
\tilde X_t=(X_t,t)\in\mathbb R^{d+1} \) where \(t\in[0,T]\) as in Equation \eqref{eq:timeaug}.



\subsection{Convergence of Discrete-time Approximations}\label{subsec:discrete_to_continuous}

The purpose of this subsection is to connect the implementable hedge in discrete time to the continuous-time trading object. We establish the discrete-to-continuous convergence result as follows.

\begin{theorem}[\(L^p\) convergence of discrete It\^o signature]
\label{thm:lpconvergence}
Under Assumptions~\ref{ass:1}--\ref{ass:2}, for every multi-index \(\Gamma\) and every \(p\in[1,\infty)\),
\[
\max_{0\le k\le n}
\left\|
S(\tilde X)_{t_k}^{\Gamma,\mathrm I,\pi_n}
-
S(\tilde X)_{t_k}^{\Gamma,\mathrm I}
\right\|_{L^p(\Omega,\mathcal F,\mathbb P)}
\longrightarrow 0
\qquad\text{as } \|\pi_n\|\to0.
\]
Here the \(L^p\)-distance
\(
d_p(Y,Z)
:=
\|Y-Z\|_{L^p(\Omega,\mathcal F,\mathbb P)}
=
\bigl(\mathbb E^{\mathbb P}|Y-Z|^p\bigr)^{1/p}.
\)
In particular,
\[
S(\tilde X)_{t_n}^{\Gamma,\mathrm I,\pi_n}
\xrightarrow{L^p(\mathbb P)}
S(\tilde X)_{T}^{\Gamma,\mathrm I}.
\]
\end{theorem}

Theorem~\ref{thm:lpconvergence} has a clear financial interpretation, and isolates the approximation error induced by trading on a discrete grid: as the trading frequency increases, the implementable discrete signature hedge approaches the continuous-time signature representation.
It is stated in the \(L^p\) form. In particular, the \(L^2\) case matches the mean-square hedging criteria, and ensures that the terminal payoff generated by the discrete signature feature converges in quadratic mean to its continuous-time counterpart (See for example, Theorem \ref{thm:approx_hedging}).

The proof of the convergence result relies on three ingredients: a uniform moment bound for continuous and discrete It\^o-signature coordinates, an \(L^p\)-continuity estimate for the continuous It\^o signature, and a continuity property of stochastic integration with respect to the augmented diffusion \(\tilde X\). We collect these results in Appendix~\ref{appendix:subseclemmas}.

\subsection{Universal Approximation for It\^o Signature}\label{subsec:ito_universality}

We now study the universal approximation of continuous It\^o signatures. The standard universal nonlinearity theorem is naturally developed for the
Stratonovich signature, because Stratonovich iterated integrals preserve the algebraic structure needed for Stone--Weierstrass type arguments \citep{lyons2022signature,cuchiero2023signature}. Our hedging framework, however, is built from the It\^o signatures, which are aligned with non-anticipative trading gains. This subsection is to connect these two objects.

We first recall the universal nonlinearity of the Stratonovich signature.

\begin{theorem}[Universal nonlinearity, {\citet[Theorem 2.12]{cuchiero2023signature}}]\label{th:UN}
Let \(\mathcal K\) be a compact subset of time-augmented paths on \([0,T]\), and let \(f:\mathcal K\to\mathbb R\) be continuous. Then, for every \(\varepsilon>0\), there exists a linear functional \(\ell\in T(\mathbb R^{d+1})\) such that
\[
\sup_{\gamma\in\mathcal K}
\left|
f(\gamma)-\ell\bigl(S(\gamma)_T^{\mathrm S}\bigr)
\right|<\varepsilon,
\]
where the tensor algebra
\(
T(\mathbb R^{d+1})
:=
\bigcup_{m\ge0}T^{(m)}(\mathbb R^{d+1})
\) is the union of finite-level tensor spaces.
\end{theorem}

To transfer this approximation result to the It\^o setting, we use a
volatility-normalizing transformation. This gives a unified treatment of both the constant-diffusion benchmark and a broader class of state-dependent diffusions. Consider the diffusion
\begin{equation}\label{eq:general_diffusion_un}
    \mathrm{d}X_t=\mu(X_t,t)\,\mathrm{d}t+\sigma(X_t,t)\,\mathrm{d}B_t,
    \qquad
    X_0=x_0\in D,
\end{equation}
where \(D\subset\mathbb R^d\) is an open convex state domain. The goal is to construct a deterministic transformation \(g\) such that the transformed process has identity diffusion matrix. This turns the transformed state process into a constant-diffusion process, to which the It\^o--Stratonovich conversion argument can be applied.

To implement this idea, we impose the following structural conditions on the diffusion matrix.

\begin{assumption}[Invertibility and regularity of the diffusion matrix]\label{ass:ito_un_1}
For every \((x,t)\in D\times[0,T]\), the matrix \(\sigma(x,t)\) is invertible. Moreover, \(\sigma^{-1}\) is continuously differentiable in \((x,t)\).
\end{assumption}

\begin{assumption}[Gradient symmetry]\label{ass:ito_un_2}
For each fixed \(t\in[0,T]\) and each \(i=1,\dots,d\), the \(i\)-th row of \(\sigma(x,t)^{-1}\) is a conservative vector field in \(x\) on \(D\). Equivalently, for all \(j,k=1,\dots,d\),
\[
\partial_{x_k}\bigl(\sigma^{-1}(x,t)\bigr)_{ij}
=
\partial_{x_j}\bigl(\sigma^{-1}(x,t)\bigr)_{ik},
\qquad (x,t)\in D\times[0,T].
\]
\end{assumption}

\begin{assumption}[Coordinate-adjusted positive definiteness]\label{ass:ito_un_3}
For each \(t\in[0,T]\), there exists an invertible matrix
\(A(t)\in\mathbb R^{d\times d}\) such that, for every \(x\in D\),
\[
\frac{
A(t)\sigma(x,t)^{-1}
+
\sigma(x,t)^{-\top}A(t)^\top
}{2}
\succ 0.
\]
\end{assumption}

\begin{remark}\label{rem:examples}
Assumptions~\ref{ass:ito_un_1}--\ref{ass:ito_un_3} are satisfied by several standard classes of diffusion models.

\begin{itemize}
    \item \textbf{Constant-volatility diffusions.}
    Suppose \(\sigma(x,t)\equiv \Sigma(t)\), where \(\Sigma(t)\) is invertible for every \(t\in[0,T]\). This class includes Bachelier-type normal diffusions \citep{bachelier1900theorie,abijaber2025frictions}, multivariate Ornstein--Uhlenbeck models \citep{vasicek1977equilibrium}, and Gaussian short-rate models \citep{hull1990pricing}. The row fields of \(\sigma(x,t)^{-1}=\Sigma(t)^{-1}\) are constant in \(x\), so Assumption~\ref{ass:ito_un_2} holds automatically. A normalizing transformation is \(g(x,t):=\Sigma(t)^{-1}x\), which satisfies \(\nabla_x g(x,t)\sigma(x,t)=I_d\). Moreover, Assumption~\ref{ass:ito_un_3} holds by taking \(A(t):=\Sigma(t)\), since \(A(t)\sigma(x,t)^{-1}=I_d\).

    \item \textbf{One-dimensional diffusions.}
    In dimension \(d=1\), Assumption~\ref{ass:ito_un_2} is automatic. Consider \(\mathrm dX_t=\mu(X_t,t)\,\mathrm dt+s(X_t,t)\,\mathrm dB_t\), where \(s(x,t)>0\). A Lamperti-type normalizing transformation is \(g(x,t):=\int_{x^\ast}^{x}s(u,t)^{-1}\,\mathrm du\), so that \(\partial_x g(x,t)s(x,t)=1\). Assumption~\ref{ass:ito_un_3} holds with \(A(t)=1\). This class includes constant-elasticity-of-variance (CEV) type
    diffusions \(s(x)=\eta x^\alpha\) \citep{cox1975notes} and the Cox--Ingersoll--Ross (CIR) square-root diffusion \(s(x)=\eta\sqrt{x}\) \citep{cox1985intertemporal} as special cases.

    \item \textbf{Multi-dimensional geometric Brownian motion.}
    Consider \(\mathrm dX_t=\mathrm{diag}(X_t)(\mu\,\mathrm dt+\Sigma\,\mathrm dB_t)\) on \(D=(0,\infty)^d\), where \(\Sigma\) is invertible \citep{black1973pricing,merton1973theory}. Then \(\sigma(x)=\mathrm{diag}(x)\Sigma\), and \(\sigma(x)^{-1}=\Sigma^{-1}\mathrm{diag}(x_1^{-1},\dots,x_d^{-1})\). The row fields of \(\sigma(x)^{-1}\) are conservative, with normalizing transformation \(g(x)=\Sigma^{-1}(\log x_1,\dots,\log x_d)^\top\). Moreover, Assumption~\ref{ass:ito_un_3} holds by taking \(A=\Sigma\), since \(A\sigma(x)^{-1}=\mathrm{diag}(x_1^{-1},\dots,x_d^{-1})\succ0\).

    \item \textbf{Multi-asset separable local-volatility models.}
    Consider \[\mathrm dX_t=\mu(X_t,t)\,\mathrm dt+\mathrm{diag}(s_1(X_{1,t},t),\dots,s_d(X_{d,t},t))\Sigma(t)\,\mathrm dB_t,\] where each \(s_i(x_i,t)>0\) and \(\Sigma(t)\) is invertible \citep{dupire1994pricing,rubinstein1994implied}. Then \(\sigma(x,t)^{-1}=\Sigma(t)^{-1}\mathrm{diag}(s_1(x_1,t)^{-1},\dots,s_d(x_d,t)^{-1})\). Define \(h_i(x_i,t):=\int_{x_i^\ast}^{x_i}s_i(u,t)^{-1}\,\mathrm du\), \(h(x,t):=(h_1(x_1,t),\dots,h_d(x_d,t))^\top\), and \(g(x,t):=\Sigma(t)^{-1}h(x,t)\). Then \(\nabla_x g(x,t)\sigma(x,t)=I_d\), and Assumption~\ref{ass:ito_un_3} holds with \(A(t):=\Sigma(t)\).
\end{itemize}

\end{remark}

By Lemma~\ref{lem:global_transform}, Assumptions~\ref{ass:ito_un_1}--\ref{ass:ito_un_3}
imply the existence of a deterministic map
\(
g\in C^{2,1}(D\times[0,T];\mathbb R^d)
\)
such that
\(
\nabla_x g(x,t)\sigma(x,t)=I_d.
\)
Define the induced path transformation
\[
\Phi_g(\tilde x)_t:=(g(x_t,t),t),\qquad t\in[0,T].
\]
The transformed It\^o signature components appearing below are therefore It\^o signature components of the volatility-normalized path \(\Phi_g(\tilde x)\). We now show the universal approximation theorem for the It\^o signature.

\begin{theorem}[Finite-order universal approximation for transformed It\^o signature]
\label{thm:ito_un_general}
Suppose Assumptions~\ref{ass:1}--\ref{ass:ito_un_3} hold. Let \(D\subset\mathbb R^d\) be the state domain specified in the assumptions, and
\(\tilde{\mathcal K}\subset C([0,T],\mathbb{R}^{d+1})\) be compact under the uniform path metric \(d_\infty(\tilde x,\tilde y):=\sup_{0\le t\le T}|\tilde x_t-\tilde y_t|\). Then, for every continuous function \(f\in C(\tilde{\mathcal K},\mathbb{R})\) and every \(\varepsilon>0\), there exist an integer \(m\ge1\) and coefficients \(\{\beta_{\Lambda}\}_{|\Lambda|\le m}\) such that
\[
\sup_{\tilde x\in\tilde{\mathcal K}}
\left|
f(\tilde x)-\sum_{|\Lambda|\le m}\beta_{\Lambda}\,S(\Phi_g(\tilde x))_T^{\Lambda,\mathrm I}
\right|
<\varepsilon.
\]
The approximation error is measured by the uniform norm \(\|h\|_{\infty,\tilde{\mathcal K}}:=\sup_{\tilde x\in\tilde{\mathcal K}}|h(\tilde x)|\).
\end{theorem}

Theorem~\ref{thm:ito_un_general} is valuable because it shows that the universal approximation of the It\^o signature is not confined to the constant-diffusion benchmark: after an appropriate state transformation, the same finite-order signature representation can cover a substantially broader class of path-dependent payoffs. To translate this approximation result into a direct hedging statement, one needs the transformed state process \(g(X)\) to be tradable or replicable through available contracts. Once \(g(X)\) is tradable, the earlier self-financing construction applies to the It\^o signature of the transformed process as well, so the transformed signature coordinates can again serve as hedge building blocks. For example, in the multi-asset geometric Brownian motion (GBM) case, \(g(X)=\Sigma^{-1}\log X\) is a linear combination of log-prices, which are closely related to log-contract baskets or variance-swap dispersion exposures \citep{neuberger1994log,carr2009volatility}.

The constant-diffusion benchmark is an important special case in which this additional tradability issue disappears. If the diffusion matrix is a constant invertible matrix \(\Sigma\), then the fixed linear transformation \(Y=\Sigma^{-1}X\) reduces the model to the identity-diffusion case, so the It\^o signature is taken directly over the tradable price process \(X\). Thus, for constant invertible diffusion matrices, the approximation result and the self-financing interpretation are directly aligned. We state this tradable special case separately because it is the setting used in the hedging-error results below.

\begin{corollary}[Finite-order universal approximation for It\^o signature under constant diffusion]
\label{cor:ito_un_constant}
Consider the normalized constant-diffusion benchmark
\(\mathrm dX_t=b(X_t,t)\,\mathrm dt+\mathrm dW_t,\)
\(X_0=x_0\in\mathbb R^d,
\)
and let \(\tilde X_t=(X_t,t)\). Let \(\tilde{\mathcal K}\subset C([0,T],\mathbb{R}^{d+1})\) be compact under the uniform path metric \(d_\infty(\tilde x,\tilde y):=\sup_{0\le t\le T}|\tilde x_t-\tilde y_t|\). Then, for every continuous function \(f\in C(\tilde{\mathcal K},\mathbb{R})\) and every \(\varepsilon>0\), there exist an integer \(m\ge1\) and coefficients \(\{\beta_{\Lambda}\}_{|\Lambda|\le m}\) such that
\[
\sup_{\tilde x\in\tilde{\mathcal K}}
\left|
f(\tilde x)-\sum_{|\Lambda|\le m}\beta_{\Lambda}\,S(\tilde x)_T^{\Lambda,\mathrm I}
\right|
<\varepsilon.
\]
The approximation error is measured by the uniform norm \(\|h\|_{\infty,\tilde{\mathcal K}}:=\sup_{\tilde x\in\tilde{\mathcal K}}|h(\tilde x)|\).
\end{corollary}

Corollary~\ref{cor:ito_un_constant} is the approximation result used in the
hedging-error analysis below. In this benchmark, the signature coordinates are computed from the tradable price process itself, so the approximation theorem can be combined directly with the discrete tradability result in
Theorem~\ref{thm:hedging strategy}.
\subsection{\texorpdfstring{\(\varepsilon\)-Hedging Error Control}{Epsilon-Hedging Error Control}}
\label{subsec:hedging_optimality}
The preceding corollary identifies the constant-diffusion benchmark as the case where the approximation result is written directly in terms of It\^o signatures of the tradable price process. We now combine this approximation result with the tradability of discrete It\^o signatures to obtain an implementable \(\varepsilon\)-hedging statement. As established in Section~\ref{sec:framework}, each discrete It\^o-signature coordinate is the terminal gain of a self-financing trading strategy, up to deterministic cash terms. Hence any finite linear combination of
such coordinates is also implementable as a self-financing portfolio.

In the spirit of the \(\varepsilon\)-arbitrage
terminology of \citet{bertsimas2001hedging}, the goal of this subsection is to show that the proposed
signature construction gives an \(\varepsilon\)-hedge: for every
\(\varepsilon>0\), one can construct a self-financing signature portfolio whose
terminal hedging error is smaller than \(\varepsilon\) in the \(L^2\) sense. 


Fix \(q>0\). Let \(\mathcal L(X)\) denote the class of predictable \(X\)-integrable processes, and define
\[
\mathcal H^q(X)
:=
\left\{
(p_0,\theta)\in \mathbb R\times \mathcal L(X):
\;
p_0+\int_0^T \theta_t^\top\,\mathrm dX_t \in L^q
\right\}.
\]
For \((p_0,\theta)\in\mathcal H^q(X)\), the corresponding terminal wealth is
\[
\mathcal W_T(p_0,\theta)
=
p_0+\int_0^T \theta_t^\top\,\mathrm dX_t.
\]

\begin{theorem}[Approximate hedging by discrete It\^o-signature portfolios]
\label{thm:approx_hedging}
Assume that the underlying process is the constant-diffusion benchmark, and let \(\tilde X\) denote its time-augmented path. Let
\(
F:\tilde{\mathcal K}\to\mathbb R
\)
be a continuous payoff functional defined on a compact set \(\tilde{\mathcal K}\subset C([0,T],\mathbb R^{d+1})\), and assume that
\(
\mathbb P(\tilde X\in\tilde{\mathcal K})=1.
\)
Let \((\pi_n)_{n\ge1}\) be any deterministic sequence of partitions of
\([0,T]\) such that \(\|\pi_n\|\to0\).
Then, for every \(\varepsilon>0\), there exist an integer \(m\ge1\), a partition $\pi_n$ with sufficiently small mesh, and coefficients
\(
\beta=\{\beta_{\Gamma}\}_{|\Gamma|\le m},
\)
such that the discrete signature portfolio
\[
H_T^{(m,\pi_n)}(\beta)
:=
\beta_0+\sum_{1\le |\Gamma|\le m}\beta_{\Gamma}\,
S(\tilde X)_{T}^{\Gamma,\mathrm I,\pi_n}
\]
is the terminal wealth of a strategy in \(\mathcal H^q(X)\) and satisfies
\(
\mathbb E^{\mathbb P}\Bigl[\bigl|F(\tilde X)-H_T^{(m,\pi_n)}\bigr|^2\Bigr]<\varepsilon.
\)

More precisely, there exists
\(
(p_0^{m,n},\theta^{m,n})\in\mathcal H^q(X)
\)
such that
\(
H_T^{(m,\pi_n)}(\beta)
=
\mathcal W_T(p_0^{m,n},\theta^{m,n}),
\)
and
\(\mathbb E^{\mathbb P}\left[
\bigl|
F(\tilde X)-\mathcal W_T(p_0^{m,n},\theta^{m,n})
\bigr|^2\right]<\varepsilon.
\)
Here the hedging error is measured by the squared \(L^2\)-distance.
\end{theorem}

Theorem~\ref{thm:approx_hedging} shows that the discrete It\^o-signature construction yields an \(\varepsilon\)-hedge for any prescribed \(\varepsilon>0\). The linear signature expansion first approximates the payoff, and the tradability result then identifies the same expansion with the terminal wealth of a self-financing strategy. 
We next show that the same construction yields a small expected loss under a broad class of economically meaningful loss criteria.

\begin{corollary}[\(\varepsilon\)-hedging under Lipschitz penalties]
\label{cor:near_optimal_signature}
Under the setting of Theorem~\ref{thm:approx_hedging}, let
\(
P:\mathbb R\to\mathbb R_+
\)
be a nonnegative Lipschitz penalty function with \(P(0)=0\). Then, for every \(\eta>0\), there exists a signature-based strategy
\(
(p_0^\eta,\theta^\eta)\in\mathcal H^q(X)
\)
such that
\[
\mathbb E^{\mathbb P}\Bigl[
P\Bigl(
F(\tilde X)-\mathcal W_T(p_0^\eta,\theta^\eta)
\Bigr)
\Bigr]
<\eta.
\]
Here performance is measured by the expected penalty loss under \(\mathbb P\).
    
\end{corollary}

Corollary~\ref{cor:near_optimal_signature} shows that the proposed It\^o-signature hedging framework produces a computationally tractable self-financing strategy whose expected hedging loss under any nonnegative Lipschitz penalty can be made arbitrarily small. The Lipschitz condition covers, for example, absolute hedging error and capped or piecewise linear shortfall-type penalties, which are often used to control downside exposure \citep{follmer1999quantile,follmer2000efficient}. For more general loss criteria, \citet{lyons2020non} studies the  Stratonovich signatures for hedging in a range of settings, including exponential utility, transaction costs, and liquidity constraints.

Note that quantile-based risk measures such as VaR are determined by tail
probabilities instead of Lipschitz transformations of the hedging error.
 We next present a complementary result to Corollary~\ref{cor:near_optimal_signature} for Value-at-Risk (VaR) and Conditional Value-at-Risk (CVaR), two standard tail-risk measures in financial risk management
\citep{rockafellar2000optimization,basak2001value,rockafellar2002conditional}.

\begin{corollary}[Control of VaR and CVaR of terminal hedging losses]
\label{cor:var_cvar_hedging_loss}
Under the setting of Theorem~\ref{thm:approx_hedging}, define the terminal hedging loss by
\(L_T(p_0,\theta):=|F(\tilde X)-\mathcal W_T(p_0,\theta)|\). For a loss random variable \(L\) and \(\alpha\in(0,1)\), define
\(\operatorname{VaR}_{\alpha}(L):=\inf\{x\in\mathbb R:\mathbb P(L\le x)\ge\alpha\}\)
and
\(\operatorname{CVaR}_{\alpha}(L):=\inf_{z\in\mathbb R}\{z+(1-\alpha)^{-1}\mathbb E^{\mathbb P}[(L-z)_+]\}\).
Then, for any fixed \(\alpha\in(0,1)\) and any \(\delta>0\), there exists a
signature-based strategy \((p_0,\theta)\in\mathcal H^q(X)\) such that
\[
\operatorname{VaR}_{\alpha}(L_T(p_0,\theta))<\delta
\quad\text{and}\quad
\operatorname{CVaR}_{\alpha}(L_T(p_0,\theta))<\delta .
\]
\end{corollary}

Although this corollary is stated for the two-sided terminal hedging loss
\(L_T(p_0,\theta)\), it also covers the terminal shortfall loss
\((F(\tilde X)-\mathcal{W}_T(p_0,\theta))_+\), since this loss is bounded above by
\(L_T(p_0,\theta)\). 

Now  we see that at the population level, finite discrete It\^o-signature portfolios can approximate the target payoff arbitrarily well,
the next  question is: when the signature coefficients are estimated from finite data, how large is the resulting out-of-sample hedging error? The next section answers this question. 

\section{Statistical Learning and Total Mean-Square Hedging Error}\label{sec:statistical_learning}
This section analyzes the statistical component of the It\^o-signature hedging error, i.e., the additional error caused by estimating
the finite-dimensional signature coefficients from data. The goal of this section is to combine this statistical error with the oracle approximation and discretization errors into a single total mean-square hedging bound.

To start, take a truncation order \(m\) and a partition \(\pi_n\), and consider the finite
discrete signature payoff
\(
H_T^{(m,\pi_n)}(\beta)
=
\beta_0+\sum_{1\le |\Gamma|\le m}\beta_{\Gamma}\,
S(\tilde X)_{T}^{\Gamma,\mathrm I,\pi_n}
\) as in Theorem \ref{thm:approx_hedging}.
Within this finite-dimensional class, choosing a hedge is equivalent to choosing
the coefficients of a linear model on a tradable feature space. Hence the
out-of-sample mean-square hedging error can be analyzed through the
generalization error of the corresponding truncated signature regression. 

Next, take the constant-volatility benchmark as in the previous analysis, where the discrete It\^o-signature coordinates of the original tradable process \(X\) are themselves hedge building blocks. 
Fix a truncation order \(m\ge1\) and a partition \(\pi_n\). Let
\(
x_{m,\pi_n}\in\mathbb R^{p_m}
\)
denote the centered vector obtained by arranging all retained nonconstant discrete It\^o-signature coordinates of \(\tilde X\) up to order \(m\) on the partition \(\pi_n\) in a fixed deterministic order, and let
\(
Y\in\mathbb R
\)
denote the centered payoff. The corresponding population linear projection is
\begin{equation}\label{eq:centered_projection}
Y=x_{m,\pi_n}^\top\beta_{m,\pi_n}^*+\varepsilon_{m,\pi_n},
\end{equation}
with
\[
\mathbb E[\varepsilon_{m,\pi_n}]=0,
\
\mathbb E[x_{m,\pi_n}\,\varepsilon_{m,\pi_n}]=0,
\
\Sigma_{m,\pi_n}:=\mathbb E[x_{m,\pi_n}x_{m,\pi_n}^\top],
\
\sigma_{m,\pi_n,*}^2:=\mathbb E[\varepsilon_{m,\pi_n}^2]
=
\mathbb E\bigl[(Y-x_{m,\pi_n}^\top\beta_{m,\pi_n}^*)^2\bigr].
\]
The intercept can be recovered separately from the original uncentered variables and corresponds to the cash position in the hedging portfolio, so it plays no role in the main estimation result below.


Note that the resulting estimation problem is both high-dimensional and heavy-tailed. As the truncation order increases, the number of retained signature coordinates grows rapidly, while the payoff is typically expected to depend materially on only a relatively small subset of them. In addition, financial returns are well known to be heavy tailed \citep{mandelbrot1963variation,cont2001empirical}, and discrete
It\^o-signature coordinates are built from iterated sums and products of return increments. 
Nevertheless, for each fixed truncation order \(m\), they have finite moments for every order.
This allows us to exploit the robust sparse regression method of \citet{fan2021shrinkage}, which is based on truncation of both the response and the regressors.

More precisely, let \((Y_i,x_{m,\pi_n,i})_{i=1}^N\) be independent and identically distributed (i.i.d.) copies of the centered pair \((Y,x_{m,\pi_n})\). For truncation levels \(\tau_1,\tau_2>0\), define
\[
\widetilde{Y}_i(\tau_1)=\operatorname{sgn}(Y_i)\,(|Y_i|\wedge \tau_1),
\qquad
\widetilde{x}_{m,\pi_n,ij}(\tau_2)=\operatorname{sgn}(x_{m,\pi_n,ij})\,(|x_{m,\pi_n,ij}|\wedge \tau_2),
\]
and write \(\widetilde{x}_{m,\pi_n,i}(\tau_2)=(\widetilde{x}_{m,\pi_n,i1}(\tau_2),\dots,\widetilde{x}_{m,\pi_n,ip_m}(\tau_2))^\top\). We also set
\[
\widehat{\Sigma}_{\widetilde{Y}\widetilde{x}_{m,\pi_n}}
=
\frac{1}{N}\sum_{i=1}^N \widetilde{Y}_i(\tau_1)\,\widetilde{x}_{m,\pi_n,i}(\tau_2),
\qquad
\widehat{\Sigma}_{\widetilde{x}_{m,\pi_n}\widetilde{x}_{m,\pi_n}}
=
\frac{1}{N}\sum_{i=1}^N \widetilde{x}_{m,\pi_n,i}(\tau_2)\widetilde{x}_{m,\pi_n,i}(\tau_2)^\top.
\]

\begin{theorem}\label{thm:robusterror}
Assume that \((Y_i,x_{m,\pi_n,i})_{i=1}^N\) are i.i.d.\ copies of the centered pair \((Y,x_{m,\pi_n})\), where \(x_{m,\pi_n}\in\mathbb R^{p_m}\) is the vector of retained discrete It\^o-signature predictors up to order \(m\). Suppose \(\|\beta_{m,\pi_n}^*\|_1\le R<\infty\) and \(\|\beta_{m,\pi_n}^*\|_s\le \rho\) for some \(0\le s\le 1\). Denote
$
K_{4,m,\pi_n}
=
\max\left\{
\mathbb E|Y|^4,\;
\max_{1\le j\le p_m}\mathbb E|x_{m,\pi_n,j}|^4
\right\}
<\infty.$
Choose \(\tau_1,\tau_2 > 0\) such that $
\tau_1,\tau_2 \asymp_R \left(\frac{K_{4,m,\pi_n}N}{\log p_m}\right)^{1/4},
$
and for any \(\xi>2\), set
$
\lambda_N = 2C\sqrt{\frac{K_{4,m,\pi_n}\xi\log p_m}{N}},
$
where \(C\) depends only on \(R\). Define
\[
\widehat{\beta}_{m,\pi_n}(\tau_1,\tau_2,\lambda_N)
\in
\arg\min_{\beta\in\mathbb R^{p_m}}
\left\{
-\widehat{\Sigma}_{\widetilde{Y}\widetilde{x}_{m,\pi_n}}^\top\beta
+\frac{1}{2}\beta^\top\widehat{\Sigma}_{\widetilde{x}_{m,\pi_n}\widetilde{x}_{m,\pi_n}}\beta
+\lambda_N\|\beta\|_1
\right\}.
\]
Then there exists a constant \(C_1\), depending only on \(R\), such that whenever
$
\rho \left( \frac{K_{4,m,\pi_n}\xi \log p_m}{N} \right)^{(1-s)/2} \leq C_1,
$
the following inequalities hold with probability at least \(1-3p_m^{-(\xi-2)}\):
\begin{align*}
\left\| \widehat{\beta}_{m,\pi_n}(\tau_1,\tau_2,\lambda_N) - \beta_{m,\pi_n}^* \right\|_2^2
&\leq C_2 \rho \left( \frac{K_{4,m,\pi_n}\xi \log p_m}{N} \right)^{1 - s/2}, \\
\left\| \widehat{\beta}_{m,\pi_n}(\tau_1,\tau_2,\lambda_N) - \beta_{m,\pi_n}^* \right\|_1
&\leq C_3 \rho \left( \frac{K_{4,m,\pi_n}\xi \log p_m}{N} \right)^{(1 - s)/2},
\end{align*}
where the constants \(C_2\) and \(C_3\) depend only on \(R\).
\end{theorem}

Theorem~\ref{thm:robusterror} gives a high-probability coefficient error bound for the truncated signature regression. We next translate this coefficient bound into a conditional prediction bound, which also has a direct hedging interpretation. 
Let
\(
\mathcal D_N:=\{(Y_i,x_{m,\pi_n,i})\}_{i=1}^N
\)
denote the training sample, and let
\(
(Y_{\mathrm{new}},x_{m,\pi_n,\mathrm{new}})
\)
be an independent test point with the same distribution as \((Y,x_{m,\pi_n})\). For any estimator \(\widehat\beta_{m,\pi_n}\) measurable with respect to \(\mathcal D_N\), define the conditional prediction risk
\[
\mathcal R_{m,\pi_n}(\widehat\beta_{m,\pi_n}\mid \mathcal D_N)
:=
\mathbb E\Bigl[
\bigl(Y_{\mathrm{new}}-x_{m,\pi_n,\mathrm{new}}^\top\widehat\beta_{m,\pi_n}\bigr)^2
\,\big|\,\mathcal D_N
\Bigr],
\]
which is the conditional out-of-sample mean-square hedging loss of the It\^o signature hedge, as the residual of the linear signature model is exactly the hedging error of the corresponding truncated signature portfolio. 

When the signature features are tradable in the sense of Theorem~\ref{thm:hedging strategy}, we have the following corollary that controls the out-of-sample hedging error bound for the corresponding truncated It\^o signature portfolio.

\begin{corollary}[Generalization and hedging error bound]\label{cor:gen_hedge_error}
Under the assumptions and notation of Theorem~\ref{thm:robusterror}, the conditional generalization error is bounded with probability at least \(1-3p_m^{-(\xi-2)}\):
\[
\mathcal R_{m,\pi_n}(\widehat\beta_{m,\pi_n}\mid \mathcal D_N)
\le
\sigma_{m,\pi_n,*}^2
+
\lambda_{\max}(\Sigma_{m,\pi_n})\,
C_2 \rho \left( \frac{K_{4,m,\pi_n}\xi \log p_m}{N} \right)^{1 - s/2}.
\]
\end{corollary}



We now combine the statistical bound above with the oracle error of the best finite discrete signature hedge. This oracle error captures the approximation and discretization components before finite-sample estimation is taken into account.
For a truncation order \(m\) and a partition \(\pi_n\), define the oracle
finite-feature hedging error by
\[
\mathcal E_{\mathrm{or}}(m,\pi_n)
:=
\inf_{\beta_0,\{\beta_\Gamma:1\le |\Gamma|\le m\}}
\mathbb E^{\mathbb P}
\left[
\left|
F(\tilde X)
-
\beta_0
-
\sum_{1\le |\Gamma|\le m}
\beta_\Gamma
S(\tilde X)_{T}^{\Gamma,\mathrm I,\pi_n}
\right|^2
\right].
\]
Equivalently, after centering the payoff and the retained signature coordinates,
\(\mathcal E_{\mathrm{or}}(m,\pi_n)=\sigma_{m,\pi_n,*}^2\). Thus
\(\mathcal E_{\mathrm{or}}(m,\pi_n)\) is the best achievable mean-square hedging
error within the class of discrete It\^o-signature portfolios of order \(m\) on
a partition \(\pi_n\), by combining both the finite-order approximation error and
the discrete implementation error.

\begin{theorem}[Total mean-square hedging error]\label{thm:total_hedging_error}
Assume the constant-diffusion benchmark and the setting of
Theorem~\ref{thm:approx_hedging}. Fix a truncation order \(m\ge1\) and a
partition \(\pi_n\). Suppose the assumptions of
Theorem~\ref{thm:robusterror} hold for the centered pair
\((Y,x_{m,\pi_n})\), with covariance matrix \(\Sigma_{m,\pi_n}\), dimension \(p_m\), and
fourth-moment bound \(K_{4,m,\pi_n}\). Then, with probability at least
\(1-3p_m^{-(\xi-2)}\),
\[
\mathbb E^{\mathbb P}
\left[
\left.
\left|
F(\tilde X_{\mathrm{new}})
-
H_{T,\mathrm{new}}^{(m,\pi_n)}(\widehat\beta_{m,\pi_n})
\right|^2
\,\right|\,\mathcal D_N
\right]
\le
\mathcal E_{\mathrm{or}}(m,\pi_n)
+
\lambda_{\max}(\Sigma_{m,\pi_n})\,
C_2\rho
\left(
\frac{K_{4,m,\pi_n}\xi\log p_m}{N}
\right)^{1-s/2},
\]
where \(H_{T,\mathrm{new}}^{(m,\pi_n)}(\widehat\beta_{m,\pi_n})\) is the terminal wealth
of the estimated discrete It\^o-signature hedge applied to an independent test
path, including the separately recovered intercept as the cash position.
Moreover,
\[
\lim_{m\to\infty}\limsup_{\|\pi_n\|\to0}
\mathcal E_{\mathrm{or}}(m,\pi_n)
=
0.
\]
\end{theorem}

Theorem~\ref{thm:total_hedging_error} combines the oracle population error with the statistical estimation error and yields a total out-of-sample mean-square hedging error bound for the estimated It\^o-signature portfolio. The first term, \(\mathcal E_{\mathrm{or}}(m,\pi_n)\), is the oracle error of the best discrete
signature hedge at truncation order \(m\) and trading grid \(\pi_n\). This term
summarizes the approximation and discretization errors studied in the earlier
subsections. The second term is the statistical estimation error caused by
learning the signature coefficients from \(N\) training paths. For a fixed truncation order \(m\), the discrete-to-continuous convergence first sends the tradable discrete signature features to their continuous-time It\^o-signature counterparts as \(\|\pi_n\|\to 0\). After this fixed-order continuous-time limit is taken, increasing \(m\) invokes the universal approximation property of the continuous signature expansion. At the same time, increasing \(m\) raises the feature dimension \(p_m\) and may enlarge the statistical estimation term. This decomposition provides the theoretical basis for selecting the signature depth and trading grid by balancing discretization, approximation, and estimation errors.

\section{Comparison with Benchmarks on Simulated Data}\label{sec:simulation}

This section evaluates the practical performance of the proposed It\^o-signature hedge on simulated data. In particular, we design the simulation study to show whether the method remains effective when training data are scarce, whether this performance can be achieved at a reasonable computational cost, and whether the fitted linear coefficients provide a financially interpretable decomposition of the hedge.

Our main findings are consistent across all simulated experiments. The It\^o-signature hedge is the most stable method in the small-sample regime and achieves strong performance with substantially less training data than the neural network benchmark. The neural network can improve markedly when the sample size becomes large and may eventually match or exceed signature-based performance, but this gain comes at a much higher data requirement. The Stratonovich-signature benchmark is at a clear practical disadvantage in the present setting because the lead--lag lift and the computation of expected future signatures generate a much larger optimization problem. We further exploit the linear structure of the It\^o-signature hedge to inspect the fitted coefficient vector and show how different payoff structures load on different tradable signature basis strategies.

\subsection{Experimental Design}\label{subsec:sim_setup}

We compare three approaches: a feedforward neural network hedge following \citet{lutkebohmert2022robust}, a Stratonovich-signature hedge following \citet{lyons2020non}, and the proposed It\^o-signature hedge. \footnote{For both It\^o-signature and Stratonovich signature methods, we consider different truncation order as a robustness check, the results are similar, and we present the results of truncation order $=6$ here.} A detailed comparison of the models is reported in Appendix~\ref{app:sim_method_details}.

We simulate the underlying asset price under the risk-neutral measure \(\mathbb Q\) using the geometric Brownian motion
\(
\mathrm dS_t=\sigma S_t\,\mathrm dW_t^{\mathbb Q},
\)
with annualized volatility \(\sigma=0.2\), maturity \(T=1\) (one year), \(n=250\)
daily rebalancing dates, initial price \(S_0=10\), strike \(K=10\), and risk-free
rate and dividend yield \(r=q=0\). We consider four benchmark products: a European call, a European put, a geometric average Asian call, and a floating-strike lookback put. For each product, we train all methods on the same simulated datasets, with training sample sizes ranging from \(2^6\) to \(2^{15}\), and evaluate them on a common test set of 10{,}000 paths. For robustness we repeat the experiments for 10 different random seeds. The GBM benchmark admits closed-form prices and delta hedges for all four products considered here; we use these formulas only for evaluation and collect them in Appendix~\ref{app:sim_benchmarks}. For all methods, we select hyperparameters via a 75/25 random train--validation split of the training data and report total time under a common accounting convention.\endnote{For the signature-based methods, the reported runtime includes both signature feature computation and the full Lasso fitting procedure, including hyperparameter search. For the neural network, the reported runtime covers the hyperparameter search phase plus the final 200-epoch retraining on the full training set.}

\subsection{Main Performance Comparison across Products}\label{subsec:sim_main_results}
We consider four benchmark contracts that span increasing levels of path dependence, from terminal-value payoffs to functionals involving the running average and the running maximum. This allows us to evaluate whether the relative performance of the competing methods is stable across payoff structures.

\begin{figure}[htbp]
    \centering

    \subfigure[European Call.\label{fig:error_mse_call}]{
        \includegraphics[width=0.38\linewidth]{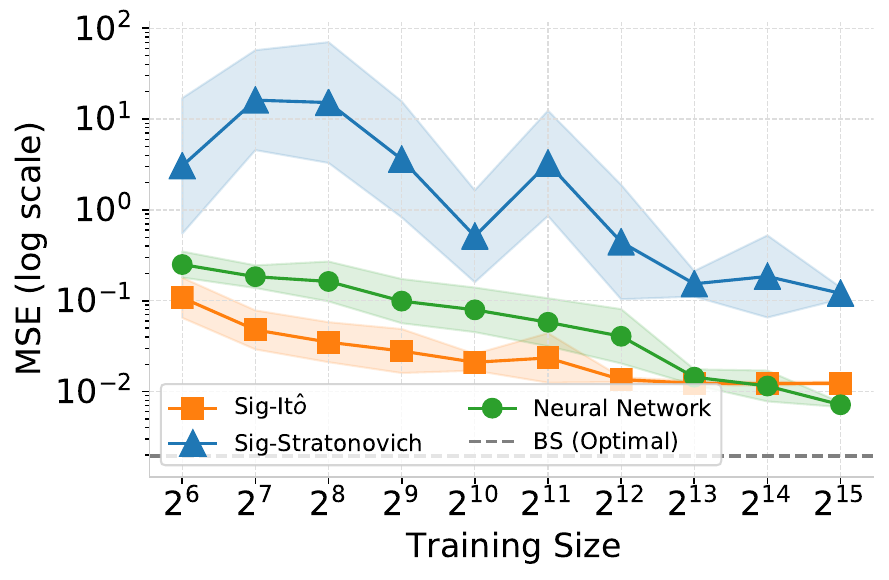}
    }
    \hspace{0.5cm}
    \subfigure[European Put.\label{fig:error_mse_put}]{
        \includegraphics[width=0.38\linewidth]{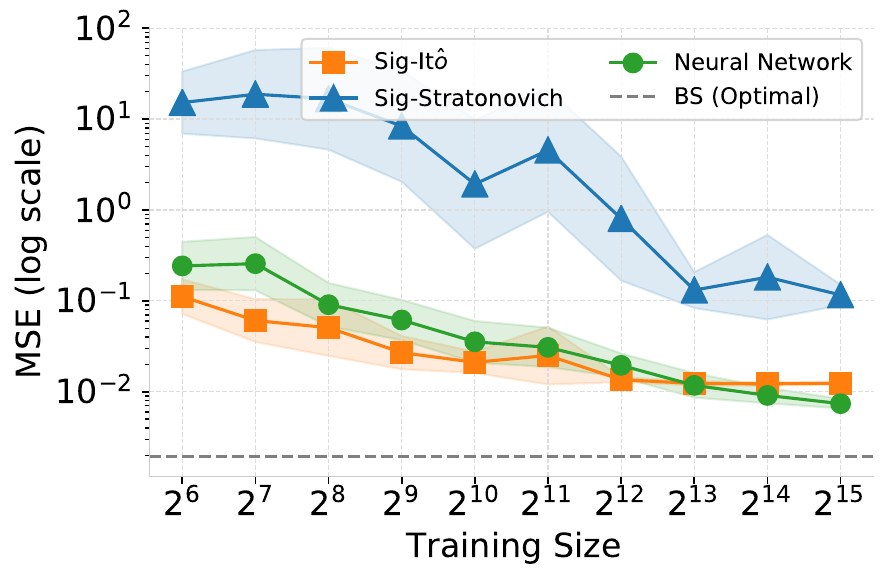}
    }

    \vspace{0.3cm}

    \subfigure[Geometric Average Asian Call.\label{fig:error_mse_asian}]{
        \includegraphics[width=0.38\linewidth]{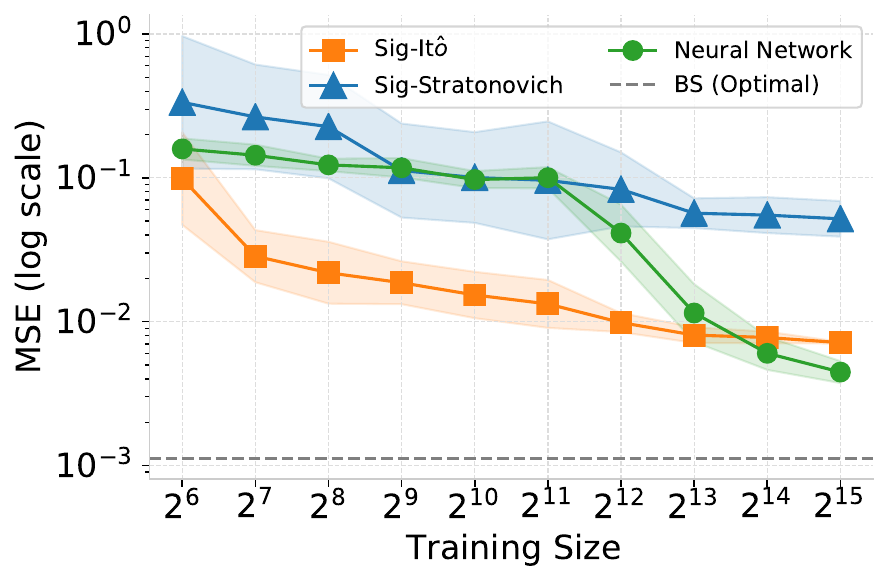}
    }
    \hspace{0.5cm}
    \subfigure[Floating-strike Lookback Put.\label{fig:error_mse_lookback}]{
        \includegraphics[width=0.38\linewidth]{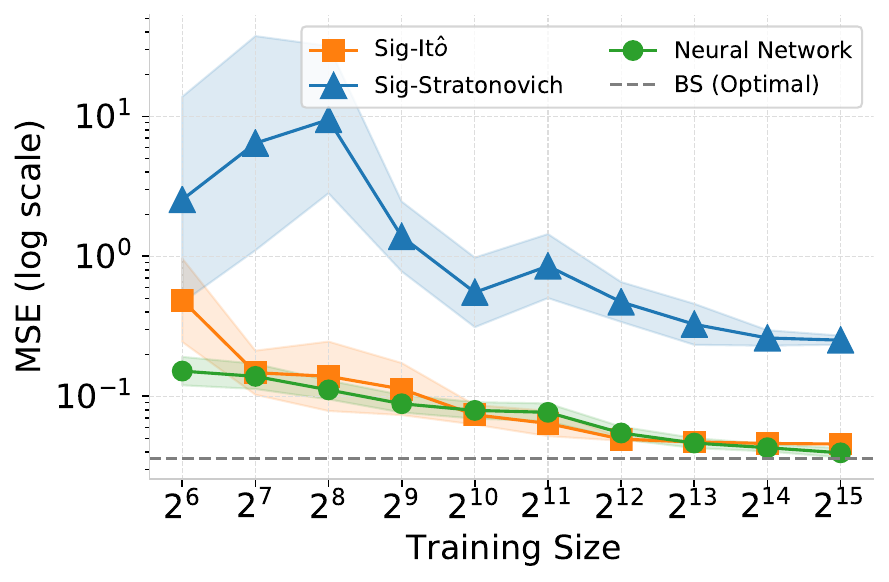}
    }
    
    \caption{Hedging Error MSE as a function of training size.
    Each panel shows the mean MSE with 95\% confidence intervals
    across 10 repeated experiments for Sig-It\^o,
    Sig-Stratonovich, and a neural network baseline.
    Dashed line: Black--Scholes benchmark optimal error.
    \label{fig:error_mse}}
\end{figure}

Figure \ref{fig:error_mse} reports the out-of-sample hedging error mean squared error (MSE) across training sample sizes\footnote{We also consider the corresponding out-of-sample hedging error mean absolute error (MAE). The pattern is consistent with the MSE results in Figure \ref{fig:error_mse}.}. Three patterns are especially clear. First, the It\^o-signature hedge performs strongly in the small-sample regime across all four products. Its hedging error stabilizes quickly as the training size increases, and it consistently dominates the Stratonovich-signature benchmark. Second, the neural network is much more sensitive to sample size. With limited data it is clearly less reliable, but with sufficiently large samples its performance improves substantially and may eventually approach, or in some cases slightly improve upon, the It\^o-signature hedge. Third, the cross-product ranking is stable: the It\^o-signature method remains competitive not only for European options but also for the more path-dependent Asian and lookback contracts. These observations are consistent with Theorem~\ref{thm:total_hedging_error}, which provides control of the total mean-square hedging error for the proposed it\^o signature hedge. In this sense, the MSE results in Figure~\ref{fig:error_mse} give empirical evidence that the theoretical bound is also reflected in finite-sample simulation performance.

From a practical perspective, the most important feature of Figure \ref{fig:error_mse} is the small-sample behavior. Financial datasets are often limited, especially when the payoff of interest is path dependent or when the hedging rule must be recalibrated over relatively short rolling windows. In this regime, the It\^o-signature hedge delivers the best overall trade-off between robustness and accuracy. By contrast, the Stratonovich-signature benchmark is consistently hampered by its much larger representation, while the neural network requires substantially more data before its flexibility becomes beneficial.

\subsection{Computational Efficiency}\label{subsec:sim_efficiency}

The performance comparisons above show that the It\^o-signature hedge is accurate in small samples. We next examine whether this accuracy comes at a computational cost. This comparison is central to the proposed framework: the It\^o-signature hedge turns signature coordinates into tradable basis strategies, so that hedging rules can be estimated through a low-dimensional and computationally light regression problem rather than through a large nonlinear optimization procedure.

\begin{table}[htbp]
\centering
\caption{Total runtime of hedging methods (seconds). Values are averaged over 10 repeated experiments. For signature-based methods, the reported runtime includes signature computation.}
\label{tab:runtime_by_option_type}
\resizebox{\textwidth}{!}{
\begin{tabular}{lrrrrrrrr}
\toprule
\textbf{Method}
& \(\mathbf{2^8}\)
& \(\mathbf{2^9}\)
& \(\mathbf{2^{10}}\)
& \(\mathbf{2^{11}}\)
& \(\mathbf{2^{12}}\)
& \(\mathbf{2^{13}}\)
& \(\mathbf{2^{14}}\)
& \(\mathbf{2^{15}}\) \\
\midrule

\multicolumn{9}{l}{\textbf{Panel A: European Call}} \\
\midrule
Neural Network 
& 1{,}286.73 & 1{,}534.03 & 1{,}890.77 & 2{,}831.45 & 5{,}639.23 & 11{,}604.05 & 19{,}575.20 & 39{,}007.69 \\
Sig-It\^o 
& \textbf{0.12} & \textbf{0.25} & \textbf{0.50} & \textbf{1.01} & \textbf{2.04} & \textbf{4.38} & \textbf{9.09} & \textbf{19.18} \\
Sig-Stratonovich
& 69.43 & 134.27 & 266.66 & 544.13 & 950.33 & 1{,}873.09 & 3{,}682.70 & 6{,}998.05 \\

\addlinespace
\midrule
\multicolumn{9}{l}{\textbf{Panel B: European Put}} \\
\midrule
Neural Network 
& 1{,}492.66 & 1{,}772.40 & 2{,}288.74 & 2{,}978.23 & 4{,}919.30 & 9{,}625.94 & 19{,}545.92 & 39{,}060.41 \\
Sig-It\^o
& \textbf{0.12} & \textbf{0.25} & \textbf{0.51} & \textbf{1.02} & \textbf{2.08} & \textbf{4.43} & \textbf{9.14} & \textbf{19.25} \\
Sig-Stratonovich
& 38.27 & 77.56 & 155.96 & 308.86 & 615.24 & 1{,}239.25 & 2{,}498.80 & 5{,}027.72 \\

\addlinespace
\midrule
\multicolumn{9}{l}{\textbf{Panel C: Geometric Average Asian Call}} \\
\midrule
Neural Network
& 1{,}548.91 & 1{,}665.64 & 2{,}021.38 & 2{,}531.95 & 5{,}394.43 & 10{,}143.04 & 25{,}033.05 & 42{,}883.52 \\
Sig-It\^o
& \textbf{0.13} & \textbf{0.25} & \textbf{0.51} & \textbf{1.04} & \textbf{2.13} & \textbf{4.39} & \textbf{8.87} & \textbf{19.38} \\
Sig-Stratonovich
& 39.16 & 77.04 & 154.83 & 303.19 & 613.20 & 1{,}236.48 & 2{,}511.59 & 4{,}964.26 \\

\addlinespace
\midrule
\multicolumn{9}{l}{\textbf{Panel D: Floating-strike Lookback Put}} \\
\midrule
Neural Network
& 1{,}544.77 & 1{,}692.19 & 1{,}974.54 & 2{,}459.03 & 5{,}042.20 & 10{,}072.72 & 20{,}259.99 & 40{,}345.93 \\
Sig-It\^o
& \textbf{0.13} & \textbf{0.26} & \textbf{0.53} & \textbf{1.07} & \textbf{2.31} & \textbf{4.62} & \textbf{9.61} & \textbf{20.37} \\
Sig-Stratonovich
& 32.19 & 66.00 & 131.17 & 261.91 & 533.00 & 1{,}073.10 & 2{,}147.24 & 4{,}346.85 \\

\bottomrule
\end{tabular}}
\vspace{0.5em}
\end{table}

Table~\ref{tab:runtime_by_option_type} reports the total computation and training time across option types and training sizes. The key pattern is that the It\^o-signature method requires substantially less computation than both the Stratonovich-signature benchmark and the neural-network baseline. This advantage is large across all products and all training sizes. Even when the training sample becomes large, the It\^o-signature hedge remains computationally inexpensive, while the Stratonovich-signature method incurs much higher cost due to the lead--lag lift and the larger signature representation. The neural network is even more time-consuming, reflecting the cost of iterative training. Thus, the strong small-sample accuracy of the It\^o-signature hedge does not come at the expense of computational efficiency.

\begin{figure}[htbp]
    \centering
    \subfigure[European Call.\label{fig:time_vs_mse_call}]{
        \includegraphics[width=0.38\linewidth]{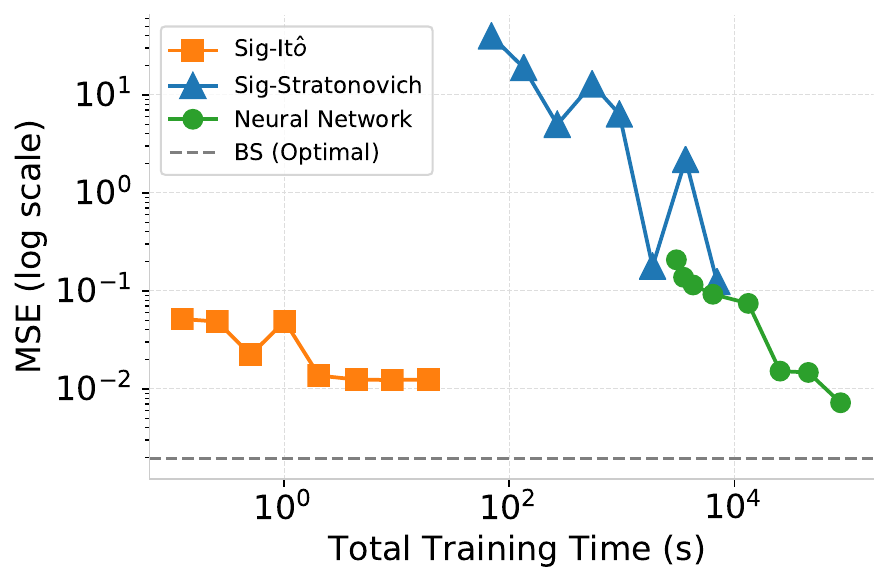}
    }
    \hspace{0.5cm}
    \subfigure[European Put.\label{fig:time_vs_mse_put}]{
        \includegraphics[width=0.38\linewidth]{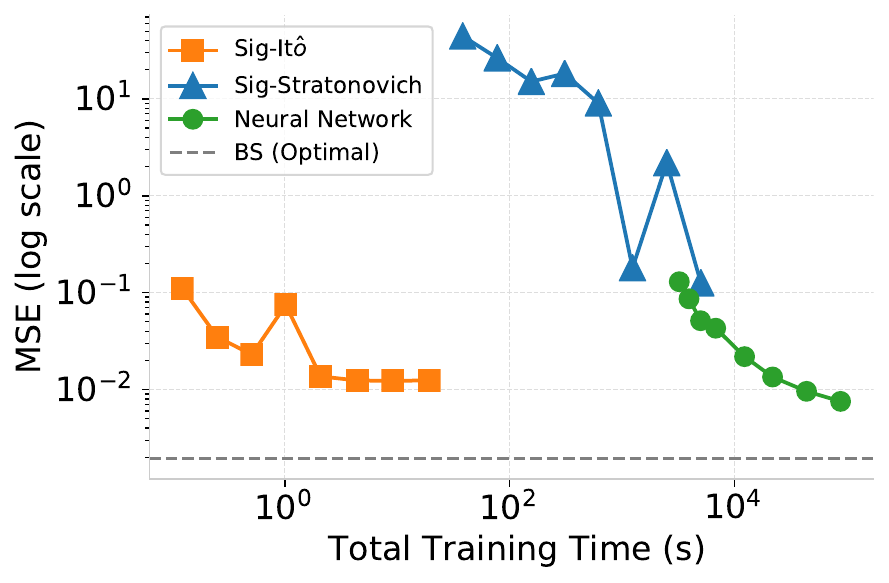}
    }

    \vspace{0.3cm}

    \subfigure[Geometric Average Asian Call.\label{fig:time_vs_mse_asian}]{
        \includegraphics[width=0.38\linewidth]{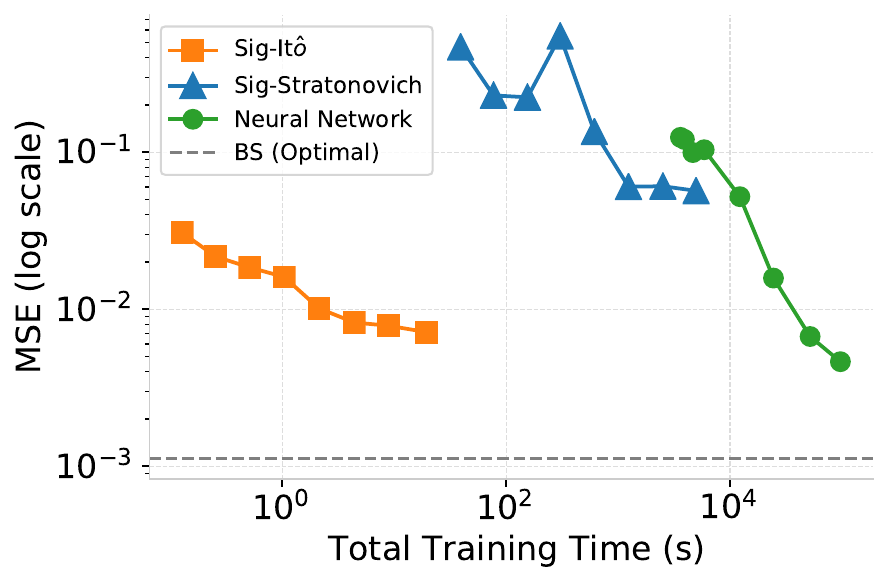}
    }
    \hspace{0.5cm}
    \subfigure[Floating-strike Lookback Put.\label{fig:time_vs_mse_lookback}]{
        \includegraphics[width=0.38\linewidth]{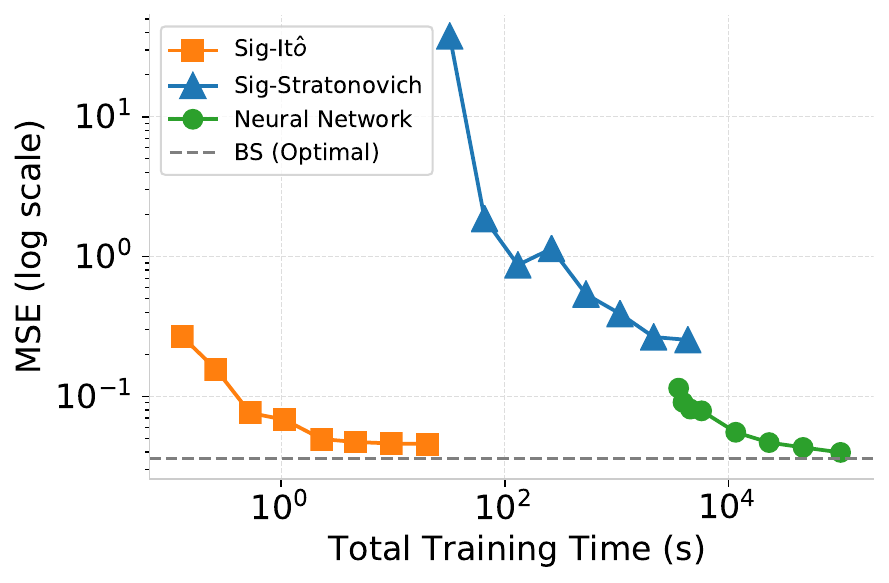}
    }
    \caption{Computational efficiency: hedging error MSE vs.\ total training
    time on a log--log scale. Each curve represents the mean over 10
    repeated experiments as training size increases from $2^6$ to $2^{15}$.
    The lower-left region corresponds to lower error and shorter runtime.
    Dashed line: Black--Scholes optimal hedge.
    \label{fig:time_vs_mse}}
\end{figure}

Figure~\ref{fig:time_vs_mse} provides a direct accuracy--efficiency comparison by plotting hedging error MSE against total training time on a log--log scale. Points closer to the lower-left corner are preferable because they achieve lower hedging error with less computation. Across the three products, the It\^o-signature curves are located close to this favorable region: they achieve competitive MSE levels while requiring only a very small amount of training time. By contrast, the Stratonovich-signature benchmark is shifted far to the right and often upward, indicating that it is both more expensive and less accurate in this simulation design. The neural network can reduce its MSE when the training size becomes large, but this improvement requires orders of magnitude more computation.

The comparison in Figure~\ref{fig:time_vs_mse} highlights the main practical advantage of the proposed method. The It\^o-signature hedge delivers a favorable accuracy--efficiency trade-off rather than merely improving one dimension of performance. This is particularly important for hedging applications in which models must be recalibrated repeatedly over rolling windows or across many contracts. In such settings, computational cost directly affects whether a method can be used in real time. The results therefore suggest that the It\^o-signature framework is not only statistically effective but also computationally scalable.

Taken together, Table~\ref{tab:runtime_by_option_type} and Figure~\ref{fig:time_vs_mse} show that the proposed It\^o-signature hedge occupies the most favorable region of the empirical accuracy--efficiency frontier. It achieves strong hedging performance with dramatically lower runtime than the Stratonovich-signature and neural-network benchmarks. This efficiency is a direct consequence of using tradable It\^o-signature bases, which convert the hedging problem into a transparent and low-cost linear estimation problem.

Additional diagnostics, including position MSE relative to the benchmark delta and the cash position difference, are reported in Appendix~\ref{app:detailed_product_results}.

\subsection{Interpreting Path Dependence through Fitted Signature Coefficients}\label{subsec:sim_interpretability}

We next use the fitted coefficient vector to examine how the hedge differs
between vanilla and path-dependent payoffs. This is a useful implication of the linear It\^o-signature formulation. The fitted hedge is not only a prediction rule, it also shows which tradable basis it\^o signature strategies are used by the model.

In the simulation setting, the underlying path is the one-dimensional
stock price \(S\) with time augmentation. Each nonconstant discrete
It\^o-signature coordinate of this time-augmented path corresponds to a tradable self-financing basis strategy by Theorem~\ref{thm:hedging strategy}. Hence the estimated coefficient on a coordinate measures the loading placed on the corresponding pathwise trading basis. This gives the signature hedge a direct basis-level decomposition, which is not available for the neural-network benchmark.

\begin{figure}[htbp]
    \centering
    \subfigure[European Call.\label{fig:top_coef_call}]{
        \includegraphics[width=0.38\linewidth]{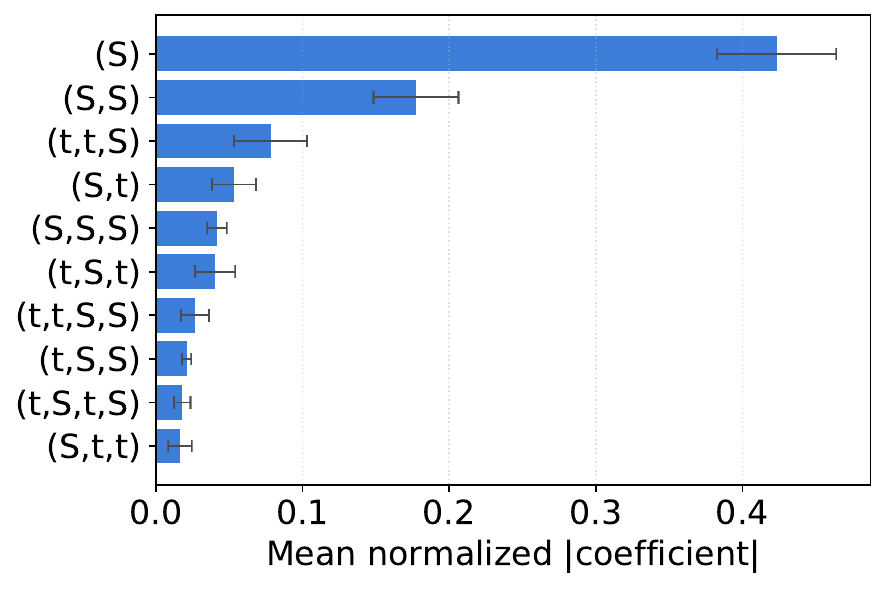}
    }
    \hspace{0.5cm}
    \subfigure[European Put.\label{fig:top_coef_put}]{
        \includegraphics[width=0.38\linewidth]{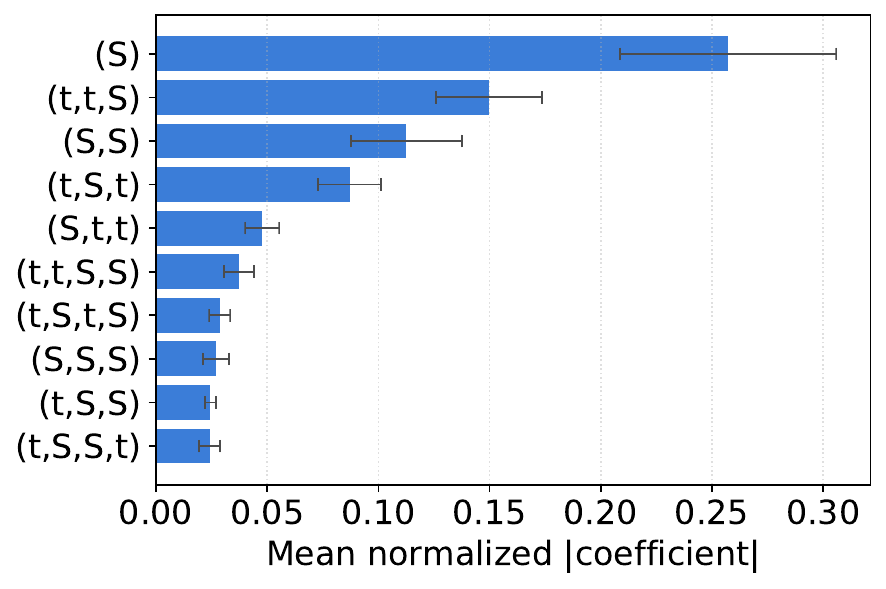}
    }

    \vspace{0.3cm}

    \subfigure[Geometric Average Asian Call.\label{fig:top_coef_asian}]{
        \includegraphics[width=0.38\linewidth]{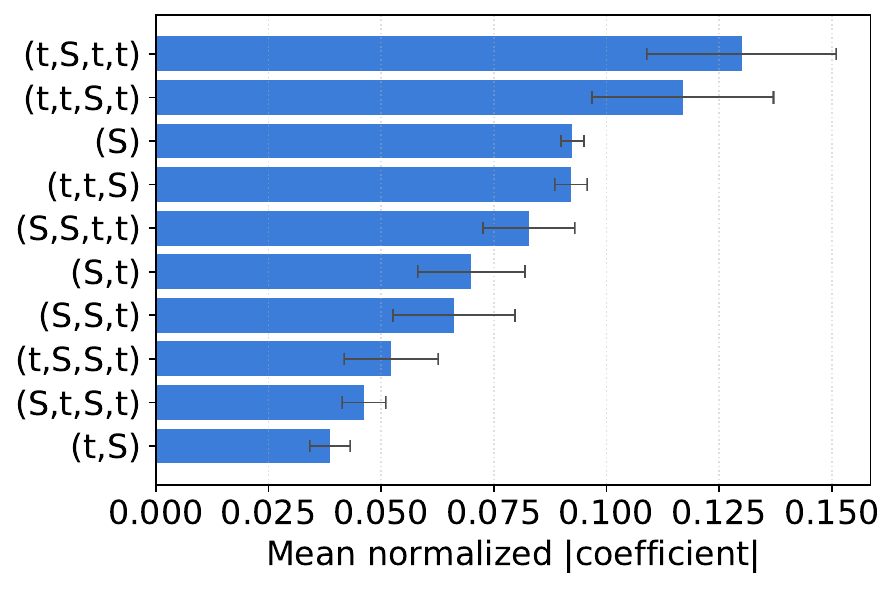}
    }
    \hspace{0.5cm}
    \subfigure[Floating-strike Lookback Put.\label{fig:top_coef_lookback}]{
        \includegraphics[width=0.38\linewidth]{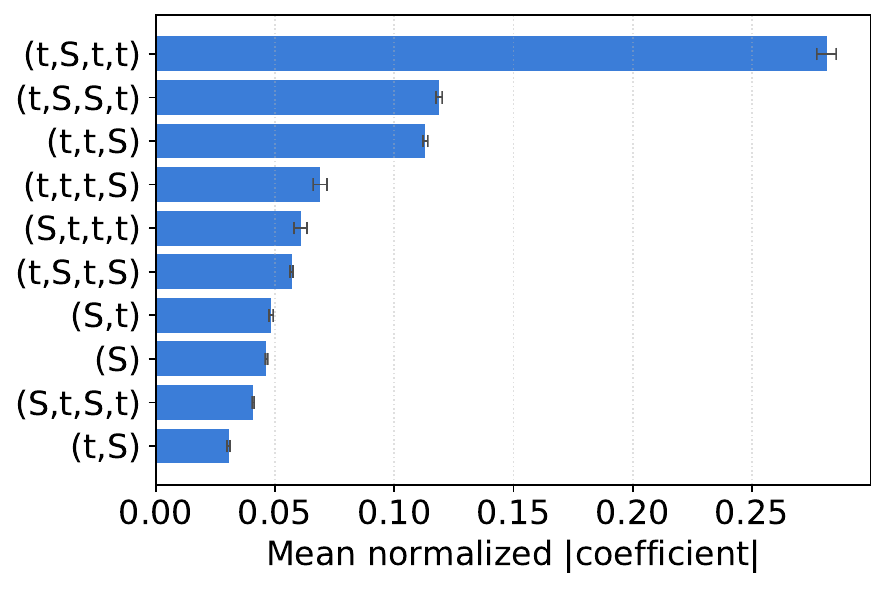}
    }
    \caption{Top signature-basis loadings of the fitted Sig-It\^o hedge.
    Each panel reports the ten largest order-4 It\^o-signature coefficients by
    mean normalized absolute magnitude, averaged across ten random seeds at
    training size \(N=2^{15}\). The coordinate labels use \(S\) for the stock
    coordinate and \(t\) for calendar time. Error bars indicate standard errors
    across seeds.}
    \label{fig:top_signature_coefficients}
\end{figure}

Figure~\ref{fig:top_signature_coefficients} reports the ten largest
coefficients for the order-4 Sig-It\^o hedge trained with \(N=2^{15}\)
simulated paths. Coefficients are ranked by mean normalized absolute magnitude
and averaged across the ten random seeds used in the simulation study. Error
bars report standard errors across seeds. In the coordinate labels, \(S\)
denotes the stock coordinate and \(t\) denotes calendar time. For example,
\((t,S,t,t)\) denotes the discrete It\^o-signature coordinate whose iterated
integration order is time, stock, time, and time.

The coefficient patterns clearly separate vanilla from path-dependent payoffs.
For the European call and put, the fitted hedge is dominated by the first-order stock coordinate \((S)\). The remaining terms are much smaller and mainly capture stock--stock or stock--time corrections. This is consistent with the standard hedging intuition for vanilla options: the leading hedge component is directional exposure to the underlying asset, while higher-order terms provide smaller nonlinear adjustments to local stock exposure \citep{black1973pricing,hull2017optimal}.

The path-dependent payoffs use a richer set of basis strategies. For the
geometric Asian call, several of the largest coefficients involve repeated
calendar-time coordinates, such as \((t,S,t,t)\), \((t,t,S,t)\), and
\((t,t,S)\). These terms are not the running average itself, but they represent polynomial time-weighted stock exposures. They therefore carry the same type of time-aggregated path information as integrals such as
\(\int_0^T S_u\,du\), which is consistent with the averaging structure of Asian payoffs \citep{geman1993bessel,rogers1995value}. For the floating-strike lookback put with path dependent payoff,  hedging strategies is more complex in the sense that the characterization in terms of  higher-order exposures becomes more dominant. This is also intuitive, because a
lookback payoff depends on the realized extremum of the path and cannot be
summarized by terminal stock exposure alone.

Overall, Figure~\ref{fig:top_signature_coefficients} shows that the fitted
signature hedge reveals the information used by the hedge. Vanilla payoffs
mainly use low-order stock exposure. Path-dependent payoffs activate more
time-interaction and higher-order signature coordinates. This difference
supports the main economic interpretation of the It\^o-signature hedge: the
linear coefficients identify which tradable pathwise basis strategies are
needed for each payoff structure.

\section{Empirical Analysis}\label{sec:emp}

This section evaluates the proposed It\^o-signature hedge on market-traded vanilla options and path-dependent exotic options. We use two signature ideas for two roles. The It\^o signature hedging framework provides tradable path-dependent hedging bases. The signature kernel further provides path-similarity weights for limited and non-stationary financial data, following the weighting idea of \citet{guTransportationMarketRate2024}.

In the main empirical tables, we report two representative implementations of the proposed It\^o-signature hedge. The first, denoted by \textbf{Sig-It\^o}, estimates a sparse linear hedge on the discretized It\^o-signature basis using an \(\ell_1\)-penalized regression. The second, denoted by \textbf{Sig-It\^o-SigKernel}, uses the same It\^o-signature basis but estimates the local linear hedge with signature-kernel path-similarity weights and an \(\ell_1\)-penalized regression. Additional details on the weighting scheme and alternative fitting specifications are reported in Appendix~\ref{app:comparable_methods}. We also examine transaction costs as a robustness check in Appendix~\ref{app:robustness}.

Since the signature-based hedge is optimized for terminal hedging performance and market prices are unavailable for the synthetic contracts, our primary empirical metric is the terminal hedging error at expiry, as in recent empirical studies of discrete-time option hedging effectiveness \citep{augustyniak2023discrete}. We scale this error by the index price and summarize performance in two complementary ways: the mean absolute scaled error across contracts and the win rate relative to the benchmark hedge. The detailed results are then reported by option type, maturity, and moneyness \citep{aretz2023moneyness}.

The empirical results show that  the strongest performance is obtained by the signature-kernel implementation. The improvement becomes more pronounced for path-dependent exotic options, especially for lookback options. Moreover, the gains are particularly strong for short-maturity contracts and deep out-of-the-money options, where local delta-based approximations are typically less reliable. \endnote{Deep out-of-the-money option prices are also closely related to tail and disaster-risk components; see, for example, \citet{fan2016does,song2016tale,barro2021rare}.}

\subsection{Market-traded Vanilla Options}\label{subsec:emp_vanilla}

We first compare the proposed It\^o-signature hedge with the traditional Black--Scholes delta hedge on real market option data. Our empirical study uses daily S\&P 500 Weekly (SPXW) option data from 2011 to 2025, sourced from OptionMetrics \endnote{\url{https://optionmetrics.com/}}. We focus on SPXW options because they are among the most widely used and empirically relevant benchmark contracts in the option-hedging literature \citep{hull2017optimal,ruf2019neural,Nian2021LearningSO,chen2026neural}.

We compare the Black--Scholes practitioners' delta, the two It\^o-signature implementations defined above, and the neural-network hedge of \citet{lutkebohmert2022robust}. The Black--Scholes delta is computed from the implied volatility backed out from observed option prices.

We evaluate all methods using the same backtesting design. At the close of each trading day, we compute the hedge position implied by each model and implement that position with a one-day lag to avoid look-ahead bias. To ensure comparability across methods, we use the midpoint of the best bid and ask quotes as the daily option price throughout the hedging path. We use the chronologically first 10{,}000 options as a development sample for hyperparameter tuning, and the remaining options form an out-of-sample test set.

\begin{table}[htbp]
\centering
\caption{Vanilla option hedging performance: error at expiry (\(\times 10^{-3}\)) and win rate against Black--Scholes (BS). NN denotes the neural-network benchmark.}
\label{tab:overall_hedging_performance}
\begin{tabular}{@{} c c c c c c c c c @{}}
\toprule
& & BS & \multicolumn{2}{c}{Sig-It\^o} & \multicolumn{2}{c}{Sig-It\^o-SigKernel} & \multicolumn{2}{c}{NN} \\
\cmidrule(lr){4-5} \cmidrule(lr){6-7} \cmidrule(lr){8-9}
Option & Samples & Mean$\downarrow$ & Mean$\downarrow$ & Win\%$\uparrow$ & Mean$\downarrow$ & Win\%$\uparrow$ & Mean$\downarrow$ & Win\%$\uparrow$ \\
\midrule
Overall & 436{,}135 & 6.007 & 5.744 & 73.3 & \textbf{5.445} & \textbf{83.9} & 8.836 & 26.1 \\
Call    & 162{,}527 & 6.178 & 6.213 & 73.1 & \textbf{5.705} & \textbf{88.9} & 9.626 & 23.3 \\
Put     & 273{,}608 & 5.906 & 5.466 & 73.4 & \textbf{5.290} & \textbf{80.9} & 8.349 & 27.8 \\
\bottomrule
\end{tabular}
\end{table}

By Table~\ref{tab:overall_hedging_performance}, it is clear that  the unweighted \textbf{Sig-It\^o} hedge is superior to  Black--Scholes, reducing the overall mean error from \(6.007\) to \(5.744\) and achieving a win rate of \(73.3\%\). Adding signature-kernel weighting further improves performance: \textbf{Sig-It\^o-SigKernel} lowers the overall mean error to \(5.445\) and raises the win rate to \(83.9\%\). These results indicate that the It\^o-signature  is useful for vanilla-option hedging, while path-similarity weighting provides  additional  improvement in real market data.

Table~\ref{tab:challenging_vanilla_performance} further reports representative short-maturity and out-of-the-money subsamples, where local delta-based approximations are typically less reliable.
\begin{table}[htbp]
\centering
\caption{Vanilla options in challenging subsamples: representative short-maturity and out-of-the-money buckets. Error at expiry (\(\times 10^{-3}\)) and win rate against Black--Scholes.}
\label{tab:challenging_vanilla_performance}
\begin{tabular}{@{}c c c c c c c c c@{}}
\toprule
& & BS & \multicolumn{2}{c}{Sig-It\^o} & \multicolumn{2}{c}{Sig-It\^o-SigKernel} & \multicolumn{2}{c}{NN} \\
\cmidrule(lr){4-5} \cmidrule(lr){6-7} \cmidrule(lr){8-9}
Bucket & Samples & Mean$\downarrow$ & Mean$\downarrow$ & Win\%$\uparrow$ & Mean$\downarrow$ & Win\%$\uparrow$ & Mean$\downarrow$ & Win\%$\uparrow$ \\
\midrule
\multicolumn{9}{c}{\textbf{Panel A: Short maturity}} \\
\midrule
Call, \(<5\) days       & 11{,}784 & 3.043 & 2.719 & 73.9 & \textbf{2.683} & \textbf{85.3} & 3.111 & 16.1 \\
Call, 5--10 days       & 15{,}360 & 3.973 & 2.946 & 71.8 & \textbf{2.684} & \textbf{96.1} & 4.541 & 27.5 \\
Call, 10--15 days      & 17{,}834 & 6.711 & 4.191 & 74.9 & \textbf{3.807} & \textbf{96.7} & 7.026 & 26.6 \\
Call, 15--20 days      & 22{,}273 & 6.974 & 4.665 & 80.2 & \textbf{4.309} & \textbf{94.3} & 8.835 & 25.8 \\
Put, \(<5\) days        & 12{,}547 & 3.440 & 2.537 & 90.5 & \textbf{2.501} & \textbf{98.8} & 2.690 & 34.0 \\
Put, 5--10 days        & 15{,}593 & 4.373 & 3.112 & 84.2 & \textbf{2.852} & \textbf{98.6} & 4.449 & 37.1 \\
Put, 10--15 days       & 22{,}251 & 4.860 & 3.493 & 81.8 & \textbf{3.150} & \textbf{95.8} & 5.910 & 34.8 \\
Put, 15--20 days       & 30{,}261 & 6.327 & 4.113 & 86.0 & \textbf{3.882} & \textbf{95.9} & 7.190 & 31.7 \\
\midrule
\multicolumn{9}{c}{\textbf{Panel B: Out-of-the-money}} \\
\midrule
Call, \(S/K<0.8\)            & 6{,}096  & 0.176 & 0.127 & 98.4 & \textbf{0.084} & \textbf{99.7} & 0.528 & 8.70  \\
Call, \(0.8\le S/K<0.9\)    & 11{,}996 & \textbf{1.300} & 1.944 & 84.1 & 1.655 & \textbf{93.6} & 2.724 & 17.7 \\
Call, \(0.9\le S/K<0.95\)   & 34{,}161 & \textbf{2.836} & 4.282 & 62.7 & 3.735 & \textbf{85.9} & 4.360 & 19.0 \\
Call, \(0.95\le S/K<1\)     & 68{,}676 & 5.887 & 5.545 & 70.4 & \textbf{5.030} & \textbf{87.8} & 7.750 & 23.6 \\
Put, \(1\le S/K<1.05\)      & 67{,}587 & 7.066 & 6.176 & 76.1 & \textbf{6.016} & \textbf{85.8} & 9.931 & 27.9 \\
Put, \(1.05\le S/K<1.1\)    & 46{,}390 & 5.246 & 4.861 & 69.4 & \textbf{4.618} & \textbf{79.0} & 7.026 & 27.2 \\
Put, \(1.1\le S/K<1.2\)     & 48{,}628 & 3.881 & 3.749 & 70.3 & \textbf{3.634} & \textbf{74.7} & 5.378 & 26.1 \\
Put, \(S/K>1.2\)            & 54{,}392 & 2.440 & 1.792 & 70.2 & \textbf{1.776} & \textbf{72.0} & 2.251 & 29.1 \\
\bottomrule
\end{tabular}
\end{table}
It highlights two aspects especially relevant for practice. First, \textbf{Sig-It\^o-SigKernel} performs particularly well for short-dated options, where hedging errors are most sensitive to timing, local convexity, and strike-crossing risk \citep{lim2019intraday,bakshi2022dark}. Second, the gains are pronounced in out-of-the-money regions, where payoff nonlinearities are stronger relative to price level. These subsample results show that the signature-kernel version is most useful when the vanilla hedging problem is locally unstable or highly nonlinear.

\subsection{Path-dependent Exotic Options}\label{subsec:emp_exotic}

We next turn to path-dependent exotic options to test  whether pathwise features improve hedging performance. Since such contracts are typically traded over the counter and are therefore difficult to observe in large quantities, we follow \citet{lutkebohmert2022robust} and construct synthetic exotic options on real S\&P 500 index data.

Our empirical study again uses daily S\&P 500 index data from 2011 to 2025. On each trading day, we initiate a new synthetic option, estimate the hedging rule using a rolling window of historical data, and evaluate the resulting terminal hedging error at expiry. We use the same two It\^o-signature implementations and neural-network benchmark as in the vanilla-option analysis.

A key difference from the vanilla-option setting is that no closed-form benchmark delta is available for these path-dependent contracts. We therefore use a Monte Carlo-based hedge as the benchmark, following \citet{broadie1996estimating} and \citet{glasserman2004monte}. To reflect realistic trading constraints, all hedge positions are implemented with a one-day delay.

We consider four path-dependent contracts: geometric Asian calls, geometric Asian puts, lookback calls, and lookback puts. These contracts capture two canonical forms of path dependence: dependence on the running average and dependence on the running extremum of the price path. We evaluate them over a grid of maturities and moneyness levels.

The development period is 2011--2013, which is used for hyperparameter selection, and the out-of-sample test period is 2014--2025. As in the vanilla-option analysis, our primary metric is the terminal hedging error at expiry, scaled by the option price. This common evaluation framework allows us to compare the exotic-option results directly with those for market-traded vanilla options. Additional details on the benchmark construction, payoff definitions, and parameter grids are provided in Appendix~\ref{app:emp_exotic_setup} and Appendix~\ref{app:emp_exotic_payoffs}.

\begin{table}[htbp]
  \centering
  \caption{Path-dependent exotic option hedging performance: error at expiry (\(\times 10^{-3}\)) and win rate against Monte Carlo (MC). NN denotes the neural-network benchmark.}
  \label{tab:synthetic_hedging_performance}
  \begin{tabular}{@{}c c c c c c c c c@{}}
    \toprule
    & & MC & \multicolumn{2}{c}{Sig-It\^o} & \multicolumn{2}{c}{Sig-It\^o-SigKernel} & \multicolumn{2}{c}{NN} \\
    \cmidrule(lr){4-5} \cmidrule(lr){6-7} \cmidrule(lr){8-9}
    Option & Samples & Mean$\downarrow$ & Mean$\downarrow$ & Win\%$\uparrow$ & Mean$\downarrow$ & Win\%$\uparrow$ & Mean$\downarrow$ & Win\%$\uparrow$ \\
    \midrule
    Asian Call    & 146{,}223 & 6.435 & 10.55 & 63.5 & \textbf{5.961} & \textbf{71.8} & 38.23 & 9.21 \\
    Asian Put     & 146{,}223 & 6.830 & 10.80 & 64.1 & \textbf{5.419} & \textbf{75.5} & 34.64 & 12.4 \\
    Lookback Call & 146{,}223 & 26.66 & 19.55 & 72.5 & \textbf{12.90} & \textbf{86.6} & 38.25 & 31.8 \\
    Lookback Put  & 146{,}223 & 16.69 & 22.62 & 61.4 & \textbf{12.46} & \textbf{77.4} & 33.97 & 29.4 \\
    \bottomrule
  \end{tabular}
\end{table}

Table~\ref{tab:synthetic_hedging_performance} shows that the advantage of It\^o-signature hedging becomes substantially more pronounced for path-dependent exotic options, especially with signature-kernel weighting. 
Indeed, \textbf{Sig-It\^o-SigKernel} achieves the lowest mean error for all four exotic payoffs. The improvement is much stronger than in the vanilla setting, particularly for lookback options. As expected,  signature features can capture relevant path information beyond the capability of local state variables.


\begin{table}[htbp]
  \centering
  \caption{Path-dependent exotic options in challenging subsamples: representative short-maturity and extreme-moneyness buckets. Error at expiry (\(\times 10^{-3}\)) and win rate against Monte Carlo.}
  \label{tab:challenging_exotic_performance}
  \begin{tabular}{@{}c c c c c c c c c@{}}
    \toprule
    & & MC & \multicolumn{2}{c}{Sig-It\^o} & \multicolumn{2}{c}{Sig-It\^o-SigKernel} & \multicolumn{2}{c}{NN} \\
    \cmidrule(lr){4-5} \cmidrule(lr){6-7} \cmidrule(lr){8-9}
    Bucket & Samples & Mean$\downarrow$ & Mean$\downarrow$ & Win\%$\uparrow$ & Mean$\downarrow$ & Win\%$\uparrow$ & Mean$\downarrow$ & Win\%$\uparrow$ \\
    \midrule
    \multicolumn{9}{c}{\textbf{Panel A: Short maturity}} \\
    \midrule
    Asian call, \(T=5\)       & 21{,}084 & \textbf{4.819}  & 5.755 & 68.4 & 4.917 & \textbf{70.3} & 15.78 & 12.6 \\
    Asian call, \(T=10\)      & 21{,}049 & 5.642  & 6.836 & 77.2 & \textbf{4.306} & \textbf{83.2} & 22.95 & 9.57 \\
    Asian put, \(T=5\)        & 21{,}084 & 5.115  & 6.089 & 67.0 & \textbf{5.005} & \textbf{71.0} & 15.23 & 14.6 \\
    Asian put, \(T=10\)       & 21{,}049 & 5.929  & 7.484 & 74.2 & \textbf{4.449} & \textbf{83.5} & 22.09 & 13.3 \\
    Lookback call, \(T=5\)    & 21{,}084 & 7.435  & 8.466 & 68.7 & \textbf{5.594} & \textbf{78.8} & 14.26 & 23.2 \\
    Lookback call, \(T=10\)   & 21{,}049 & 13.92  & 11.81 & 79.1 & \textbf{5.913} & \textbf{95.8} & 23.13 & 30.2 \\
    Lookback put, \(T=5\)     & 21{,}084 & 7.731  & 7.900 & 67.9 & \textbf{6.102} & \textbf{75.7} & 15.26 & 25.8 \\
    Lookback put, \(T=10\)    & 21{,}049 & 12.56  & 11.96 & 74.2 & \textbf{6.428} & \textbf{92.3} & 22.54 & 31.8 \\
    \midrule
    \multicolumn{9}{c}{\textbf{Panel B: Extreme moneyness}} \\
    \midrule
    Asian call, \(S/K=0.8\)   & 20{,}889 & 0.123 & 0.003 & 99.9 & \textbf{0.001} & \textbf{100}   & 37.40 & 0.024 \\
    Asian call, \(S/K=0.9\)   & 20{,}889 & 2.089  & 0.337   & 99.1 & \textbf{0.079}   & \textbf{100}   & 41.73 & 0.82 \\
    Asian put, \(S/K=1.1\)    & 20{,}889 & 2.472  & 1.596   & 94.6 & \textbf{0.113}    & \textbf{100}   & 41.17 & 2.00 \\
    Asian put, \(S/K=1.2\)    & 20{,}889 & 0.733 & \textbf{0.026}  & \textbf{100} & \textbf{0.026}  & \textbf{100}   & 41.84 & 0.22 \\
    Lookback call, \(S/K=0.8\)& 20{,}889 & 3.784  & 0.234   & 99.9 & \textbf{0.161}    & \textbf{100}   & 44.41 & 0.99 \\
    Lookback call, \(S/K=0.9\)& 20{,}889 & 16.63  & 9.348   & 88.5 & \textbf{3.687}    & \textbf{98.0}  & 46.03 & 8.49 \\
    Lookback put, \(S/K=1.1\) & 20{,}889 & 10.06  & 11.90   & 79.2 & \textbf{4.036}    & \textbf{97.0}  & 33.94 & 12.9 \\
    Lookback put, \(S/K=1.2\) & 20{,}889 & 3.090  & 4.284   & 88.2 & \textbf{0.919}    & \textbf{99.5}  & 42.55 & 2.17 \\
    \bottomrule
  \end{tabular}
\end{table}

Table~\ref{tab:challenging_exotic_performance} focuses on representative short-maturity and extreme-moneyness subsamples.
It reinforces the same conclusion in the more challenging exotic subsamples: the signature-kernel version is especially effective in short-maturity lookback contracts. The gains are also significant in extreme-moneyness regions. For example, for lookback puts with \(S/K=1.2\), \textbf{Sig-It\^o-SigKernel} reduces the mean error from \(3.090\) under Monte Carlo to \(0.919\) and wins on \(99.5\%\) of contracts. These path-dependent results are stronger than the corresponding vanilla-option gains, due to the capability of signature  to capture pathwise nonlinear patterns.

Note that with real market data, the rolling estimation problem is affected by non-stationarity, volatility regimes, and changes in the local relation between paths and hedging errors. The signature kernel  provides a natural way to address this issue by comparing historical paths in signature space and giving more weight to observations whose recent path histories are similar to the current one. Our implementation is inspired by \citet{guTransportationMarketRate2024} which shows that  signature-based similarity comparison  can help identify seasonality and regime switching in non-stationary time-series forecasting problems. In our setting, the incorporation of \textbf{Sig-It\^o-SigKernel} allows for  a data-adaptive localization of the It\^o-signature hedge for non-stationary financial time series data analysis.

In summary, for vanilla options, \textbf{Sig-It\^o} is competitive with the Black--Scholes delta, and \textbf{Sig-It\^o-SigKernel} presents a clear advantage. On path-dependent exotic options, this advantage is substantially stronger. \textbf{Sig-It\^o-SigKernel} improves on the Monte Carlo benchmark for all four payoff types, with the largest gains for lookback options, where the mean error is reduced by more than one half for calls and by about one quarter for puts. Overall, the results support the proposed It\^o-signature framework as a sample-efficient and practically effective approach to dynamic hedging, especially in path-dependent settings.
(Detailed maturity-, moneyness-, and year-sorted results for vanilla and exotic options, including the full set of alternative signature-based specifications, are collected in Appendix~\ref{app:emp_vanilla_detail} and Appendix~\ref{app:emp_exotic_detail}).

\section{Conclusion}
\label{sec:conclusion}

This paper develops an interpretable and implementable framework for dynamic hedging based on the It\^o signature of asset-price paths, turning nonlinear path-dependent hedging into a linear allocation problem over transparent and directly tradable hedging bases.

It establishes three theoretical results: implementable discrete signature features converge to their continuous-time counterparts, finite It\^o-signature expansions provide payoff population approximation and \(\varepsilon\)-hedging error control, and the finite sample estimation problem admits high-dimensional robust-regression guarantees for coefficient estimation and out-of-sample hedging error.

 In simulations, the It\^o-signature hedge performs strongly against neural-network and Stratonovich-signature benchmarks, especially when training data are limited. It achieves a favorable balance among accuracy, sample efficiency, and computational cost. In the empirical study of S\&P 500 index options, this approach performs robustly across a broad range of contracts, especially for short-maturity and deep out-of-the-money options, and for path-dependent exotic options. The signature-kernel weighted implementation shows that path-similarity technique significantly improves empirical hedging performance when financial time-series data are limited and non-stationary.

Several directions remain open. Theoretically, it would be valuable to extend the approximation results further beyond the constant-volatility benchmark while preserving direct tradability in the original market.  Empirically, it would be natural to study high-frequency data \citep{fan2012vast}, portfolio-level hedging with multiple underlyings, and adaptive model selection for signature depth in rolling real-time applications . It would also be useful to connect the proposed framework more closely with risk attribution \citep{wu2025common}.
The present work is a small step towards bridging signature methods, statistical analysis, and practical hedging, with more transparent machine-learning tools for quantitative finance.
\theendnotes
\putbib
\end{bibunit}

\newpage
\begin{bibunit}
\begin{APPENDICES}
\renewcommand{\thepage}{\arabic{page}}
\setcounter{page}{1}
\counterwithin{figure}{section}
\makeatletter
\renewcommand\p@subfigure{\thefigure}
\makeatother
\counterwithin{equation}{section}
\counterwithin{table}{section}
\counterwithin{theorem}{section}
\counterwithin{corollary}{section}
\counterwithin{proposition}{section}
\counterwithin{definition}{section}
\counterwithin{example}{section}
\counterwithin{lemma}{section}
\counterwithin{remark}{section}
\counterwithin{assumption}{section}

\section*{\centering Electronic Companion}
\addcontentsline{toc}{section}{Electronic Companion}
\section{Comparison with Existing Signature-Based Hedging Methods}
\label{app:lit_comparison}

This appendix provides a brief comparison between our It\^o-signature hedging
framework and those closely related signature-based hedging methods. 

Table~\ref{tab:signature_hedging_comparison} highlights the major differences:
existing methods are primarily based on Stratonovich or rough-path
signatures, while our method uses discretized It\^o-signature coordinates as directly tradable self-financing building blocks. Our approach enables  transparent hedging implementation in both simulation
studies and empirical applications, and supports population-level approximation results in Section~\ref{sec: theoretical analysis} and the finite-sample statistical guarantees in Section~\ref{sec:statistical_learning}. 

\begin{table}[htbp]
\centering
\caption{Comparison of signature-based hedging models.}
\label{tab:signature_hedging_comparison}
\small
\renewcommand{\arraystretch}{1.15}
\resizebox{\textwidth}{!}{%
\begin{tabular}{
    >{\centering\arraybackslash}p{0.28\textwidth}
    >{\centering\arraybackslash}p{0.14\textwidth}
    >{\centering\arraybackslash}p{0.18\textwidth}
    >{\centering\arraybackslash}p{0.14\textwidth}
    >{\centering\arraybackslash}p{0.14\textwidth}
    >{\centering\arraybackslash}p{0.14\textwidth}
}
\toprule
\textbf{Reference} 
& \textbf{Method} 
& \textbf{Model Role} 
& \textbf{Statistical Analysis\(^\dagger\)} 
& \textbf{Sim./num.} 
& \textbf{Market data} \\
\midrule

Ours
& It\^o
& Model-free
& \(\checkmark\)
& \(\checkmark\)
& \(\checkmark\) \\

\midrule
\citet{lyons2020non}
& Stratonovich
& Model-free
& --
& \(\checkmark\)
& \(\checkmark\) \\

\midrule
\citet{abijaber2025a}
& Stratonovich
& Model-specific
& --
& \(\checkmark\)
& -- \\

\midrule
\citet{abijaber2025b}
& Stratonovich
& Model-specific
& --
& \(\checkmark\)
& -- \\

\midrule
\citet{abijaber2025frictions}
& Stratonovich
& Model-specific
& --
& \(\checkmark\)
& -- \\

\midrule
\citet{cirone2025roughkernelhedging}
& Rough path
& Model-free
& --
& \(\checkmark\)
& -- \\

\midrule
\citet{cuchiero2025universal}
& Rough path
& Model-free
& --
& --
& -- \\

\bottomrule
\\[0.05em]
\end{tabular}%
}
\begin{minipage}{0.96\textwidth}
\footnotesize
\emph{Note.} \(^\dagger\) ``Statistical Analysis'' refers specifically to
finite-sample estimation-error or out-of-sample hedging-error bounds for the
learned hedge. A dash indicates that the cited work does not provide this
particular type of finite-sample statistical guarantee, although it may contain
other analytical results.
\end{minipage}
\end{table}

\section{Computational Environment}
\label{app:computational_environment}

We run all simulation and empirical experiments on a Linux server with Ubuntu
20.04.4 LTS. The server is equipped with two Intel(R) Xeon(R) Gold 6238 CPUs at
2.10GHz, providing 88 logical CPU threads in total, and 628 GiB of system memory.
The server also has two NVIDIA GeForce RTX 2080 Ti GPUs, each with 11 GB of GPU
memory. The CUDA version is \texttt{12.8}.

We implement the experiments in Python \texttt{3.9.7}. The main numerical
packages include NumPy \texttt{1.22.4}, pandas \texttt{2.0.1}, SciPy
\texttt{1.10.1}, scikit-learn \texttt{1.2.2}, PyTorch \texttt{2.6.0},
Matplotlib \texttt{3.7.1}, and statsmodels \texttt{0.13.5}. The runtime results
reported in Table~\ref{tab:runtime_by_option_type} are measured on this
computational environment.

\section{Additional simulation details and results}\label{app:simulation_details}
\subsection{Method details}\label{app:sim_method_details}

This appendix summarizes the methods used for simulation comparison. Table~\ref{tab:sim_methods} describes the input information, number of trainable or fitted parameters, and main computational burden for each method. 

\begin{table}[htbp]
\centering
\caption{Methods included in the simulation comparison.}
\label{tab:sim_methods}
\resizebox{\textwidth}{!}{%
\begin{tabular}{m{0.25\textwidth} m{0.33\textwidth} >{\centering\arraybackslash}m{0.12\textwidth} m{0.45\textwidth}}
\toprule
\multicolumn{1}{c}{\textbf{Method}} 
& \multicolumn{1}{c}{\textbf{Inputs}}  
& \multicolumn{1}{c}{\textbf{\# Param}} 
& \multicolumn{1}{c}{\textbf{Main computational burden}} \\
\midrule
Neural network
& Observed price path together with hand-crafted path features 
& 151 
& Nonlinear training over 200 epochs \\
\midrule
Sig-It\^o
& Discrete It\^o-signature coordinates of the observed price path 
& 127 
& Signature computation and linear regression \\
\midrule
Sig-Stratonovich
& Lead--lag path and expected future Stratonovich signature 
& 5{,}461 
& Lead--lag lift, future-signature computation, and high-dimensional optimization \\
\bottomrule
\end{tabular}%
}
\end{table}

\subsection{Closed-form benchmark formulas}\label{app:sim_benchmarks}
In Section \ref{sec:simulation}, we use the following closed-form benchmark formulas under the Black--Scholes setting with \(r=q=0\). These formulas provide the benchmark prices and delta hedges used to evaluate the learned hedging strategies.
Here, \(N(\cdot)\) denotes the standard normal cumulative distribution function and \(n(\cdot)\) denotes its density.

\paragraph{European call option.}
The payoff is
$\max(S_T-K,0).$
The Black--Scholes price at time \(t\) is
$
P_t=S_t N(d_1(t)) - K N(d_2(t)),
$
where
$d_1(t) = \frac{\ln(S_t/K) + \frac{\sigma^2 (T-t)}{2}}{\sigma \sqrt{T-t}},  d_2(t) = d_1(t) - \sigma \sqrt{T-t}.$
The corresponding delta hedge is
$
\Delta_t=N(d_1(t)).
$

\paragraph{European put option.}
The payoff is $
\max(K-S_T,0).
$ The Black--Scholes price at time \(t\) is $
P_t=K N(-d_2(t)) - S_t N(-d_1(t)),
$ with \(d_1(t)\) and \(d_2(t)\) defined as in the European call case.
The corresponding delta hedge is $
\Delta_t=N(d_1(t))-1=-N(-d_1(t)).
$

\paragraph{Geometric average Asian call option.}

The payoff is $
 \max\left(\exp\left(
\frac{1}{T}\int_0^T \log S_u\,du
\right) - K, 0\right).
$ At time \(t\), given the historical path \(\{S_u:0\le u\le t\}\), the price is
$
C_t
=
A_t \exp\left(\mu_X B+\frac{1}{2}\sigma_X^2 B^2\right) N(d_1)-K N(d_2).
$
Here
$
G_t = \exp\left(\frac{1}{t} \int_0^t \ln S_u du\right), 
\quad A_t = G_t^{t/T}, B = \frac{T-t}{T}, 
\mu_X = \ln S_t - \frac{\sigma^2(T-t)}{4}, 
 \sigma_X^2 = \frac{\sigma^2(T-t)}{3}, 
 d_1 = \frac{\ln(A_t/K) + \mu_X B + \sigma_X^2 B^2}{\sigma_X B},
 d_2 = d_1 - \sigma_X B.
$
The corresponding delta hedge at time $t$ is
\[
\Delta_t = {G_t}^{t/T} \cdot S_t^{\frac{T-t}{T}-1} \cdot \exp\left[-\frac{\sigma^2(T-t)^2}{4T} + \frac{\sigma^2(T-t)^3}{6T^2}\right] \cdot \frac{T-t}{T} \cdot N(d_1),
\]
with the same parameters as the price formula.

\paragraph{Floating-strike lookback put option.}
The payoff is
$
\max\!\left(\max_{0\le \tau\le T} S_\tau - S_T,\,0\right)
=
\max_{0\le \tau\le T} S_\tau - S_T.
$
Its pricing formula can be found in \citet{goldman1979,hull2018}. We focus on the special case of $r=q=0$.
Let $
M_t:=\max_{0\le u\le t} S_u,
\tau:=T-t,
$
and define
$
b_1=\frac{\ln(M_t/S_t)+\frac{1}{2}\sigma^2\tau}{\sigma\sqrt{\tau}},
b_2=b_1-\sigma\sqrt{\tau}.
$ The price at time $t$ is given by:
\[
V_t = M_t N(b_1) - S_t N(b_2) + S_t \sigma\sqrt{\tau} n(-b_2) + S_t \frac{\sigma^2\tau}{2} N(-b_2) - S_t \ln(M_t/S_t) N(-b_2),
\]

The delta is piecewise defined based on the relationship between $S_t$ and $M_t$:

Case 1: $S_t < M_t$
\[
\Delta_t = \left(1 + \frac{\sigma^2\tau}{2} - \ln(M_t/S_t)\right) - \left(2 + \frac{\sigma^2\tau}{2} - \ln(M_t/S_t)\right) N(b_2) + \sigma\sqrt{\tau} n(-b_2),
\]

Case 2: $S_t = M_t$
\[
\Delta_t = N\left(\frac{\sigma\sqrt{\tau}}{2}\right) - N\left(-\frac{\sigma\sqrt{\tau}}{2}\right) + \sigma\sqrt{\tau} n\left(\frac{\sigma\sqrt{\tau}}{2}\right) + \frac{\sigma^2\tau}{2} N\left(\frac{\sigma\sqrt{\tau}}{2}\right),
\]
where the notation is the same as above.

\subsection{Detailed performance}\label{app:detailed_product_results}

This appendix reports two additional diagnostics for the simulated hedging experiments.
Complemented to the main text which focuses on terminal hedging errors measured by MSE and MAE, we examine here whether the learned strategies can recover the benchmark trading positions and whether they require a reasonable initial cash account. 

Figure~\ref{fig:position_mse} reports the MSE between the learned trading position and the benchmark Black--Scholes hedge. The pattern is broadly consistent with the terminal-error results in the main text. Specifically,  the It\^o-signature hedge achieves smaller position errors with relatively fewer training samples and remains stable as the sample size increases. The Stratonovich-signature benchmark is less stable and generally has larger position errors, especially with small and medium samples. The neural network improves with more data, yet  is more sensitive to the training size. These results suggest that the It\^o-signature method achieves superior terminal hedging performance in terms of hedging errors, and is capable of learning trading strategies close to the benchmark hedge.

\begin{figure}[htbp]
    \centering
    \subfigure[European Call.\label{fig:position_mse_call}]{
        \includegraphics[width=0.38\linewidth]{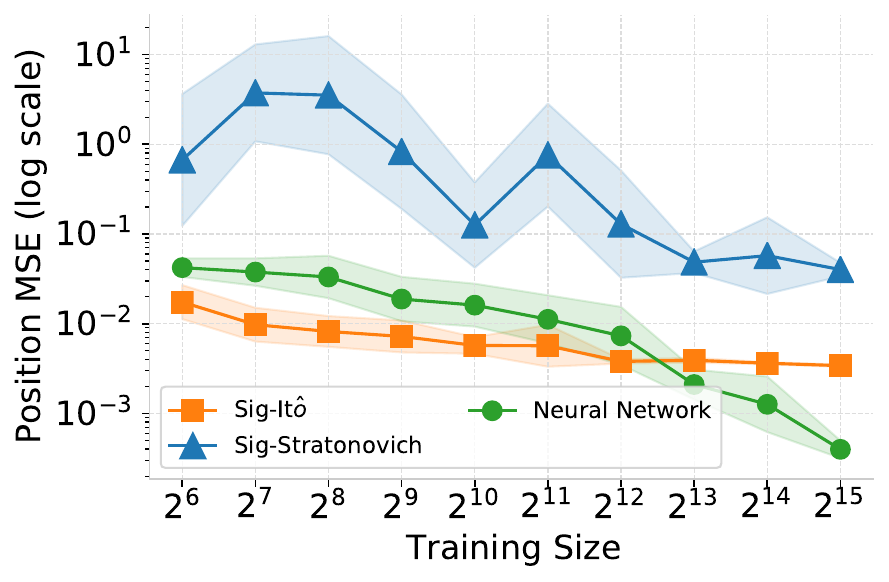}
    }
    \hspace{0.3cm}
    \subfigure[European Put.\label{fig:position_mse_put}]{
        \includegraphics[width=0.38\linewidth]{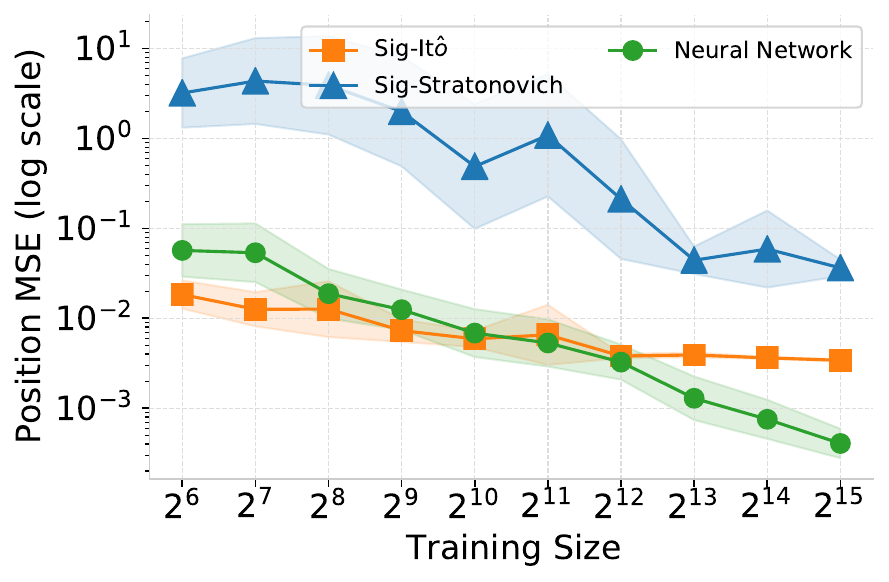}
    }

    \vspace{0.2cm}

    \subfigure[Geometric Average Asian Call.\label{fig:position_mse_asian}]{
        \includegraphics[width=0.38\linewidth]{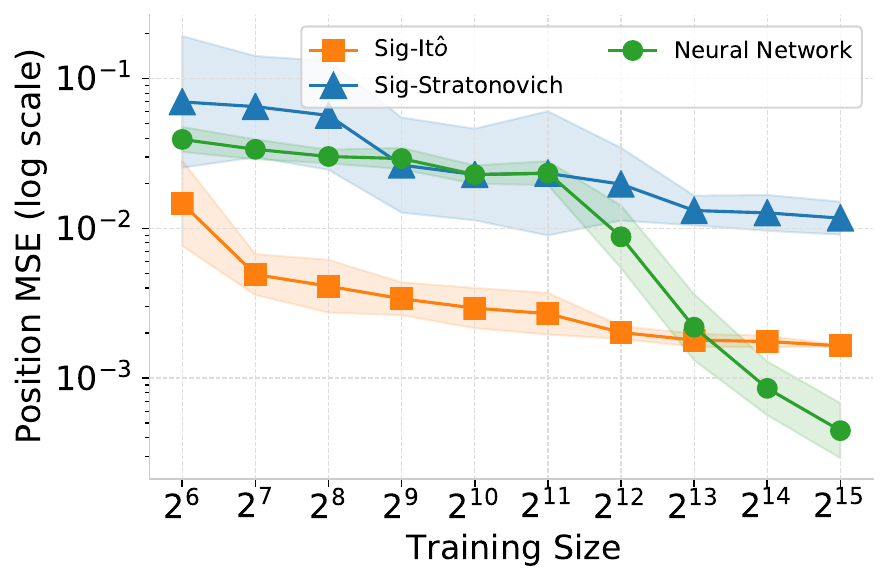}
    }
    \hspace{0.3cm}
    \subfigure[Floating-strike Lookback Put.\label{fig:position_mse_lookback}]{
        \includegraphics[width=0.38\linewidth]{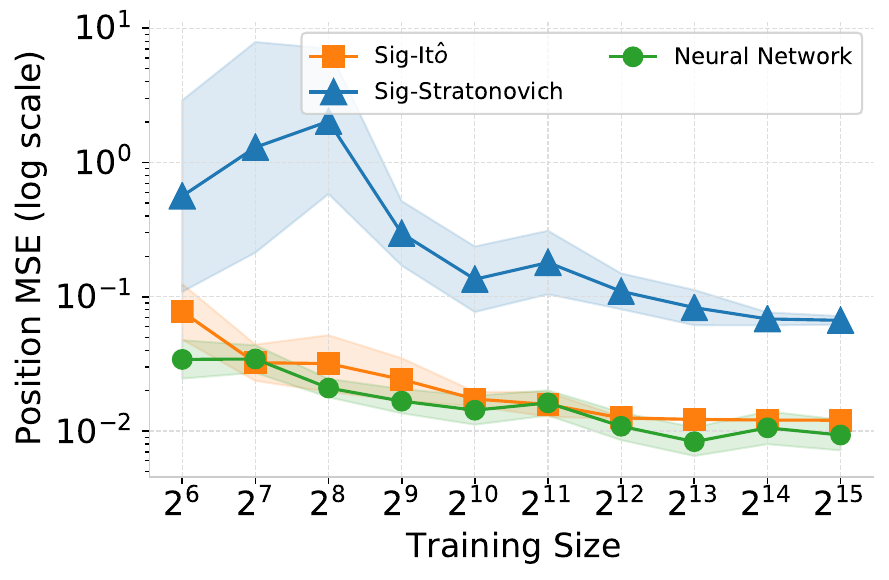}
    }
    \caption{Position MSE as a function of training size.
    Each panel shows the mean MSE with 95\% confidence intervals
    across 10 repeated experiments for Sig-It\^o,
    Sig-Stratonovich, and a neural network baseline.
    Dashed line: Black--Scholes optimal hedge.
    \label{fig:position_mse}}
\end{figure}

We next examine the initial cash account required by learned self-financing hedges for which such a cash component is directly identified. Note that this quantity is not the same as the option price \citep{bertsimas2001hedging}. Here the reported initial cash difference measures the gap between the initial cash component of the learned hedge and that of the benchmark Black--Scholes hedge. For the It\^o-signature hedge, the intercept term in the fitted linear signature portfolio is precisely the initial cash position of the corresponding self-financing strategy. By contrast, the Stratonovich-signature benchmark does not directly identify a self-financing cash account, its initial cash position would have to be assigned externally. We therefore do not report an initial cash difference for the Stratonovich-based hedging method. Figure~\ref{fig:cash_diff} reports the initial cash difference relative to the benchmark Black--Scholes hedge for the It\^o-signature hedge and the neural-network baseline. The It\^o-signature hedge exhibits relatively fast stabilization of the estimated initial cash position as the training size increases, and its initial cash requirement is already close to the benchmark level in small-sample regimes.

\begin{figure}[htbp]
    \centering
    \subfigure[European Call.\label{fig:cash_diff_call}]{
        \includegraphics[width=0.38\linewidth]{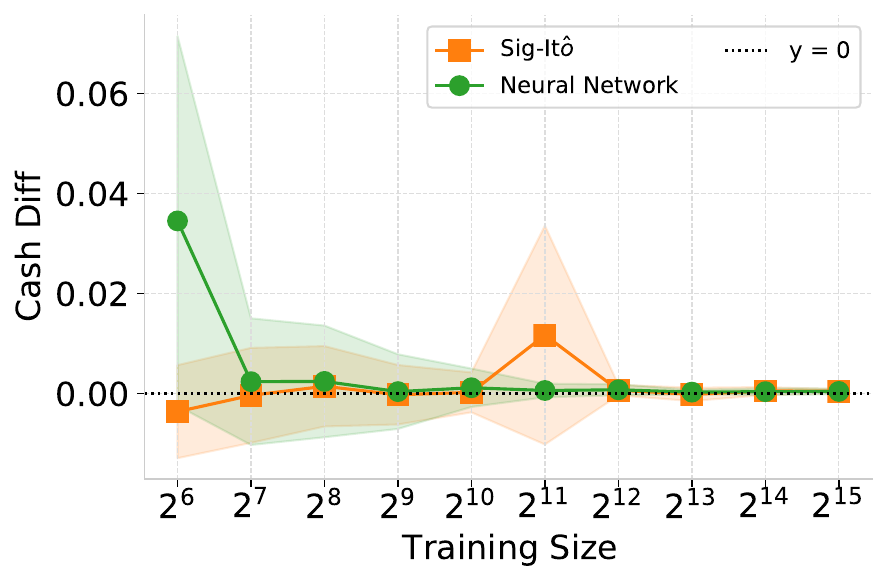}
    }
    \hspace{0.5cm}
    \subfigure[European Put.\label{fig:cash_diff_put}]{
        \includegraphics[width=0.38\linewidth]{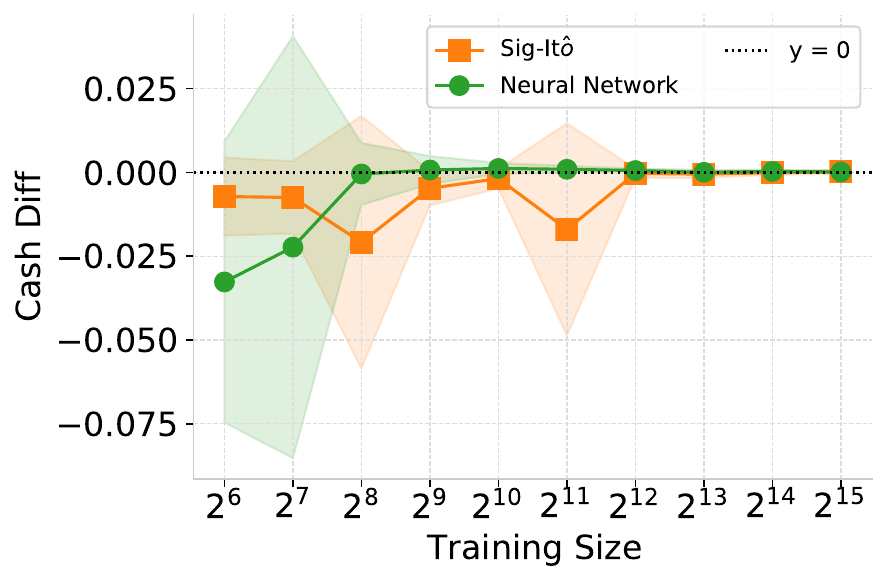}
    }

    \vspace{0.3cm}

    \subfigure[Geometric Average Asian Call.\label{fig:cash_diff_asian}]{
        \includegraphics[width=0.38\linewidth]{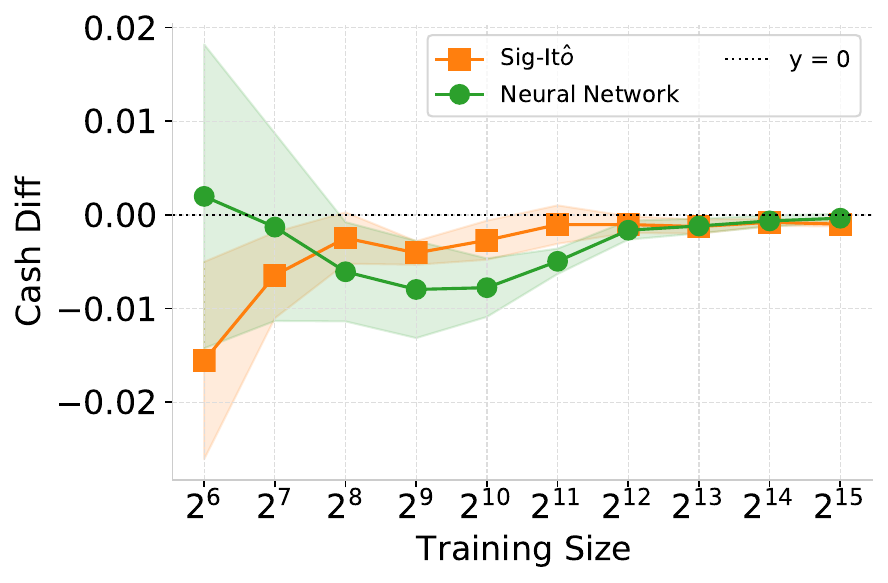}
    }
    \hspace{0.5cm}
    \subfigure[Floating-strike Lookback Put.\label{fig:cash_diff_lookback}]{
        \includegraphics[width=0.38\linewidth]{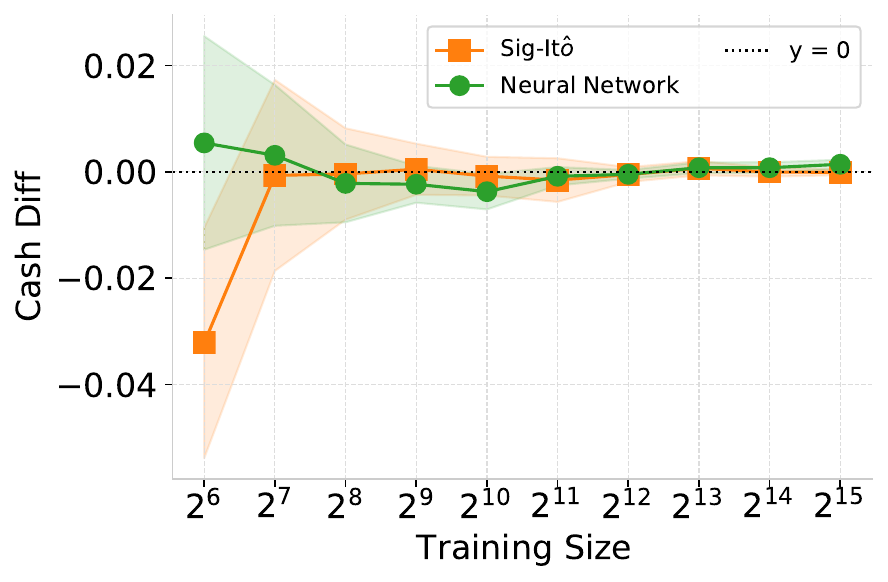}
    }
    \caption{Initial cash difference with the optimal Black--Scholes hedge as a function of training size.
    Each panel shows the mean difference with 95\% confidence intervals
    across 10 repeated experiments for Sig-It\^o and a neural network baseline.
    \label{fig:cash_diff}}
\end{figure}

\section{Additional empirical details and results}\label{app:emp}
\subsection{Comparable signature-based specifications}\label{app:comparable_methods}

This appendix provides additional implementation details and supplementary empirical results for the hedging experiments in Section~\ref{sec:emp}. We first describe the signature-based specifications used throughout the empirical analysis. We then provide further details on the construction of synthetic path-dependent exotic options. Finally, we report heterogeneity results by maturity, moneyness, and calendar year for both vanilla and exotic options.

Table~\ref{tab:comparable_methods} summarizes the signature-based specifications considered in the empirical analysis. All specifications use the same discretized It\^o-signature hedging basis, but differ in how the local linear relation between signature components and option payoffs is estimated. We consider two unweighted linear specifications, OlsSig and LassoSig, and three weighted linear specifications, ExpSig, WlsSig, and WlassoSig. The weighted specifications differ in whether the weights are based on exponential decay or on signature-kernel path similarity. In the main text, we report LassoSig as the representative unweighted signature hedge, denoted by \textbf{Sig-It\^o}, and WlassoSig as the representative signature-kernel-weighted hedge, denoted by \textbf{Sig-It\^o-SigKernel}. The appendix reports the full set of specifications to examine which linear fitting and weighting choices are most effective within the same signature-hedging framework.

For the signature-kernel-weighted specifications, let \(X_t\) denote the current time-augmented return path and let \(X_i\) denote the \(i\)-th historical training path. We compute the signature-kernel distance
\[
d_i
=
\left[
K_{\mathrm{Sig}}(X_t,X_t)
+
K_{\mathrm{Sig}}(X_i,X_i)
-
2K_{\mathrm{Sig}}(X_t,X_i)
\right]^{1/2}.
\]
The weight assigned to the \(i\)-th training observation is
\(
w_i
=
\frac{\exp(-\gamma d_i)}
{\sum_{j\in\mathcal I_t}\exp(-\gamma d_j)},
\)
where \(\mathcal I_t\) is the training window and \(\gamma>0\) controls how strongly the regression focuses on similar paths. Thus, historical paths closer to the current market path receive larger weights, following the path-similarity weighting idea of \citet{guTransportationMarketRate2024}.

\begin{table}[htbp]
\centering
\caption{Comparable signature-based specifications used in the empirical analysis.}
\label{tab:comparable_methods}
\resizebox{\textwidth}{!}{%
\begin{tabular}{m{0.18\textwidth} m{0.82\textwidth}}
\toprule
\multicolumn{1}{c}{\textbf{Method}}  &
\multicolumn{1}{c}{\textbf{Description}} \\
\midrule
\multicolumn{1}{c}{\textbf{ExpSig}}
&
The linear regression is estimated using weights that decay exponentially with the age of the training path:
\[
w_i \propto \exp(-\lambda \cdot \mathrm{age}_i).
\]
This specification gives more weight to recent observations and is designed to capture time variation in the local hedging relation. \\
\midrule
\multicolumn{1}{c}{\textbf{OlsSig}}
&
The linear regression is estimated by ordinary least squares with equal weights for all training paths. This specification serves as a baseline without any weights.
\\
\midrule
\multicolumn{1}{c}{\begin{tabular}{@{}c@{}}\textbf{LassoSig}\\ \textbf{(Sig-It\^o)}\end{tabular}}
&
The linear regression is estimated by an \(\ell_1\)-penalized model with equal weights for all training paths. This specification serves as a simple Lasso baseline without any weights. \\
\midrule
\multicolumn{1}{c}{\textbf{WlsSig}}
&
The linear regression is estimated by weighted least squares using the signature-kernel weights defined above. These weights are larger for historical return paths that are closer to the current market path. This specification follows the path-similarity weighting idea of \citet{guTransportationMarketRate2024,kiraly2019kernels}. \\
\midrule
\multicolumn{1}{c}{\begin{tabular}{@{}c@{}}\textbf{WlassoSig}\\ \textbf{(Sig-It\^o-SigKernel)}\end{tabular}}
&
The linear regression uses the same signature-kernel weights but adds an \(\ell_1\) penalty. This specification combines path-similarity weighting with automatic variable selection. \\
\bottomrule
\end{tabular}%
}
\end{table}

Overall, OlsSig and LassoSig test the It\^o-signature basis without weighting. ExpSig tests simple recency weighting. WlsSig and WlassoSig test whether signature-kernel path similarity improves the local hedge. The main text reports LassoSig as \textbf{Sig-It\^o} and WlassoSig as \textbf{Sig-It\^o-SigKernel}.

\subsection{Additional setup for path-dependent exotic options}
\label{app:emp_exotic_setup}

We construct synthetic exotic options on the S\&P 500 index using a
rolling-window procedure. On each trading day \(t\), we initiate a new virtual
option with maturity \(t+n\), fit the competing hedging rules using historical
data available up to time \(t\), and then evaluate the terminal hedging error at
expiry. This procedure generates one out-of-sample hedging error observation for
each trading day.

More precisely, the procedure is as follows:
\begin{enumerate}
    \item On each trading day \(t\), we initiate a new virtual option with
    maturity \(t+n\).
    \item We train the hedging model using a rolling window of historical price
    data over the interval \([t+1-L-n,t]\), and then use the fitted hedge to
    hedge the virtual option initiated at date \(t\).
    \item At the option's expiration date \(t+n\), we compute the hedging error
    as the difference between the terminal value of the hedging portfolio and
    the option payoff.
\end{enumerate}

Because no closed-form hedge is available for these path-dependent contracts,
we use a Monte Carlo-based hedge as the benchmark, following
\citet{glasserman2004monte} and \citet{broadie1996estimating}. The benchmark
simulates future price paths from a stochastic-volatility model with jumps (SVJ) \citep{bates1996jumps,bakshi1997empirical}. The simulated log-price and variance dynamics are
\begin{align*}
\mathrm d\log S_u
&=
\left(
-\frac12 v_u
-
\lambda_J\bigl(e^{\mu_J+\sigma_J^2/2}-1\bigr)
\right)\mathrm du
+
\sqrt{v_u}\,\mathrm dW_u^S
+
J_u\,\mathrm dN_u, \\
\mathrm dv_u
&=
\kappa(\theta-v_u)\,\mathrm du
+
\xi\sqrt{v_u}\,\mathrm dW_u^v,
\qquad
\mathrm{corr}(\mathrm dW_u^S,\mathrm dW_u^v)=\rho,
\end{align*}
where the jump sizes are i.i.d. \(J_k\sim N(\mu_J,\sigma_J^2)\) and \(N_u\) is a Poisson process with
intensity \(\lambda_J\). The jump compensator is included in the drift so that
the simulation is under the risk-neutral normalization with zero interest rate.

The SVJ parameters are calibrated in a parsimonious rolling manner. At the
initiation date \(t\), we estimate the annualized volatility scale from the most
recent \(m\) daily observations,
\[
\widehat\sigma_t
=
\frac{\operatorname{sd}(\Delta\log S)}{\sqrt{1/252}},
\]
with \(\widehat\sigma_t\) truncated to \([0.01,2]\) for numerical stability. We
then set
\[
v_0=\widehat\sigma_t^2,
\qquad
\theta=\widehat\sigma_t^2,
\qquad
\kappa=3,
\qquad
\xi=\max\{0.05,0.6\widehat\sigma_t\},
\qquad
\rho=-0.5,
\]
and use fixed jump parameters
\[
\lambda_J=3,
\qquad
\mu_J=-0.02,
\qquad
\sigma_J=0.08.
\]
Thus the rolling calibration matches the local realized volatility level, while
the remaining SVJ parameters specify a stable benchmark jump-volatility
environment. The calibrated parameters are held fixed along the hedging path of
the virtual option, while the Monte Carlo simulations are updated conditional on
the realized price history at each rebalancing date.

At each rebalancing date, the benchmark price is estimated as the sample mean of
simulated payoffs over \(N_{\mathrm{MC}}\) paths conditional on the current
realized history. Since the interest rate is set to zero, no additional discount
factor is applied. The benchmark delta is computed by a central finite
difference with relative bump \(\epsilon=1\%\):
\[
\Delta_t
=
\frac{\widehat V(S_t+\delta S)-\widehat V(S_t-\delta S)}
{2\delta S},
\qquad
\delta S=\epsilon S_t.
\]
To reduce the Monte Carlo noise, the up- and down-shifted simulations use common
random numbers: the same Gaussian price shocks, variance shocks, jump shocks,
and Poisson jump counts are used in the two bumped scenarios. The initial
benchmark price uses a fixed random seed, and the dynamic delta simulations use
date-specific fixed seeds, so that the benchmark is reproducible across methods. All hedge positions, including the Monte Carlo benchmark delta and the learned
hedges, are implemented with a one-day delay.

We use 2011--2013 as the development period for hyperparameter selection and
2014--2025 as the genuine out-of-sample test period. The development-period
tuning covers the signature-based hedges, the neural-network benchmark, and the
Monte Carlo benchmark. For the signature methods, we tune the recency-decay parameter \(\lambda\) in ExpSig, the signature-kernel similarity parameter \(\gamma\) in WlsSig and WlassoSig, and the regularization strength in LassoSig and WlassoSig. For the neural-network benchmark, we tune the network
architecture and training parameters. For the Monte Carlo benchmark, we select
the rolling-window length used to estimate the local volatility scale and the
number of simulated paths used to compute benchmark prices and deltas. All
selected hyperparameters are fixed after the development period and are then
used without further adjustment throughout the 2014--2025 out-of-sample
evaluation.

\subsection{Payoff definitions for path-dependent exotic options}\label{app:emp_exotic_payoffs}

We consider four path-dependent contracts: geometric Asian calls, geometric Asian puts, lookback calls, and lookback puts. Let \(S_0,S_1,\ldots,S_T\) denote the observed discrete price path and let \(K\) denote the strike price.

\paragraph{Geometric Asian options.}
Define the geometric average
$
\bar S_{\mathrm{geo}}
=
\exp\left(\frac{1}{T+1}\sum_{i=0}^{T}\ln S_i\right).
$
The payoffs are
$
h_T^{\text{Asian call}}
=
\max\!\left(\bar S_{\mathrm{geo}}-K,0\right), \quad
h_T^{\text{Asian put}}
=
\max\!\left(K-\bar S_{\mathrm{geo}},0\right).
$

\paragraph{Lookback options.}
The payoffs depend on the running extremum of the path:
$
h_T^{\text{Lookback call}}
=
\max\!\left(\max_{0\le i\le T}S_i-K,0\right),\quad
h_T^{\text{Lookback put}}
=
\max\!\left(K-\min_{0\le i\le T}S_i,0\right).
$

We evaluate each exotic option type over the following grids:
$
\text{Maturity grid}=\{5,10,15,20,30,50,100\},$
and 
$\text{Moneyness grid}=\{0.80,0.90,0.95,1.00,1.05,1.10,1.20\}.
$

\subsection{Vanilla-option detailed results}\label{app:emp_vanilla_detail}

This subsection reports additional heterogeneity results for vanilla options. The benchmark is the Black--Scholes delta hedge. We split the empirical sample by maturity, moneyness \(S/K\), and calendar year to examine whether the gains from signature-based hedging are concentrated in specific contract types or remain stable across different option characteristics and market regimes. In addition to the two specifications emphasized in the main text, LassoSig (\textbf{Sig-It\^o}) and WlassoSig (\textbf{Sig-It\^o-SigKernel}), the appendix reports the full set of signature-based linear specifications.

Table~\ref{tab:overall_hedging_performance_total} reports the aggregate vanilla-option results. The signature-kernel-weighted specifications provide the strongest overall performance. WlassoSig achieves the lowest mean terminal hedging error in all cases. WlsSig and WlassoSig also deliver the highest winning percentages against Black--Scholes. Comparing the unweighted linear estimators, LassoSig is very close to OlsSig and slightly improves the overall mean error from \(5.755\) to \(5.744\). Comparing the weighting schemes, exponential decay improves the win rate relative to the benchmark but is much weaker than signature-kernel weighting. The large increase in win rate from about \(73\%\) for the unweighted OLS/Lasso specifications to \(83.9\%\) for the signature-kernel-weighted specifications suggests that selecting historically similar paths is more informative than simply emphasizing recent observations.

\begin{table}[htbp]
\centering
\caption{Vanilla option hedging performance: error at expiry (\(\times 10^{-3}\)) and win rate against Black--Scholes.}
\label{tab:overall_hedging_performance_total}
\resizebox{\textwidth}{!}{%
\begin{tabular}{@{} c c c c c c c c c c c c c c c @{}}
\toprule
& & BS & \multicolumn{2}{c}{ExpSig} & \multicolumn{2}{c}{OlsSig} & \multicolumn{2}{c}{LassoSig} & \multicolumn{2}{c}{WlsSig} & \multicolumn{2}{c}{WlassoSig} & \multicolumn{2}{c}{NN} \\
\cmidrule(lr){4-5} \cmidrule(lr){6-7} \cmidrule(lr){8-9} \cmidrule(lr){10-11} \cmidrule(lr){12-13} \cmidrule(lr){14-15}
Option & Samples & Mean$\downarrow$ & Mean$\downarrow$ & Win\%$\uparrow$ & Mean$\downarrow$ & Win\%$\uparrow$ & Mean$\downarrow$ & Win\%$\uparrow$ & Mean$\downarrow$ & Win\%$\uparrow$ & Mean$\downarrow$ & Win\%$\uparrow$ & Mean$\downarrow$ & Win\%$\uparrow$ \\
\midrule
Overall & 436{,}135 & 6.007 & 5.968 & 70.7 & 5.755 & 73.4 & 5.744 & 73.3 & 5.463 & \textbf{83.9} & \textbf{5.445} & \textbf{83.9} & 8.836 & 26.1 \\
Call    & 162{,}527 & 6.178 & 6.491 & 70.8 & 6.253 & 73.2 & 6.213 & 73.1 & 5.722 & \textbf{88.9} & \textbf{5.705} & \textbf{88.9} & 9.626 & 23.3 \\
Put     & 273{,}608 & 5.906 & 5.658 & 70.6 & 5.460 & 73.5 & 5.466 & 73.4 & 5.310 & \textbf{80.9} & \textbf{5.290} & \textbf{80.9} & 8.349 & 27.8 \\
\bottomrule
\end{tabular}%
}
\end{table}

Tables~\ref{tab:vanilla_by_maturity_panel}--\ref{tab:vanilla_by_year_panel} provide the heterogeneity analysis by maturity, moneyness, and calendar year. The maturity split shows that the gains from signature-kernel weighting are strongest for short- and medium-maturity contracts. For call options with maturities below 30 days, WlsSig or WlassoSig almost always has the lowest signature-based mean error and the winning percentage is typically above \(90\%\) for maturities from 5 to 30 days, a similar pattern also appears for puts. For longer maturities, especially beyond 50 days, Black--Scholes sometimes has the lowest mean error, but WlsSig and WlassoSig still deliver the highest winning percentages even in those long-maturity buckets, indicating that the path-similarity weighting remains useful for the majority of observations. The moneyness split leads to a similar conclusion. Around the near-the-money region, WlassoSig reduces mean errors relative to the unweighted specifications. In extreme moneyness buckets, the signature-kernel-weighted specifications still usually obtain the highest winning percentages. The year-by-year results further show that the advantage of signature-kernel weighting is persistent across market regimes. WlsSig or WlassoSig achieves the highest winning percentage in every call and put year, including volatile periods such as 2020 and 2022.

\begin{table}[htbp]
\centering
\caption{Vanilla option hedging performance by maturity: error at expiry (\(\times 10^{-3}\)) and win rate against Black--Scholes.}
\label{tab:vanilla_by_maturity_panel}
\resizebox{\textwidth}{!}{%
\begin{tabular}{@{} c c c c c c c c c c c c c c c @{}}
\toprule
& & BS & \multicolumn{2}{c}{ExpSig} & \multicolumn{2}{c}{OlsSig} & \multicolumn{2}{c}{LassoSig} & \multicolumn{2}{c}{WlsSig} & \multicolumn{2}{c}{WlassoSig} & \multicolumn{2}{c}{NN} \\
\cmidrule(lr){4-5} \cmidrule(lr){6-7} \cmidrule(lr){8-9} \cmidrule(lr){10-11} \cmidrule(lr){12-13} \cmidrule(lr){14-15}
Maturity (days) & Samples & Mean$\downarrow$ & Mean$\downarrow$ & Win\%$\uparrow$ & Mean$\downarrow$ & Win\%$\uparrow$ & Mean$\downarrow$ & Win\%$\uparrow$ & Mean$\downarrow$ & Win\%$\uparrow$ & Mean$\downarrow$ & Win\%$\uparrow$ & Mean$\downarrow$ & Win\%$\uparrow$ \\
\midrule
\multicolumn{15}{c}{\textbf{Panel A: Call options}} \\
\midrule
$<$5 & 11{,}784 & 3.043 & 2.783 & 72.0 & 2.734 & 74.2 & 2.719 & 73.9 & \textbf{2.679} & \textbf{85.4} & 2.683 & 85.3 & 3.111 & 16.1 \\
5-10 & 15{,}360 & 3.973 & 3.040 & 72.0 & 2.952 & 71.3 & 2.946 & 71.8 & 2.725 & \textbf{96.1} & \textbf{2.684} & \textbf{96.1} & 4.541 & 27.5 \\
10-15 & 17{,}834 & 6.711 & 4.494 & 72.4 & 4.164 & 74.8 & 4.191 & 74.9 & \textbf{3.780} & \textbf{96.7} & 3.807 & \textbf{96.7} & 7.026 & 26.6 \\
15-20 & 22{,}273 & 6.974 & 4.939 & 76.1 & 4.658 & 80.0 & 4.665 & 80.2 & 4.311 & \textbf{94.3} & \textbf{4.309} & \textbf{94.3} & 8.835 & 25.8 \\
20-25 & 30{,}452 & 5.567 & 4.453 & 75.4 & 4.250 & 78.0 & 4.233 & 77.9 & \textbf{3.725} & 93.8 & \textbf{3.725} & \textbf{93.9} & 7.605 & 24.7 \\
25-30 & 19{,}497 & 6.824 & 6.290 & 76.9 & 6.160 & 79.6 & 6.174 & 79.8 & \textbf{5.193} & \textbf{94.0} & 5.238 & \textbf{94.0} & 9.468 & 23.2 \\
30-50 & 21{,}350 & 5.804 & 6.201 & 68.9 & 6.186 & 70.8 & 6.167 & 71.1 & \textbf{5.451} & \textbf{84.2} & 5.454 & \textbf{84.2} & 9.813 & 22.1 \\
50-100 & 13{,}523 & \textbf{7.815} & 13.43 & 55.8 & 12.86 & 59.1 & 12.90 & 58.7 & 12.0 & \textbf{73.8} & 12.0 & \textbf{73.8} & 14.51 & 22.1 \\
$>$100 & 10{,}454 & \textbf{9.595} & 20.53 & 51.8 & 19.78 & 55.5 & 19.14 & 54.9 & 19.26 & 62.9 & 18.91 & \textbf{63.0} & 29.34 & 14.6 \\
\midrule
\multicolumn{15}{c}{\textbf{Panel B: Put options}} \\
\midrule
$<$5 & 12{,}547 & 3.440 & 2.571 & 89.7 & 2.532 & 90.8 & 2.537 & 90.5 & 2.513 & \textbf{98.9} & \textbf{2.501} & 98.8 & 2.690 & 34.0 \\
5-10 & 15{,}593 & 4.373 & 3.217 & 82.5 & 3.111 & 84.1 & 3.112 & 84.2 & \textbf{2.852} & \textbf{98.6} & \textbf{2.852} & \textbf{98.6} & 4.449 & 37.1 \\
10-15 & 22{,}251 & 4.860 & 3.725 & 78.7 & 3.473 & 81.9 & 3.493 & 81.8 & 3.183 & \textbf{95.8} & \textbf{3.150} & \textbf{95.8} & 5.910 & 34.8 \\
15-20 & 30{,}261 & 6.327 & 4.332 & 81.1 & 4.099 & 85.8 & 4.113 & 86.0 & 3.888 & \textbf{95.9} & \textbf{3.882} & \textbf{95.9} & 7.190 & 31.7 \\
20-25 & 52{,}988 & 4.960 & 3.567 & 81.1 & 3.374 & 84.5 & 3.380 & 84.5 & 3.318 & \textbf{92.2} & \textbf{3.306} & \textbf{92.2} & 6.633 & 30.1 \\
25-30 & 38{,}666 & 4.882 & 3.374 & 74.9 & 3.235 & 77.1 & 3.242 & 76.9 & 3.065 & \textbf{82.5} & \textbf{3.060} & \textbf{82.5} & 5.244 & 28.0 \\
30-50 & 45{,}519 & 5.764 & 5.019 & 62.4 & 4.805 & 66.0 & 4.825 & 65.6 & 4.765 & \textbf{70.7} & \textbf{4.746} & \textbf{70.7} & 7.786 & 25.2 \\
50-100 & 27{,}342 & \textbf{8.157} & 8.770 & 56.6 & 8.663 & 58.3 & 8.732 & 57.7 & 8.506 & \textbf{62.1} & 8.496 & \textbf{62.1} & 12.15 & 23.3 \\
$>$100 & 28{,}441 & \textbf{9.427} & 16.34 & 39.4 & 15.94 & 41.7 & 15.85 & 41.3 & 15.65 & \textbf{46.6} & 15.58 & \textbf{46.6} & 21.66 & 12.9 \\
\bottomrule
\end{tabular}%
}
\end{table}

\begin{table}[htbp]
\centering
\caption{Vanilla option hedging performance by moneyness: error at expiry (\(\times 10^{-3}\)) and win rate against Black--Scholes.}
\label{tab:vanilla_by_moneyness_panel}
\resizebox{\textwidth}{!}{%
\begin{tabular}{@{} c c c c c c c c c c c c c c c @{}}
\toprule
& & BS & \multicolumn{2}{c}{ExpSig} & \multicolumn{2}{c}{OlsSig} & \multicolumn{2}{c}{LassoSig} & \multicolumn{2}{c}{WlsSig} & \multicolumn{2}{c}{WlassoSig} & \multicolumn{2}{c}{NN} \\
\cmidrule(lr){4-5} \cmidrule(lr){6-7} \cmidrule(lr){8-9} \cmidrule(lr){10-11} \cmidrule(lr){12-13} \cmidrule(lr){14-15}
Moneyness (S/K) & Samples & Mean$\downarrow$ & Mean$\downarrow$ & Win\%$\uparrow$ & Mean$\downarrow$ & Win\%$\uparrow$ & Mean$\downarrow$ & Win\%$\uparrow$ & Mean$\downarrow$ & Win\%$\uparrow$ & Mean$\downarrow$ & Win\%$\uparrow$ & Mean$\downarrow$ & Win\%$\uparrow$ \\
\midrule
\multicolumn{15}{c}{\textbf{Panel A: Call options}} \\
\midrule
$<$0.8      & 6{,}096  & 0.176 & 0.104 & 98.5 & 0.101 & 98.3 & 0.127 & 98.4 & \textbf{0.083} & \textbf{99.8} & 0.084 & 99.7 & 0.528 & 8.70  \\
0.8-0.9     & 11{,}996 & \textbf{1.300} & 2.044 & 83.3 & 2.030 & 83.1 & 1.944 & 84.1 & 1.697 & 93.4 & 1.655 & \textbf{93.6} & 2.724 & 17.7 \\
0.9-0.95    & 34{,}161 & \textbf{2.836} & 4.501 & 61.9 & 4.374 & 61.8 & 4.282 & 62.7 & 3.795 & \textbf{85.9} & 3.735 & \textbf{85.9} & 4.360 & 19.0 \\
0.95-1      & 68{,}676 & 5.887 & 5.822 & 66.8 & 5.572 & 70.7 & 5.545 & 70.4 & 5.033 & \textbf{87.9} & \textbf{5.030} & 87.8 & 7.750 & 23.6 \\
1-1.05      & 27{,}712 & 10.77 & 10.06 & 76.3 & 9.501 & 81.0 & 9.501 & 80.8 & 8.772 & \textbf{92.3} & \textbf{8.771} & \textbf{92.3} & 15.67 & 30.4 \\
1.05-1.1    & 9{,}147  & \textbf{12.93} & 14.00 & 78.7 & 13.80 & 79.5 & 13.78 & 78.7 & 13.35 & \textbf{88.6} & 13.43 & \textbf{88.6} & 22.76 & 27.7 \\
1.1-1.2     & 4{,}143  & \textbf{14.49} & 14.98 & 79.5 & 15.06 & 80.0 & 14.96 & 80.0 & 14.74 & 83.3 & 14.57 & \textbf{83.7} & 31.51 & 25.4 \\
1.2-1.3     & 504     & 14.85 & 11.58 & 79.5 & 11.58 & 79.5 & \textbf{11.56} & 79.5 & 11.95 & 82.0 & 11.96 & \textbf{82.8} & 61.27 & 27.9 \\
1.3-1.5     & 63      & 18.19 & 14.67 & 59.1 & 14.70 & 54.5 & 14.72 & 54.5 & \textbf{14.65} & 63.6 & \textbf{14.65} & 63.6 & 20.58 & \textbf{64.7} \\
$>$1.5      & 29      & \textbf{17.57} & 18.02 & \textbf{10.3} & 18.02 & \textbf{10.3} & 18.02 & \textbf{10.3} & 18.02 & \textbf{10.3} & 18.02 & \textbf{10.3} & 27.34 & 0.0  \\
\midrule
\multicolumn{15}{c}{\textbf{Panel B: Put options}} \\
\midrule
0.5-0.7     & 22      & 56.94 & \textbf{7.914} & \textbf{100} & \textbf{7.914} & \textbf{100} & 7.926 & \textbf{100} & \textbf{7.914} & \textbf{100} & \textbf{7.914} & \textbf{100} & 105.3 & 0.0  \\
0.7-0.8     & 102     & 27.04 & \textbf{13.51} & \textbf{80.6} & 13.53 & \textbf{80.6} & 13.53 & \textbf{80.6} & 13.52 & \textbf{80.6} & 13.53 & \textbf{80.6} & 23.56 & 52.4 \\
0.8-0.9     & 836     & 30.18 & 31.64 & 72.8 & 30.95 & 75.2 & 30.18 & 75.2 & \textbf{28.30} & 83.5 & \textbf{28.30} & \textbf{84.0} & 48.75 & 35.7 \\
0.9-0.95    & 6{,}818 & \textbf{15.36} & 19.42 & 67.9 & 18.14 & 72.6 & 18.00 & 72.5 & 17.51 & 82.0 & 17.45 & \textbf{82.4} & 28.17 & 28.1 \\
0.95-1      & 48{,}833 & 9.023 & 9.174 & 74.4 & 8.750 & 80.2 & 8.701 & 79.7 & 8.468 & \textbf{91.6} & \textbf{8.410} & \textbf{91.6} & 13.96 & 28.3 \\
1-1.05      & 67{,}587 & 7.066 & 6.404 & 70.4 & 6.113 & 76.4 & 6.176 & 76.1 & 6.018 & \textbf{85.9} & \textbf{6.016} & 85.8 & 9.931 & 27.9 \\
1.05-1.1    & 46{,}390 & 5.246 & 4.999 & 67.3 & 4.836 & 69.7 & 4.861 & 69.4 & 4.646 & \textbf{79.0} & \textbf{4.618} & \textbf{79.0} & 7.026 & 27.2 \\
1.1-1.2     & 48{,}628 & 3.881 & 3.708 & 70.6 & 3.752 & 70.1 & 3.749 & 70.3 & 3.642 & \textbf{74.7} & \textbf{3.634} & \textbf{74.7} & 5.378 & 26.1 \\
$>$1.2      & 54{,}392 & 2.440 & \textbf{1.771} & 70.8 & 1.785 & 70.3 & 1.792 & 70.2 & 1.781 & \textbf{72.0} & 1.776 & \textbf{72.0} & 2.251 & 29.1 \\
\bottomrule
\end{tabular}%
}
\end{table}

\begin{table}[htbp]
\centering
\caption{Vanilla option hedging performance by year: error at expiry (\(\times 10^{-3}\)) and win rate against Black--Scholes.}
\label{tab:vanilla_by_year_panel}
\resizebox{\textwidth}{!}{%
\begin{tabular}{@{} c c c c c c c c c c c c c c c @{}}
\toprule
& & BS & \multicolumn{2}{c}{ExpSig} & \multicolumn{2}{c}{OlsSig} & \multicolumn{2}{c}{LassoSig} & \multicolumn{2}{c}{WlsSig} & \multicolumn{2}{c}{WlassoSig} & \multicolumn{2}{c}{NN} \\
\cmidrule(lr){4-5} \cmidrule(lr){6-7} \cmidrule(lr){8-9} \cmidrule(lr){10-11} \cmidrule(lr){12-13} \cmidrule(lr){14-15}
Year & Samples & Mean$\downarrow$ & Mean$\downarrow$ & Win\%$\uparrow$ & Mean$\downarrow$ & Win\%$\uparrow$ & Mean$\downarrow$ & Win\%$\uparrow$ & Mean$\downarrow$ & Win\%$\uparrow$ & Mean$\downarrow$ & Win\%$\uparrow$ & Mean$\downarrow$ & Win\%$\uparrow$ \\
\midrule
\multicolumn{15}{c}{\textbf{Panel A: Call options}} \\
\midrule
2014 & 1{,}906 & 3.261 & 4.104 & 65.7 & 3.919 & 69.1 & \textbf{3.014} & 72.9 & 3.510 & 82.2 & 3.180 & \textbf{82.6} & 7.594 & 18.4 \\
2015 & 2{,}702 & 8.724 & 7.687 & 69.5 & 7.255 & 71.6 & 7.480 & 70.9 & 6.901 & 86.5 & \textbf{6.818} & \textbf{87.3} & 13.34 & 38.4 \\
2016 & 4{,}798 & 3.327 & 3.122 & 70.1 & 3.283 & 66.1 & 3.278 & 65.6 & 2.991 & \textbf{85.6} & \textbf{2.990} & 85.4 & 4.222 & 25.5 \\
2017 & 6{,}672 & 2.617 & 2.107 & 79.6 & 2.130 & 76.0 & 2.050 & 75.1 & 1.825 & \textbf{95.9} & \textbf{1.780} & \textbf{95.9} & 3.426 & 24.1 \\
2018 & 11{,}291 & 6.224 & 6.108 & 75.5 & 5.463 & 81.3 & 5.514 & 80.7 & \textbf{4.911} & \textbf{95.0} & 4.932 & 94.8 & 9.105 & 33.2 \\
2019 & 11{,}447 & 4.936 & 5.299 & 57.9 & 4.699 & 60.9 & 4.615 & 61.3 & 3.801 & 89.0 & \textbf{3.799} & \textbf{89.1} & 6.133 & 18.2 \\
2020 & 18{,}203 & \textbf{10.23} & 14.04 & 60.0 & 13.01 & 63.1 & 13.01 & 63.6 & 11.59 & \textbf{83.2} & 11.65 & \textbf{83.2} & 15.13 & 27.4 \\
2021 & 22{,}247 & \textbf{2.772} & 7.183 & 56.4 & 7.564 & 54.0 & 7.541 & 54.8 & 7.255 & 69.4 & 7.258 & \textbf{69.5} & 8.794 & 12.7 \\
2022 & 23{,}449 & 10.03 & 7.562 & 80.3 & 7.340 & 85.6 & 7.345 & 85.1 & \textbf{6.794} & \textbf{94.4} & \textbf{6.794} & \textbf{94.4} & 15.60 & 25.0 \\
2023 & 19{,}491 & 3.654 & 2.573 & 78.9 & 2.636 & 80.9 & 2.642 & 80.5 & \textbf{2.562} & \textbf{92.4} & \textbf{2.562} & 92.3 & 6.207 & 17.9 \\
2024 & 23{,}927 & 3.759 & 3.232 & 79.4 & 3.129 & 82.8 & 3.076 & 83.0 & 3.025 & \textbf{96.9} & \textbf{2.971} & 96.8 & 7.198 & 23.6 \\
2025 & 16{,}394 & 10.36 & 9.002 & 69.5 & 8.558 & 74.6 & 8.386 & 74.6 & 7.634 & \textbf{93.4} & \textbf{7.523} & \textbf{93.4} & 10.28 & 28.9 \\
\midrule
\multicolumn{15}{c}{\textbf{Panel B: Put options}} \\
\midrule
2014 & 5{,}502 & \textbf{2.356} & 4.759 & 17.6 & 4.460 & 19.3 & 4.628 & 17.0 & 4.361 & 21.0 & 4.361 & \textbf{21.1} & 6.543 & 11.8 \\
2015 & 6{,}464 & 8.163 & \textbf{5.894} & 91.5 & 5.960 & 91.4 & 6.175 & 90.5 & 5.916 & \textbf{92.3} & 5.929 & \textbf{92.3} & 9.019 & 49.5 \\
2016 & 11{,}333 & 4.864 & 4.417 & 71.7 & 4.182 & 74.0 & 4.210 & 74.0 & 4.043 & 80.5 & \textbf{4.042} & \textbf{80.6} & 4.860 & 38.3 \\
2017 & 16{,}281 & \textbf{2.556} & 3.889 & 34.7 & 3.747 & 35.7 & 3.677 & 34.9 & 3.735 & \textbf{38.3} & 3.704 & \textbf{38.3} & 6.308 & 12.1 \\
2018 & 20{,}410 & 5.172 & 4.610 & 76.7 & 4.611 & 78.1 & 4.728 & 77.3 & 4.477 & 81.5 & \textbf{4.451} & \textbf{81.6} & 6.679 & 26.4 \\
2019 & 22{,}506 & 4.738 & 3.456 & 72.5 & 3.189 & 74.0 & 3.194 & 74.3 & 3.133 & 84.5 & \textbf{3.124} & \textbf{84.6} & 5.922 & 28.7 \\
2020 & 27{,}585 & 12.35 & 11.18 & 72.8 & 10.74 & 74.6 & 10.74 & 74.5 & 10.46 & \textbf{88.0} & \textbf{10.36} & 87.9 & 14.63 & 34.8 \\
2021 & 29{,}884 & \textbf{4.472} & 9.257 & 51.7 & 9.069 & 54.7 & 9.067 & 54.6 & 9.088 & \textbf{59.3} & 9.088 & \textbf{59.3} & 9.916 & 18.9 \\
2022 & 31{,}320 & 6.729 & 4.929 & 81.1 & 4.725 & 86.4 & 4.733 & 86.2 & \textbf{4.540} & \textbf{96.2} & 4.541 & \textbf{96.2} & 9.954 & 28.3 \\
2023 & 35{,}459 & 3.988 & 3.935 & 77.4 & 3.687 & 80.9 & 3.690 & 81.1 & \textbf{3.461} & \textbf{90.5} & 3.463 & \textbf{90.5} & 5.683 & 29.5 \\
2024 & 41{,}584 & 5.055 & 3.981 & 74.9 & 3.936 & 77.7 & 3.946 & 77.6 & 3.888 & \textbf{82.1} & \textbf{3.878} & \textbf{82.1} & 7.520 & 22.1 \\
2025 & 25{,}280 & 8.091 & 6.097 & 83.4 & 5.808 & 88.5 & 5.694 & 88.5 & 5.358 & \textbf{98.1} & \textbf{5.312} & \textbf{98.1} & 8.829 & 40.6 \\
\bottomrule
\end{tabular}%
}
\end{table}

Across all three splits, LassoSig remains the strongest unweighted specification or is very close to OlsSig. This supports its use as the main unweighted signature hedge, because the \(\ell_1\) penalty reduces sensitivity to noisy signature coordinates without sacrificing predictive power. The comparison across weighting schemes is clearer: exponential recency weighting is useful but weaker, while signature-kernel weighting provides the most stable improvement, especially when combined with Lasso in WlassoSig.

\subsection{Exotic-option detailed results}\label{app:emp_exotic_detail}

This subsection reports additional heterogeneity results for the synthetic path-dependent exotic options. The benchmark is the Monte Carlo hedge described in Section~\ref{app:emp_exotic_setup}. Since Asian and lookback payoffs depend directly on the realized price path, these contracts provide a direct test of whether the It\^o-signature features capture path information that is useful for dynamic hedging. As in the vanilla-option analysis, the appendix reports the full set of signature-based linear specifications. The comparison is particularly informative in this setting because path-dependent payoffs generate a richer and more collinear signature feature space, making the distinction between different linear models and between different weighting schemes more important.

Table~\ref{tab:synthetic_hedging_performance_total} reports the aggregate results for the path-dependent exotic options. WlassoSig is the strongest specification in mean-error terms for all four payoff types. The gains are especially large for lookback options, where the payoff depends on the running maximum or minimum and therefore contains stronger path dependence. The comparison between OlsSig and LassoSig is much sharper than in the vanilla-option case, which shows that the \(\ell_1\) penalty is especially valuable for path-dependent hedging, where many signature coordinates are potentially relevant but only a sparse subset is useful in a local rolling window. Comparing the weighting schemes, WlsSig substantially improves the winning percentage and WlassoSig delivers the best balance between low mean error and high win rate. This indicates that signature-kernel path-similarity weighting is more effective than simple recency weighting, and that combining it with sparse coefficient selection is crucial for stable exotic-option hedging.

\begin{table}[htbp]
  \centering
  \caption{Path-dependent exotic option hedging performance: error at expiry (\(\times 10^{-3}\)) and win rate against Monte Carlo.}
  \label{tab:synthetic_hedging_performance_total}
  \resizebox{\textwidth}{!}{%
    \begin{tabular}{@{}c c c c c c c c c c c c c c c@{}}
      \toprule
      & & MC & \multicolumn{2}{c}{ExpSig} & \multicolumn{2}{c}{OlsSig} & \multicolumn{2}{c}{LassoSig} & \multicolumn{2}{c}{WlsSig} & \multicolumn{2}{c}{WlassoSig} & \multicolumn{2}{c}{NN} \\
      \cmidrule(lr){4-5} \cmidrule(lr){6-7} \cmidrule(lr){8-9} \cmidrule(lr){10-11} \cmidrule(lr){12-13} \cmidrule(lr){14-15}
      Option & Samples & Mean$\downarrow$ & Mean$\downarrow$ & Win\%$\uparrow$ & Mean$\downarrow$ & Win\%$\uparrow$ & Mean$\downarrow$ & Win\%$\uparrow$ & Mean$\downarrow$ & Win\%$\uparrow$ & Mean$\downarrow$ & Win\%$\uparrow$ & Mean$\downarrow$ & Win\%$\uparrow$ \\
      \midrule
      Asian Call    & 146{,}223 & 6.435 & 23.39 & 49.4 & 22.67 & 49.7 & 10.55 & 63.5 & 7.087 & \textbf{72.6} & \textbf{5.961} & 71.8 & 38.23 & 9.21 \\
      Asian Put     & 146{,}223 & 6.830 & 21.16 & 53.4 & 20.27 & 53.5 & 10.80 & 64.1 & 6.833 & 75.1 & \textbf{5.419} & \textbf{75.5} & 34.64 & 12.4 \\
      Lookback Call & 146{,}223 & 26.66 & 44.77 & 58.4 & 39.87 & 57.4 & 19.55 & 72.5 & 27.04 & 80.2 & \textbf{12.90} & \textbf{86.6} & 38.25 & 31.8 \\
      Lookback Put  & 146{,}223 & 16.69 & 48.18 & 49.3 & 39.67 & 47.8 & 22.62 & 61.4 & 21.82 & 71.1 & \textbf{12.46} & \textbf{77.4} & 33.97 & 29.4 \\
      \bottomrule
    \end{tabular}%
  }
\end{table}

Tables~\ref{tab:asian_by_maturity_panel}--\ref{tab:lookback_by_year_panel} report the heterogeneity results for Asian and lookback options by maturity, moneyness, and calendar year. Across these splits, the main pattern is that signature-kernel weighting is more effective than both equal weighting and exponential recency weighting. For Asian options, WlsSig and WlassoSig dominate the unweighted specifications in most maturity and moneyness buckets. The year-by-year results show the same ranking: LassoSig improves over OlsSig, while the signature-kernel-weighted specifications deliver the most stable win rates across market regimes. The evidence is stronger for lookback options, where path dependence is more pronounced. WlassoSig is consistently the most stable specification across maturity, moneyness, and year splits, whereas WlsSig alone can occasionally be noisy. This contrast suggests that the signature kernel helps identify historically relevant paths, while the \(\ell_1\) penalty further controls estimation noise and removes redundant signature coordinates. Overall, the split results show that both components matter: Lasso improves the high-dimensional signature regression, and signature-kernel weighting provides the most effective local sample selection for path-dependent hedging.

\begin{table}[htbp]
  \centering
  \caption{Asian option hedging performance by maturity: error at expiry (\(\times 10^{-3}\)) and win rate against Monte Carlo.}
  \label{tab:asian_by_maturity_panel}
  \resizebox{\textwidth}{!}{%
    \begin{tabular}{@{}c c c c c c c c c c c c c c c@{}}
      \toprule
      & & MC & \multicolumn{2}{c}{ExpSig} & \multicolumn{2}{c}{OlsSig} & \multicolumn{2}{c}{LassoSig} & \multicolumn{2}{c}{WlsSig} & \multicolumn{2}{c}{WlassoSig} & \multicolumn{2}{c}{NN} \\
      \cmidrule(lr){4-5} \cmidrule(lr){6-7} \cmidrule(lr){8-9} \cmidrule(lr){10-11} \cmidrule(lr){12-13} \cmidrule(lr){14-15}
      Maturity (days) & Samples & Mean$\downarrow$ & Mean$\downarrow$ & Win\%$\uparrow$ & Mean$\downarrow$ & Win\%$\uparrow$ & Mean$\downarrow$ & Win\%$\uparrow$ & Mean$\downarrow$ & Win\%$\uparrow$ & Mean$\downarrow$ & Win\%$\uparrow$ & Mean$\downarrow$ & Win\%$\uparrow$ \\
      \midrule
      \multicolumn{15}{c}{\textbf{Panel A: Call options}} \\
      \midrule
      5   & 21{,}084 & \textbf{4.819} & 14.16 & 65.5 & 7.770 & 66.7 & 5.755 & 68.4 & 12.73 & \textbf{77.4} & 4.917 & 70.3 & 15.78 & 12.6 \\
      10  & 21{,}049 & 5.642 & 11.74 & 59.0 & 12.79 & 58.7 & 6.836 & 77.2 & \textbf{3.598} & \textbf{84.7} & 4.306 & 83.2 & 22.95 & 9.57 \\
      15  & 21{,}014 & 5.545 & 15.02 & 51.1 & 14.90 & 51.6 & 8.052 & 69.1 & \textbf{4.082} & \textbf{77.9} & 4.619 & 77.3 & 30.00 & 7.07 \\
      20  & 20{,}979 & 6.130 & 16.92 & 49.8 & 17.79 & 50.5 & 9.419 & 65.3 & \textbf{4.585} & \textbf{77.4} & 5.069 & 74.6 & 34.46 & 11.1 \\
      30  & 20{,}909 & 6.271 & 22.55 & 46.4 & 22.86 & 46.1 & 12.08 & 58.9 & \textbf{5.709} & \textbf{71.1} & 6.301 & 68.2 & 39.85 & 10.4 \\
      50  & 20{,}769 & \textbf{7.023} & 30.90 & 39.3 & 30.53 & 39.3 & 14.34 & 53.5 & 8.176 & 61.0 & 7.384 & \textbf{63.7} & 51.89 & 6.89 \\
      100 & 20{,}419 & 9.720 & 53.39 & 34.3 & 53.07 & 34.1 & 17.63 & 51.4 & 10.83 & 58.3 & \textbf{9.246} & \textbf{64.8} & 73.93 & 6.81 \\
      \midrule
      \multicolumn{15}{c}{\textbf{Panel B: Put options}} \\
      \midrule
      5   & 21{,}084 & 5.115 & 14.85 & 63.6 & 8.826 & 64.8 & 6.089 & 67.0 & 12.97 & \textbf{78.2} & \textbf{5.005} & 71.0 & 15.23 & 14.6 \\
      10  & 21{,}049 & 5.929 & 11.10 & 61.4 & 12.33 & 59.7 & 7.484 & 74.2 & \textbf{3.705} & \textbf{84.1} & 4.449 & 83.5 & 22.09 & 13.3 \\
      15  & 21{,}014 & 5.905 & 14.48 & 54.3 & 13.79 & 54.6 & 7.999 & 69.0 & \textbf{4.295} & 77.8 & 4.456 & \textbf{79.4} & 26.04 & 13.2 \\
      20  & 20{,}979 & 6.108 & 17.16 & 50.6 & 17.46 & 51.5 & 10.27 & 62.7 & 4.674 & 77.1 & \textbf{4.619} & \textbf{77.7} & 29.30 & 11.4 \\
      30  & 20{,}909 & 6.657 & 20.36 & 49.6 & 21.26 & 48.2 & 12.79 & 58.1 & \textbf{5.682} & \textbf{72.0} & 5.930 & \textbf{72.0} & 34.65 & 12.4 \\
      50  & 20{,}769 & 7.664 & 27.73 & 44.5 & 26.67 & 46.2 & 16.02 & 54.2 & 8.103 & 65.6 & \textbf{6.702} & \textbf{69.2} & 50.49 & 10.4 \\
      100 & 20{,}419 & 10.56 & 43.16 & 49.6 & 42.33 & 49.2 & 15.18 & 63.2 & 8.441 & 70.5 & \textbf{6.832} & \textbf{75.9} & 65.84 & 11.5 \\
      \bottomrule
    \end{tabular}%
  }
\end{table}

\begin{table}[htbp]
  \centering
  \caption{Asian option hedging performance by moneyness: error at expiry (\(\times 10^{-3}\)) and win rate against Monte Carlo.}
  \label{tab:asian_by_moneyness_panel}
  \resizebox{\textwidth}{!}{%
    \begin{tabular}{@{}c c c c c c c c c c c c c c c@{}}
      \toprule
      & & MC & \multicolumn{2}{c}{ExpSig} & \multicolumn{2}{c}{OlsSig} & \multicolumn{2}{c}{LassoSig} & \multicolumn{2}{c}{WlsSig} & \multicolumn{2}{c}{WlassoSig} & \multicolumn{2}{c}{NN} \\
      \cmidrule(lr){4-5} \cmidrule(lr){6-7} \cmidrule(lr){8-9} \cmidrule(lr){10-11} \cmidrule(lr){12-13} \cmidrule(lr){14-15}
      Moneyness (S/K) & Samples & Mean$\downarrow$ & Mean$\downarrow$ & Win\%$\uparrow$ & Mean$\downarrow$ & Win\%$\uparrow$ & Mean$\downarrow$ & Win\%$\uparrow$ & Mean$\downarrow$ & Win\%$\uparrow$ & Mean$\downarrow$ & Win\%$\uparrow$ & Mean$\downarrow$ & Win\%$\uparrow$ \\
      \midrule
      \multicolumn{15}{c}{\textbf{Panel A: Call options}} \\
      \midrule
      0.80 & 20{,}889 & 0.1234 & 0.02339 & 99.9 & 0.01738 & 99.9 & 0.002979 & 99.9 & 0.000552 & \textbf{100} & \textbf{0.000509} & \textbf{100} & 37.40 & 0.0241 \\
      0.90 & 20{,}889 & 2.089  & 3.472    & 97.1 & 3.225    & 97.2 & 0.3372   & 99.1 & 0.09156  & 99.9 & \textbf{0.07915}   & \textbf{100} & 41.73 & 0.8199 \\
      0.95 & 20{,}889 & 4.913  & 24.47    & 53.7 & 23.96    & 53.8 & 11.17    & 74.1 & 3.079    & 90.0 & \textbf{1.332}     & \textbf{97.2} & 41.10 & 3.376 \\
      1.00 & 20{,}889 & \textbf{9.089}  & 41.22    & 19.9 & 35.48    & 19.8 & 23.44    & 26.7 & 20.02    & 42.6 & 9.831     & \textbf{48.0} & 34.34 & 15.58 \\
      1.05 & 20{,}889 & 9.580  & 33.02    & 23.5 & 33.79    & 23.5 & 14.99    & 44.7 & \textbf{8.898}    & \textbf{57.8} & 10.53     & 50.5 & 37.88 & 15.26 \\
      1.10 & 20{,}889 & 9.540  & 31.02    & 26.1 & 31.07    & 27.2 & 12.03    & 50.0 & \textbf{8.827}    & \textbf{59.3} & 9.933     & 54.1 & 37.44 & 15.58 \\
      1.20 & 20{,}889 & 9.711  & 30.47    & 25.9 & 31.17    & 26.3 & 11.86    & 49.8 & \textbf{8.694}    & \textbf{58.9} & 10.02     & 52.7 & 37.71 & 13.87 \\
      \midrule
      \multicolumn{15}{c}{\textbf{Panel B: Put options}} \\
      \midrule
      0.80 & 20{,}889 & 9.905  & 30.72    & 26.2 & 31.21    & 26.3 & 12.28    & 49.0 & \textbf{8.888}    & \textbf{58.1} & 10.13     & 52.0 & 29.34 & 20.33 \\
      0.90 & 20{,}889 & 10.30  & 31.56    & 26.1 & 32.51    & 26.0 & 12.73    & 50.0 & \textbf{9.230}    & \textbf{60.0} & 10.21     & 54.6 & 29.68 & 20.40 \\
      0.95 & 20{,}889 & \textbf{10.78}  & 35.11    & 24.5 & 33.52    & 25.1 & 19.01    & 42.1 & 11.79     & \textbf{54.7} & 11.24     & 52.9 & 28.49 & 22.16 \\
      1.00 & 20{,}889 & 8.558  & 40.06    & 19.2 & 33.72    & 20.4 & 24.43    & 27.4 & 17.16     & 53.9 & \textbf{5.865}    & \textbf{69.5} & 32.78 & 16.01 \\
      1.05 & 20{,}889 & 5.058  & 8.866    & 80.5 & 9.323    & 79.2 & 5.539    & 85.6 & 0.6212    & 98.7 & \textbf{0.3424}   & \textbf{99.7} & 39.19 & 5.836 \\
      1.10 & 20{,}889 & 2.472  & 1.768    & 97.5 & 1.619    & 97.5 & 1.596    & 94.6 & 0.1143    & \textbf{100} & \textbf{0.1132}   & \textbf{100} & 41.17 & 2.001 \\
      1.20 & 20{,}889 & 0.7334 & \textbf{0.02602} & \textbf{100} & \textbf{0.02602} & \textbf{100} & \textbf{0.02602} & \textbf{100} & \textbf{0.02602} & \textbf{100} & \textbf{0.02602} & \textbf{100} & 41.84 & 0.217 \\
      \bottomrule
    \end{tabular}%
  }
\end{table}

\begin{table}[htbp]
\centering
\caption{Asian option hedging performance by year: error at expiry (\(\times 10^{-3}\)) and win rate against Monte Carlo.}
\label{tab:asian_by_year_panel}
\resizebox{\textwidth}{!}{%
\begin{tabular}{@{}c c c c c c c c c c c c c c c@{}}
\toprule
& & MC & \multicolumn{2}{c}{ExpSig} & \multicolumn{2}{c}{OlsSig} & \multicolumn{2}{c}{LassoSig} & \multicolumn{2}{c}{WlsSig} & \multicolumn{2}{c}{WlassoSig} & \multicolumn{2}{c}{NN} \\
\cmidrule(lr){4-5} \cmidrule(lr){6-7} \cmidrule(lr){8-9} \cmidrule(lr){10-11} \cmidrule(lr){12-13} \cmidrule(lr){14-15}
Year & Samples & Mean$\downarrow$ & Mean$\downarrow$ & Win\%$\uparrow$ & Mean$\downarrow$ & Win\%$\uparrow$ & Mean$\downarrow$ & Win\%$\uparrow$ & Mean$\downarrow$ & Win\%$\uparrow$ & Mean$\downarrow$ & Win\%$\uparrow$ & Mean$\downarrow$ & Win\%$\uparrow$ \\
\midrule
\multicolumn{15}{c}{\textbf{Panel A: Call options}} \\
\midrule
2014 & 12{,}348 & 4.084 & 21.02 & 48.0 & 20.39 & 47.8 & 5.861 & 67.2 & 4.174 & \textbf{73.7} & \textbf{3.826} & 73.4 & 26.59 & 8.08 \\
2015 & 12{,}348 & 6.423 & 20.21 & 51.2 & 21.32 & 52.4 & 9.049 & 71.0 & 7.250 & \textbf{80.2} & \textbf{4.632} & 78.9 & 33.10 & 12.1 \\
2016 & 12{,}348 & 4.984 & 19.07 & 49.0 & 17.67 & 47.2 & 6.347 & 68.9 & 5.468 & 76.7 & \textbf{3.846} & \textbf{77.9} & 29.35 & 8.98 \\
2017 & 12{,}299 & 3.021 & 19.43 & 54.0 & 11.31 & 54.0 & 4.147 & 69.1 & 3.268 & 74.1 & \textbf{2.805} & \textbf{75.0} & 20.84 & 10.0 \\
2018 & 12{,}299 & 6.423 & 19.32 & 53.9 & 19.52 & 54.0 & 10.64 & 64.2 & \textbf{5.162} & \textbf{74.0} & 5.552 & 72.4 & 35.51 & 10.2 \\
2019 & 12{,}348 & \textbf{5.480} & 18.51 & 49.2 & 17.92 & 49.2 & 8.408 & 60.7 & 15.65 & \textbf{67.3} & 5.611 & 67.1 & 33.72 & 8.24 \\
2020 & 12{,}397 & 15.20 & 31.97 & 53.0 & 33.05 & 53.7 & 20.89 & 60.9 & \textbf{10.47} & \textbf{74.7} & 10.58 & 72.0 & 68.92 & 10.5 \\
2021 & 12{,}348 & \textbf{4.953} & 24.01 & 45.3 & 24.05 & 47.8 & 8.561 & 62.8 & 5.815 & \textbf{69.8} & 5.653 & 69.0 & 38.39 & 7.27 \\
2022 & 12{,}299 & 9.079 & 40.56 & 45.1 & 40.46 & 44.8 & 20.76 & 56.5 & \textbf{8.925} & \textbf{72.0} & 9.881 & 69.3 & 58.23 & 10.7 \\
2023 & 12{,}250 & \textbf{4.813} & 20.13 & 46.2 & 20.03 & 47.0 & 9.857 & 60.4 & 6.746 & 64.9 & 6.120 & \textbf{67.7} & 39.47 & 6.33 \\
2024 & 12{,}348 & \textbf{4.888} & 23.14 & 46.0 & 22.71 & 46.4 & 10.33 & 55.8 & 5.370 & \textbf{69.6} & 5.618 & 66.3 & 35.54 & 7.76 \\
2025 & 10{,}591 & 8.126 & 23.26 & 52.9 & 23.82 & 52.1 & 11.94 & 64.1 & \textbf{6.686} & \textbf{75.0} & 7.661 & 72.4 & 39.21 & 10.7 \\
\midrule
\multicolumn{15}{c}{\textbf{Panel B: Put options}} \\
\midrule
2014 & 12{,}348 & 4.274 & 17.94 & 54.4 & 18.68 & 53.3 & 5.562 & 71.5 & 3.876 & \textbf{77.2} & \textbf{3.581} & 76.7 & 21.06 & 13.8 \\
2015 & 12{,}348 & 7.712 & 18.45 & 54.4 & 19.43 & 53.3 & 9.005 & 68.5 & 7.431 & \textbf{79.3} & \textbf{4.852} & 78.5 & 30.58 & 14.9 \\
2016 & 12{,}348 & 4.914 & 17.09 & 53.0 & 15.00 & 52.6 & 6.058 & 71.7 & 5.162 & 78.6 & \textbf{3.526} & \textbf{79.8} & 29.04 & 9.43 \\
2017 & 12{,}299 & 2.829 & 18.54 & 56.9 & 10.66 & 56.6 & 3.930 & 70.9 & 2.827 & 75.9 & \textbf{2.177} & \textbf{78.2} & 17.14 & 10.4 \\
2018 & 12{,}299 & 6.940 & 18.05 & 53.3 & 18.68 & 53.4 & 10.64 & 63.2 & \textbf{5.083} & \textbf{75.2} & 5.133 & 74.5 & 29.51 & 15.9 \\
2019 & 12{,}348 & 5.145 & 15.33 & 54.0 & 15.22 & 53.8 & 7.719 & 62.4 & 15.62 & 68.9 & \textbf{4.809} & \textbf{71.9} & 33.53 & 11.6 \\
2020 & 12{,}397 & 16.75 & 34.32 & 51.4 & 34.98 & 52.4 & 25.84 & 56.4 & 11.11 & 76.5 & \textbf{9.900} & \textbf{79.4} & 70.07 & 14.4 \\
2021 & 12{,}348 & 5.310 & 19.06 & 54.2 & 19.87 & 55.6 & 9.093 & 63.6 & \textbf{4.900} & 76.1 & 4.962 & \textbf{76.4} & 34.42 & 11.7 \\
2022 & 12{,}299 & 9.477 & 37.39 & 44.9 & 35.36 & 45.9 & 19.19 & 55.9 & \textbf{9.466} & \textbf{71.4} & 9.583 & 70.8 & 46.38 & 13.2 \\
2023 & 12{,}250 & \textbf{4.980} & 18.64 & 54.1 & 17.40 & 54.9 & 10.08 & 62.3 & 5.672 & 70.5 & 5.194 & \textbf{72.9} & 40.33 & 7.55 \\
2024 & 12{,}348 & 5.045 & 19.12 & 55.0 & 18.16 & 55.1 & 9.786 & 60.4 & \textbf{4.721} & \textbf{74.2} & 5.026 & 71.2 & 26.17 & 13.0 \\
2025 & 10{,}591 & 8.907 & 19.74 & 55.9 & 19.77 & 55.4 & 13.08 & 62.0 & \textbf{5.998} & \textbf{77.3} & 6.440 & 75.8 & 38.00 & 13.5 \\
\bottomrule
\end{tabular}%
}
\end{table}

\begin{table}[htbp]
  \centering
  \caption{Lookback option hedging performance by maturity: error at expiry (\(\times 10^{-3}\)) and win rate against Monte Carlo.}
  \label{tab:lookback_by_maturity_panel}
  \resizebox{\textwidth}{!}{%
    \begin{tabular}{@{}c c c c c c c c c c c c c c c@{}}
      \toprule
      & & MC & \multicolumn{2}{c}{ExpSig} & \multicolumn{2}{c}{OlsSig} & \multicolumn{2}{c}{LassoSig} & \multicolumn{2}{c}{WlsSig} & \multicolumn{2}{c}{WlassoSig} & \multicolumn{2}{c}{NN} \\
      \cmidrule(lr){4-5} \cmidrule(lr){6-7} \cmidrule(lr){8-9} \cmidrule(lr){10-11} \cmidrule(lr){12-13} \cmidrule(lr){14-15}
      Maturity (days) & Samples & Mean$\downarrow$ & Mean$\downarrow$ & Win\%$\uparrow$ & Mean$\downarrow$ & Win\%$\uparrow$ & Mean$\downarrow$ & Win\%$\uparrow$ & Mean$\downarrow$ & Win\%$\uparrow$ & Mean$\downarrow$ & Win\%$\uparrow$ & Mean$\downarrow$ & Win\%$\uparrow$ \\
      \midrule
      \multicolumn{15}{c}{\textbf{Panel A: Call options}} \\
      \midrule
      5   & 21{,}084 & 7.435 & 80.83 & 56.5 & 33.19 & 55.6 & 8.466 & 68.7 & 82.42 & 64.3 & \textbf{5.594} & \textbf{78.8} & 14.26 & 23.2 \\
      10  & 21{,}049 & 13.92 & 17.35 & 66.8 & 26.81 & 60.6 & 11.81 & 79.1 & 7.851 & 90.8 & \textbf{5.913} & \textbf{95.8} & 23.13 & 30.2 \\
      15  & 21{,}014 & 16.59 & 20.87 & 60.7 & 20.00 & 61.3 & 15.06 & 72.7 & 9.474 & 85.5 & \textbf{7.547} & \textbf{89.9} & 29.91 & 31.1 \\
      20  & 20{,}979 & 21.23 & 25.71 & 60.9 & 23.71 & 61.4 & 19.18 & 67.9 & 12.05 & 82.9 & \textbf{10.84} & \textbf{87.4} & 32.29 & 33.0 \\
      30  & 20{,}909 & 26.46 & 29.11 & 61.6 & 28.78 & 59.9 & 22.00 & 70.6 & 13.83 & 84.2 & \textbf{13.20} & \textbf{87.2} & 38.87 & 34.5 \\
      50  & 20{,}769 & 39.35 & 45.61 & 54.3 & 48.51 & 54.7 & 29.19 & 68.8 & 21.09 & 81.7 & \textbf{17.13} & \textbf{86.8} & 54.58 & 34.6 \\
      100 & 20{,}419 & 62.87 & 95.16 & 47.7 & 99.84 & 48.4 & 31.68 & 79.7 & 42.71 & 72.0 & \textbf{30.65} & \textbf{80.3} & 76.10 & 36.6 \\
      \midrule
      \multicolumn{15}{c}{\textbf{Panel B: Put options}} \\
      \midrule
      5   & 21{,}084 & 7.731 & 99.26 & 54.2 & 31.69 & 49.0 & 7.900 & 67.9 & 58.62 & 63.1 & \textbf{6.100} & \textbf{75.7} & 15.26 & 25.8 \\
      10  & 21{,}049 & 12.56 & 19.07 & 60.1 & 26.71 & 58.3 & 11.96 & 74.2 & 7.326 & 87.4 & \textbf{6.428} & \textbf{92.3} & 22.54 & 31.8 \\
      15  & 21{,}014 & 13.97 & 21.96 & 51.5 & 23.74 & 52.9 & 16.23 & 64.7 & 9.495 & 80.0 & \textbf{8.160} & \textbf{86.2} & 26.17 & 33.8 \\
      20  & 20{,}979 & 14.94 & 25.26 & 52.2 & 26.91 & 45.9 & 20.14 & 59.8 & 12.06 & 72.7 & \textbf{10.60} & \textbf{77.4} & 29.21 & 31.2 \\
      30  & 20{,}909 & 18.19 & 31.88 & 48.4 & 31.78 & 49.7 & 25.23 & 60.6 & 14.17 & 72.9 & \textbf{12.57} & \textbf{77.4} & 36.44 & 29.5 \\
      50  & 20{,}769 & 22.51 & 43.23 & 45.6 & 44.31 & 45.2 & 29.50 & 59.2 & 18.77 & 69.1 & \textbf{15.47} & \textbf{74.5} & 45.31 & 28.5 \\
      100 & 20{,}419 & \textbf{27.31} & 97.69 & 32.4 & 94.13 & 33.2 & 48.26 & 43.1 & 32.39 & 52.2 & 28.44 & \textbf{57.9} & 63.92 & 25.2 \\
      \bottomrule
    \end{tabular}%
  }
\end{table}

\begin{table}[htbp]
  \centering
  \caption{Lookback option hedging performance by moneyness: error at expiry (\(\times 10^{-3}\)) and win rate against Monte Carlo.}
  \label{tab:lookback_by_moneyness_panel}
  \resizebox{\textwidth}{!}{%
    \begin{tabular}{@{}c c c c c c c c c c c c c c c@{}}
      \toprule
      & & MC & \multicolumn{2}{c}{ExpSig} & \multicolumn{2}{c}{OlsSig} & \multicolumn{2}{c}{LassoSig} & \multicolumn{2}{c}{WlsSig} & \multicolumn{2}{c}{WlassoSig} & \multicolumn{2}{c}{NN} \\
      \cmidrule(lr){4-5} \cmidrule(lr){6-7} \cmidrule(lr){8-9} \cmidrule(lr){10-11} \cmidrule(lr){12-13} \cmidrule(lr){14-15}
      Moneyness (S/K) & Samples & Mean$\downarrow$ & Mean$\downarrow$ & Win\%$\uparrow$ & Mean$\downarrow$ & Win\%$\uparrow$ & Mean$\downarrow$ & Win\%$\uparrow$ & Mean$\downarrow$ & Win\%$\uparrow$ & Mean$\downarrow$ & Win\%$\uparrow$ & Mean$\downarrow$ & Win\%$\uparrow$ \\
      \midrule
      \multicolumn{15}{c}{\textbf{Panel A: Call options}} \\
      \midrule
      0.80 & 20{,}889 & 3.784 & 0.2191 & 99.9 & 0.210 & 99.9 & 0.234 & 99.9 & 0.165 & \textbf{100} & \textbf{0.161} & \textbf{100} & 44.41 & 0.99 \\
      0.90 & 20{,}889 & 16.63 & 16.89  & 79.8 & 16.82  & 79.8 & 9.348  & 88.5 & 5.241  & 95.9 & \textbf{3.687}  & \textbf{98.0} & 46.03 & 8.49 \\
      0.95 & 20{,}889 & 25.81 & 42.94  & 38.7 & 43.96  & 35.0 & 25.29  & 54.5 & 13.97  & 84.5 & \textbf{10.85}  & \textbf{94.0} & 39.72 & 22.0 \\
      \(\geq 1.00\) & 83{,}556 & 34.82 & 63.39  & 47.1 & 54.45  & 46.5 & 25.47  & 65.2 & 42.41  & 69.8 & \textbf{18.82}  & \textbf{78.0} & 34.24 & 47.5 \\
      \midrule
      \multicolumn{15}{c}{\textbf{Panel B: Put options}} \\
      \midrule
      \(\leq 1.00\) & 83{,}556 & 22.27 & 70.35 & 34.8 & 54.74 & 35.0 & 30.27 & 49.6 & 34.31 & 55.6 & \textbf{18.39} & \textbf{64.4} & 33.17 & 41.6 \\
      1.05 & 20{,}889 & 14.32 & 35.92  & 36.0 & 37.06  & 29.8 & 20.72  & 67.1 & 9.534  & 83.7 & \textbf{8.325}  & \textbf{89.0} & 28.16 & 24.8 \\
      1.10 & 20{,}889 & 10.06 & 16.28  & 73.1 & 16.62  & 69.3 & 11.90  & 79.2 & 4.756  & 93.0 & \textbf{4.036}  & \textbf{97.0} & 33.94 & 12.9 \\
      1.20 & 20{,}889 & 3.090 & 3.840  & 95.3 & 3.950  & 95.8 & 4.284  & 88.2 & 1.140  & 98.2 & \textbf{0.919} & \textbf{99.5} & 42.55 & 2.17 \\
      \bottomrule
    \end{tabular}%
  }
\end{table}

\begin{table}[htbp]
\centering
\caption{Lookback option hedging performance by year: error at expiry (\(\times 10^{-3}\)) and win rate against Monte Carlo.}
\label{tab:lookback_by_year_panel}
\resizebox{\textwidth}{!}{%
\begin{tabular}{@{}c c c c c c c c c c c c c c c@{}}
\toprule
& & MC & \multicolumn{2}{c}{ExpSig} & \multicolumn{2}{c}{OlsSig} & \multicolumn{2}{c}{LassoSig} & \multicolumn{2}{c}{WlsSig} & \multicolumn{2}{c}{WlassoSig} & \multicolumn{2}{c}{NN} \\
\cmidrule(lr){4-5} \cmidrule(lr){6-7} \cmidrule(lr){8-9} \cmidrule(lr){10-11} \cmidrule(lr){12-13} \cmidrule(lr){14-15}
Year & Samples & Mean$\downarrow$ & Mean$\downarrow$ & Win\%$\uparrow$ & Mean$\downarrow$ & Win\%$\uparrow$ & Mean$\downarrow$ & Win\%$\uparrow$ & Mean$\downarrow$ & Win\%$\uparrow$ & Mean$\downarrow$ & Win\%$\uparrow$ & Mean$\downarrow$ & Win\%$\uparrow$ \\
\midrule
\multicolumn{15}{c}{\textbf{Panel A: Call options}} \\
\midrule
2014 & 12{,}348 & 12.88 & 56.56 & 48.7 & 59.16 & 48.5 & 14.28 & 69.0 & 19.02 & 71.6 & \textbf{8.271} & \textbf{84.0} & 31.60 & 13.7 \\
2015 & 12{,}348 & 17.90 & 38.18 & 54.8 & 41.80 & 55.6 & 16.63 & 69.8 & 37.32 & 80.0 & \textbf{8.510} & \textbf{89.0} & 26.50 & 41.9 \\
2016 & 12{,}348 & 22.69 & 51.19 & 64.8 & 33.24 & 63.2 & 12.53 & 84.2 & 20.82 & 89.1 & \textbf{7.348} & \textbf{95.3} & 28.63 & 43.6 \\
2017 & 12{,}299 & 11.88 & 89.83 & 59.4 & 26.41 & 56.8 & 9.001 & 79.0 & 10.80 & 79.0 & \textbf{6.270} & \textbf{89.0} & 29.62 & 11.6 \\
2018 & 12{,}299 & 25.46 & 44.00 & 59.8 & 55.30 & 58.4 & 20.15 & 73.5 & 15.86 & 80.7 & \textbf{14.25} & \textbf{86.4} & 39.15 & 33.1 \\
2019 & 12{,}348 & 26.30 & 28.47 & 59.6 & 29.83 & 59.8 & 13.24 & 81.6 & 99.64 & 83.2 & \textbf{9.618} & \textbf{92.9} & 34.13 & 43.8 \\
2020 & 12{,}397 & 88.78 & 60.71 & 75.8 & 63.63 & 73.4 & 39.80 & 82.9 & 40.95 & 93.5 & \textbf{30.57} & \textbf{96.9} & 59.64 & 67.3 \\
2021 & 12{,}348 & 18.69 & 34.80 & 51.2 & 31.57 & 50.7 & 17.24 & 67.7 & 14.30 & 74.8 & \textbf{11.08} & \textbf{83.8} & 51.94 & 6.04 \\
2022 & 12{,}299 & 23.30 & 41.50 & 52.4 & 43.65 & 50.0 & 32.78 & 56.6 & 17.08 & 76.0 & \textbf{18.33} & \textbf{75.6} & 38.92 & 37.8 \\
2023 & 12{,}250 & 16.28 & 32.59 & 45.8 & 31.34 & 45.3 & 21.84 & 54.7 & 15.17 & 68.6 & \textbf{14.59} & \textbf{71.7} & 40.88 & 15.5 \\
2024 & 12{,}348 & 14.00 & 28.32 & 51.1 & 29.86 & 51.3 & 17.48 & 65.9 & 13.00 & 72.9 & \textbf{11.19} & \textbf{79.5} & 40.72 & 13.1 \\
2025 & 10{,}591 & 44.38 & 28.64 & 81.0 & 31.33 & 79.8 & 19.68 & 87.1 & 19.36 & 95.7 & \textbf{15.10} & \textbf{97.0} & 37.14 & 58.7 \\
\midrule
\multicolumn{15}{c}{\textbf{Panel B: Put options}} \\
\midrule
2014 & 12{,}348 & \textbf{7.981} & 67.93 & 46.1 & 59.67 & 46.3 & 12.48 & 64.7 & 16.64 & 65.7 & 8.488 & \textbf{72.6} & 23.58 & 25.4 \\
2015 & 12{,}348 & 19.76 & 44.24 & 48.8 & 53.03 & 47.8 & 20.37 & 65.6 & 32.63 & 70.0 & \textbf{12.58} & \textbf{79.1} & 28.63 & 41.2 \\
2016 & 12{,}348 & 14.59 & 60.29 & 49.4 & 37.18 & 47.3 & 13.98 & 68.4 & 22.53 & 78.1 & \textbf{7.801} & \textbf{86.7} & 23.76 & 39.1 \\
2017 & 12{,}299 & 8.015 & 103.2 & 58.9 & 22.00 & 55.8 & 6.250 & 81.0 & 10.06 & 74.8 & \textbf{6.198} & \textbf{81.4} & 15.90 & 39.0 \\
2018 & 12{,}299 & 19.05 & 36.90 & 51.6 & 37.50 & 51.6 & 26.82 & 60.3 & 15.09 & 75.0 & \textbf{12.02} & \textbf{80.2} & 34.25 & 30.8 \\
2019 & 12{,}348 & 13.48 & 27.35 & 46.3 & 26.55 & 46.7 & 18.13 & 61.2 & 63.78 & 66.2 & \textbf{11.03} & \textbf{76.3} & 31.78 & 24.2 \\
2020 & 12{,}397 & 41.62 & 53.35 & 51.9 & 55.46 & 50.7 & 46.96 & 59.1 & 28.40 & 76.3 & \textbf{28.34} & \textbf{80.2} & 72.51 & 29.4 \\
2021 & 12{,}348 & \textbf{11.70} & 35.59 & 48.4 & 35.50 & 44.6 & 22.34 & 55.3 & 12.93 & 68.6 & 12.58 & \textbf{72.5} & 27.93 & 27.3 \\
2022 & 12{,}299 & 21.57 & 55.84 & 39.6 & 55.04 & 41.0 & 39.77 & 46.0 & 20.27 & 65.5 & \textbf{16.62} & \textbf{74.1} & 39.80 & 29.7 \\
2023 & 12{,}250 & \textbf{9.649} & 30.03 & 45.3 & 30.34 & 40.3 & 23.52 & 48.4 & 12.49 & 62.8 & 10.19 & \textbf{68.2} & 31.17 & 21.8 \\
2024 & 12{,}348 & 9.734 & 28.15 & 45.9 & 28.42 & 44.9 & 18.46 & 56.1 & 10.22 & 69.2 & \textbf{9.101} & \textbf{71.7} & 33.28 & 17.6 \\
2025 & 10{,}591 & 24.23 & 32.91 & 60.6 & 34.57 & 58.1 & 22.31 & 73.1 & 15.86 & 83.2 & \textbf{14.98} & \textbf{87.5} & 47.02 & 27.3 \\
\bottomrule
\end{tabular}%
}
\end{table}

\subsection{Robustness}\label{app:robustness}
This appendix examines the robustness of the hedging results after incorporating transaction costs. Specifically, we impose a two-sided 1 bp proportional transaction cost on rebalancing trades and recompute the hedging error at expiry. The purpose of this exercise is not to model all market frictions, but to verify whether the relative performance of the proposed It\^o-signature hedge remains stable once a simple trading-cost adjustment is included.

Table~\ref{tab:overall_hedging_performance_total_tc_1bp} reports the results for vanilla options. The overall ranking is  similar to the main results without transaction costs. WlassoSig continues to deliver the lowest mean hedging error, while WlsSig and WlassoSig achieve the highest winning percentages against Black--Scholes. The unweighted LassoSig remains the strongest non-weighted signature specification.

\begin{table}[htbp]
\centering
\caption{Vanilla option hedging performance with two-sided 1 bp transaction cost: error at expiry (\(\times 10^{-3}\)) and win rate against Black--Scholes.}
\label{tab:overall_hedging_performance_total_tc_1bp}
\resizebox{\textwidth}{!}{%
\begin{tabular}{@{} c c c c c c c c c c c c c c c @{}}
\toprule
& & BS & \multicolumn{2}{c}{ExpSig} & \multicolumn{2}{c}{OlsSig} & \multicolumn{2}{c}{LassoSig} & \multicolumn{2}{c}{WlsSig} & \multicolumn{2}{c}{WlassoSig} & \multicolumn{2}{c}{NN} \\
\cmidrule(lr){4-5} \cmidrule(lr){6-7} \cmidrule(lr){8-9} \cmidrule(lr){10-11} \cmidrule(lr){12-13} \cmidrule(lr){14-15}
Option & Samples & Mean$\downarrow$ & Mean$\downarrow$ & Win\%$\uparrow$ & Mean$\downarrow$ & Win\%$\uparrow$ & Mean$\downarrow$ & Win\%$\uparrow$ & Mean$\downarrow$ & Win\%$\uparrow$ & Mean$\downarrow$ & Win\%$\uparrow$ & Mean$\downarrow$ & Win\%$\uparrow$ \\
\midrule
Overall & 436{,}135 & 5.993 & 5.956 & 70.5 & 5.745 & 73.3 & 5.733 & 73.2 & 5.446 & \textbf{83.6} & \textbf{5.426} & \textbf{83.6} & 8.738 & 26.3 \\
Call    & 162{,}527 & 6.162 & 6.484 & 70.7 & 6.242 & 73.2 & 6.200 & 73.3 & 5.694 & \textbf{88.7} & \textbf{5.676} & \textbf{88.7} & 9.426 & 23.7 \\
Put     & 273{,}608 & 5.893 & 5.643 & 70.4 & 5.450 & 73.4 & 5.456 & 73.0 & 5.299 & \textbf{80.6} & \textbf{5.279} & \textbf{80.6} & 8.310 & 27.9 \\
\bottomrule
\end{tabular}%
}
\end{table}

Table~\ref{tab:path_dependent_hedging_performance_total_tc_1bp} reports the corresponding results for path-dependent options, where the transaction-cost adjustment leaves the main ranking essentially unchanged. WlassoSig achieves the lowest mean error for all four payoff types, while WlsSig and WlassoSig deliver the strongest win rates. This confirms that the advantage of the signature-kernel-weighted Lasso specification is not driven by ignoring proportional trading costs.

\begin{table}[htbp]
\centering
\caption{Path-dependent option hedging performance with two-sided 1 bp transaction cost: error at expiry (\(\times 10^{-3}\)) and win rate against Monte Carlo.}
\label{tab:path_dependent_hedging_performance_total_tc_1bp}
\resizebox{\textwidth}{!}{%
\begin{tabular}{@{}c c c c c c c c c c c c c c c@{}}
\toprule
& & MC & \multicolumn{2}{c}{ExpSig} & \multicolumn{2}{c}{OlsSig} & \multicolumn{2}{c}{LassoSig} & \multicolumn{2}{c}{WlsSig} & \multicolumn{2}{c}{WlassoSig} & \multicolumn{2}{c}{NN} \\
\cmidrule(lr){4-5} \cmidrule(lr){6-7} \cmidrule(lr){8-9} \cmidrule(lr){10-11} \cmidrule(lr){12-13} \cmidrule(lr){14-15}
Option & Samples & Mean$\downarrow$ & Mean$\downarrow$ & Win\%$\uparrow$ & Mean$\downarrow$ & Win\%$\uparrow$ & Mean$\downarrow$ & Win\%$\uparrow$ & Mean$\downarrow$ & Win\%$\uparrow$ & Mean$\downarrow$ & Win\%$\uparrow$ & Mean$\downarrow$ & Win\%$\uparrow$ \\
\midrule
Asian call    & 146{,}223 & 6.414 & 23.48 & 49.5 & 22.75 & 49.5 & 10.55 & 63.4 & 7.103 & \textbf{72.5} & \textbf{5.957} & 71.7 & 38.24 & 9.13 \\
Asian put     & 146{,}223 & 6.839 & 21.13 & 53.7 & 20.22 & 53.9 & 10.85 & 64.4 & 6.845 & 75.3 & \textbf{5.396} & \textbf{76.0} & 34.80 & 12.0 \\
Lookback call & 146{,}223 & 26.33 & 44.99 & 58.1 & 39.75 & 57.1 & 19.80 & 71.8 & 27.19 & 79.6 & \textbf{13.17} & \textbf{86.0} & 38.25 & 31.7 \\
Lookback put  & 146{,}223 & 16.60 & 48.29 & 49.4 & 39.53 & 47.7 & 22.58 & 61.4 & 21.81 & 71.1 & \textbf{12.46} & \textbf{77.3} & 34.03 & 29.3 \\
\bottomrule
\end{tabular}%
}
\end{table}
\section{Lemmas and Proofs \label{appendix:lemmaandproofs}}

This appendix provides the proofs of all technical results and lemmas used in the proofs.

\subsection{Lemmas}\label{appendix:subseclemmas}
\begin{lemma}[Uniform moment bound for It\^o-signature components]\label{lem:moment_bound_sig_main}
Under Assumptions~\ref{ass:1}--\ref{ass:2}, for every fixed multi-index \(\Gamma\) and \(p\ge 2\),
\begin{equation}\label{eq:strong_moment_claim}
\sup_{0\le t\le T}\mathbb E\bigl[|S(\tilde X)_t^{\Gamma,\mathrm I}|^p\bigr]
+
\sup_{n\ge 1}\max_{0\le k\le n}\mathbb E\bigl[|S(\tilde X)_{t_k}^{\Gamma,\mathrm I,\pi_n}|^p\bigr]
<\infty.
\end{equation}
\end{lemma}

\proof{Proof of Lemma \ref{lem:moment_bound_sig_main}.}
We first recall a standard consequence of Assumptions~\ref{ass:1}--\ref{ass:2}: for every \(q\ge 1\), there exists a constant \(C_q\) such that
\begin{equation}\label{eq:X_moment_bound}
\sup_{0\le t\le T}\mathbb E|X_t|^q \le C_q .
\end{equation}
Hence, by the linear-growth condition,
\begin{equation}\label{eq:mu_sigma_moment_bound}
\sup_{0\le t\le T}\mathbb E|\mu(X_t,t)|^q
+
\sup_{0\le t\le T}\mathbb E\|\sigma(X_t,t)\|^q
<\infty
\qquad\text{for every }q\ge 1.
\end{equation}

\medskip
\noindent
\textbf{Step 1: base case \(|\Gamma|=1\).}
Let \(\Gamma=(j)\).
If \(1\le j\le d\), then
$
S(\tilde X)_t^{(j),\mathrm I}=X_t^j-X_0^j,
S(\tilde X)_{t_k}^{(j),\mathrm I,\pi_n}=X_{t_k}^j-X_0^j.$
Therefore, by \eqref{eq:X_moment_bound},
\(
\sup_{0\le t\le T}\mathbb E\bigl[|S(\tilde X)_t^{(j),\mathrm I}|^p\bigr]
+
\sup_{n\ge1}\max_{0\le k\le n}\mathbb E\bigl[|S(\tilde X)_{t_k}^{(j),\mathrm I,\pi_n}|^p\bigr]
<\infty.
\)
If \(j=d+1\), then
$
S(\tilde X)_t^{(d+1),\mathrm I}=t,
S(\tilde X)_{t_k}^{(d+1),\mathrm I,\pi_n}=t_k,$
so the bound is trivial.

\medskip
\noindent
\textbf{Step 2: Induction step.}
Assume that, for all multi-indices of length \(m-1\), the bound
\eqref{eq:strong_moment_claim} holds for every exponent \(p\ge2\), and let $
\Gamma=(\Gamma^-,j),|\Gamma|=m.
$

\medskip
\noindent
\textbf{Continuous signature.}

\emph{Case 1: \(j=d+1\).}
Then
$S(\tilde X)_t^{\Gamma,\mathrm I}
=
\int_0^t S(\tilde X)_u^{\Gamma^-,\mathrm I}\,\mathrm du.
$
By Hölder's inequality for integrals,
\[
\Big|\int_0^t f(u)\,\mathrm du\Big|^p
\le
t^{p-1}\int_0^t |f(u)|^p\,\mathrm du
\le
T^{p-1}\int_0^T |f(u)|^p\,\mathrm du.
\]
Applying this with \(f(u)=S(\tilde X)_u^{\Gamma^-,\mathrm I}\), we obtain $
|S(\tilde X)_t^{\Gamma,\mathrm I}|^p
\le
T^{p-1}\int_0^T |S(\tilde X)_u^{\Gamma^-,\mathrm I}|^p\,\mathrm du.
$
Taking expectations and using the induction hypothesis gives
\(
\sup_{0\le t\le T}\mathbb E\bigl[|S(\tilde X)_t^{\Gamma,\mathrm I}|^p\bigr]
\le
T^p \sup_{0\le u\le T}\mathbb E\bigl[|S(\tilde X)_u^{\Gamma^-,\mathrm I}|^p\bigr]
<\infty.
\)

\emph{Case 2: \(1\le j\le d\).}
Write $
S(\tilde X)_t^{\Gamma,\mathrm I}
=
A_t+M_t,
$
where
\[
A_t:=\int_0^t S(\tilde X)_u^{\Gamma^-,\mathrm I}\mu^j(X_u,u)\,\mathrm du, 
M_t:=\sum_{\ell=1}^d \int_0^t S(\tilde X)_u^{\Gamma^-,\mathrm I}\sigma^{j\ell}(X_u,u)\,\mathrm dB_u^\ell.
\]
By $(|a+b|^p\le 2^{p-1}(|a|^p+|b|^p)$, we see 
$\mathbb E|S(\tilde X)_t^{\Gamma,\mathrm I}|^p
\le
2^{p-1}\bigl(\mathbb E|A_t|^p+\mathbb E|M_t|^p\bigr).
$

For the drift term, Hölder's inequality for integrals yields
\(
|A_t|^p
\le
T^{p-1}\int_0^t
|S(\tilde X)_u^{\Gamma^-,\mathrm I}|^p |\mu^j(X_u,u)|^p
\,\mathrm du.
\)
Hence
\(
\mathbb E|A_t|^p
\le
T^{p-1}\int_0^t
\mathbb E\Big[
|S(\tilde X)_u^{\Gamma^-,\mathrm I}|^p |\mu^j(X_u,u)|^p
\Big]
\,\mathrm du.
\)
Applying Hölder's inequality in expectation,
\(
\mathbb E\Big[
|S(\tilde X)_u^{\Gamma^-,\mathrm I}|^p |\mu^j(X_u,u)|^p
\Big]
\le
\Big(\mathbb E|S(\tilde X)_u^{\Gamma^-,\mathrm I}|^{2p}\Big)^{1/2}
\Big(\mathbb E|\mu^j(X_u,u)|^{2p}\Big)^{1/2}.
\)
The first factor is finite by the induction hypothesis applied with exponent \(2p\), and the second is finite by \eqref{eq:mu_sigma_moment_bound}. Therefore $
\sup_{0\le t\le T}\mathbb E|A_t|^p<\infty.$

For the martingale term, the Burkholder--Davis--Gundy inequality gives
\(
\mathbb E|M_t|^p
\le
C_p\,
\mathbb E\Big[
\Big(
\int_0^t |S(\tilde X)_u^{\Gamma^-,\mathrm I}|^2
\|\sigma^{j\cdot}(X_u,u)\|^2
\,\mathrm du
\Big)^{p/2}
\Big].
\)
Applying Hölder's inequality for integrals to the quantity inside the expectation,
\(
\Big(
\int_0^t g(u)\,\mathrm du
\Big)^{p/2}
\le
T^{\frac p2-1}\int_0^t g(u)^{p/2}\,\mathrm du,
\)
with
\(
g(u)=|S(\tilde X)_u^{\Gamma^-,\mathrm I}|^2
\|\sigma^{j\cdot}(X_u,u)\|^2,
\)
we obtain
\(
\mathbb E|M_t|^p
\le
C_p T^{\frac p2-1}
\int_0^t
\mathbb E\Big[
|S(\tilde X)_u^{\Gamma^-,\mathrm I}|^p
\|\sigma^{j\cdot}(X_u,u)\|^p
\Big]
\,\mathrm du.
\)
Again by Hölder's inequality in expectation,
\(
\mathbb E\Big[
|S(\tilde X)_u^{\Gamma^-,\mathrm I}|^p
\|\sigma^{j\cdot}(X_u,u)\|^p
\Big]
\le
\Big(\mathbb E|S(\tilde X)_u^{\Gamma^-,\mathrm I}|^{2p}\Big)^{1/2}
\Big(\mathbb E\|\sigma^{j\cdot}(X_u,u)\|^{2p}\Big)^{1/2},
\)
which is finite by the induction hypothesis and \eqref{eq:mu_sigma_moment_bound}. Thus
\(
\sup_{0\le t\le T}\mathbb E|M_t|^p<\infty,
\)
and therefore
\(
\sup_{0\le t\le T}\mathbb E\bigl[|S(\tilde X)_t^{\Gamma,\mathrm I}|^p\bigr]<\infty.
\)

\medskip
\noindent
\textbf{Discrete signature.}

\emph{Case 1: \(j=d+1\).}
Then
\(
S(\tilde X)_{t_k}^{\Gamma,\mathrm I,\pi_n}
= \sum_{r=0}^{k-1}
S(\tilde X)_{t_r}^{\Gamma^-,\mathrm I,\pi_n}\Delta t_r,
\)
where $\Delta t_r=t_{r+1}-t_r$.
Applying Hölder's inequality to the discrete average,
\(
\left|\sum_{r=0}^{k-1} a_r\Delta t_r\right|^p
\le
\left(\sum_{r=0}^{k-1}\Delta t_r\right)^{p-1}
\sum_{r=0}^{k-1}|a_r|^p\Delta t_r
\le
T^{p-1}\sum_{r=0}^{n-1}|a_r|^p\Delta t_r,
\)
we obtain
\(
\bigl|S(\tilde X)_{t_k}^{\Gamma,\mathrm I,\pi_n}\bigr|^p
\le
T^{p-1} \sum_{r=0}^{n-1}
\bigl|S(\tilde X)_{t_r}^{\Gamma^-,\mathrm I,\pi_n}\bigr|^p\Delta t_r.
\)
Taking expectations and using the induction hypothesis,
\(
\sup_{n\ge1}\max_{0\le k\le n}
\mathbb E\bigl[|S(\tilde X)_{t_k}^{\Gamma,\mathrm I,\pi_n}|^p\bigr]
<\infty.
\)

\emph{Case 2: \(1\le j\le d\).}
Define the predictable step process
\(
H_u^{\pi_n}
:=
\sum_{r=0}^{n-1}
S(\tilde X)_{t_r}^{\Gamma^-,\mathrm I,\pi_n}\mathbf 1_{(t_r,t_{r+1}]}(u).
\) Then, for every \(k\),
\(
S(\tilde X)_{t_k}^{\Gamma,\mathrm I,\pi_n}
=
\int_0^{t_k} H_u^{\pi_n}\,\mathrm dX_u^j.
\)
Write
\(
S(\tilde X)_{t_k}^{\Gamma,\mathrm I,\pi_n}
=
A_k^{\pi_n}+M_k^{\pi_n},
\)
where
\[
A_k^{\pi_n}
:=
\int_0^{t_k} H_u^{\pi_n}\mu^j(X_u,u)\,\mathrm du,
M_k^{\pi_n}
:=
\sum_{\ell=1}^d\int_0^{t_k} H_u^{\pi_n}\sigma^{j\ell}(X_u,u)\,\mathrm dB_u^\ell,
\]
then $
\mathbb E\bigl[|S(\tilde X)_{t_k}^{\Gamma,\mathrm I,\pi_n}|^p\bigr]
\le
2^{p-1}\Big(
\mathbb E|A_k^{\pi_n}|^p+\mathbb E|M_k^{\pi_n}|^p
\Big).
$

For the drift part,
\(
|A_k^{\pi_n}|^p
\le
T^{p-1}\int_0^{t_k}
|H_u^{\pi_n}|^p |\mu^j(X_u,u)|^p\,\mathrm du,
\)
hence
\(
\mathbb E|A_k^{\pi_n}|^p
\le
T^{p-1}\int_0^{t_k}
\mathbb E\bigl[|H_u^{\pi_n}|^p |\mu^j(X_u,u)|^p\bigr]\,\mathrm du.
\)
For each \(u\in(t_r,t_{r+1}]\), we have
\(
H_u^{\pi_n}=S(\tilde X)_{t_r}^{\Gamma^-,\mathrm I,\pi_n}.
\)
Therefore, by the induction hypothesis,
\(
\sup_{n\ge1}\sup_{0\le u\le T}\mathbb E|H_u^{\pi_n}|^{2p}
\le
\sup_{n\ge1}\max_{0\le r\le n}\mathbb E\bigl[|S(\tilde X)_{t_r}^{\Gamma^-,\mathrm I,\pi_n}|^{2p}\bigr]
<\infty.
\)
Using Hölder's inequality together with \eqref{eq:mu_sigma_moment_bound}, we conclude that
\(
\sup_{n\ge1}\max_{0\le k\le n}\mathbb E|A_k^{\pi_n}|^p<\infty.
\)

For the martingale part, the Burkholder--Davis--Gundy inequality yields
\[
\mathbb E|M_k^{\pi_n}|^p
\le
C_p\,
\mathbb E\Big[
\Big(
\int_0^{t_k} |H_u^{\pi_n}|^2\|\sigma^{j\cdot}(X_u,u)\|^2\,\mathrm du
\Big)^{p/2}
\Big].
\]
Arguing exactly as in the continuous case,
\[
\mathbb E|M_k^{\pi_n}|^p
\le
C_p T^{\frac p2-1}
\int_0^{t_k}
\mathbb E\bigl[|H_u^{\pi_n}|^p\|\sigma^{j\cdot}(X_u,u)\|^p\bigr]\,\mathrm du,
\]
and Hölder's inequality, together with the bound on \(\sup_{n,u}\mathbb E|H_u^{\pi_n}|^{2p}\), implies
\(
\sup_{n\ge1}\max_{0\le k\le n}\mathbb E|M_k^{\pi_n}|^p<\infty.
\)
Therefore,
\(
\sup_{n\ge1}\max_{0\le k\le n}
\mathbb E\bigl[|S(\tilde X)_{t_k}^{\Gamma,\mathrm I,\pi_n}|^p\bigr]
<\infty.
\)

This completes the induction, and proves \eqref{eq:strong_moment_claim} and in particular the statement of the lemma  for \(p=4\).
\Halmos\endproof

\begin{lemma}\label{lem:L2_cont_sig_main}
Under Assumptions~\ref{ass:1}--\ref{ass:2}, for every multi-index \(\Gamma\) there exists a constant \(C_\Gamma>0\) such that
\[
\mathbb E\bigl[|S(\tilde X)_t^{\Gamma,\mathrm I}-S(\tilde X)_s^{\Gamma,\mathrm I}|^2\bigr]
\le C_\Gamma |t-s|,
\qquad
0\le s\le t\le T.
\]
\end{lemma}

\proof{Proof of Lemma \ref{lem:L2_cont_sig_main}.}
We argue by induction on \(|\Gamma|\).

\noindent
\textbf{Step 1: base case \(|\Gamma|=1\).} Write \(\Gamma=(j)\). If \(1\le j\le d\), then
\[
S(\tilde X)_t^{(j),\mathrm I}-S(\tilde X)_s^{(j),\mathrm I}
=
X_t^j-X_s^j
=
\int_s^t \mu^j(X_u,u)\,\mathrm du
+
\sum_{\ell=1}^d \int_s^t \sigma^{j\ell}(X_u,u)\,\mathrm dB_u^\ell.
\]
Hence
\(
\mathbb E\bigl[|S(\tilde X)_t^{(j),\mathrm I}-S(\tilde X)_s^{(j),\mathrm I}|^2\bigr]
\le
2\,\mathbb E\Big|\int_s^t \mu^j(X_u,u)\,\mathrm du\Big|^2
+
2\,\mathbb E\Big|\sum_{\ell=1}^d \int_s^t \sigma^{j\ell}(X_u,u)\,\mathrm dB_u^\ell\Big|^2.
\)

By Cauchy--Schwarz and It\^o isometry,
\(
\mathbb E\Big|\int_s^t \mu^j(X_u,u)\,\mathrm du\Big|^2
\le
(t-s)\int_s^t \mathbb E|\mu^j(X_u,u)|^2\,\mathrm du,
\)
and
\(
\mathbb E\Big|\sum_{\ell=1}^d \int_s^t \sigma^{j\ell}(X_u,u)\,\mathrm dB_u^\ell\Big|^2
=
\int_s^t \mathbb E\|\sigma^{j\cdot}(X_u,u)\|^2\,\mathrm du.
\)
By the linear-growth condition and the standard second-moment bound for \(X\),
\(
\sup_{0\le u\le T}\mathbb E|\mu(X_u,u)|^2
+
\sup_{0\le u\le T}\mathbb E\|\sigma(X_u,u)\|^2
<\infty,
\)
so
\(
\mathbb E\bigl[|S(\tilde X)_t^{(j),\mathrm I}-S(\tilde X)_s^{(j),\mathrm I}|^2\bigr]
\le C|t-s|.
\)

If \(j=d+1\), then
\(
S(\tilde X)_t^{(d+1),\mathrm I}-S(\tilde X)_s^{(d+1),\mathrm I}=t-s,
\)
so the same bound is immediate.

\noindent
\textbf{Step 2: Induction step.}
Now assume the claim holds for all multi-indices of length \(m-1\), and let \(\Gamma=(\Gamma^-,j)\) with \(|\Gamma|=m\). By definition,
\(
S(\tilde X)_t^{\Gamma,\mathrm I}-S(\tilde X)_s^{\Gamma,\mathrm I}
=
\int_s^t S(\tilde X)_u^{\Gamma^-,\mathrm I}\,\mathrm d\tilde X_u^j.
\)

If \(j=d+1\), then
\(
S(\tilde X)_t^{\Gamma,\mathrm I}-S(\tilde X)_s^{\Gamma,\mathrm I}
=
\int_s^t S(\tilde X)_u^{\Gamma^-,\mathrm I}\,\mathrm du,
\)
and therefore by Lemma \ref{lem:moment_bound_sig_main},
\[
\mathbb E\bigl[|S(\tilde X)_t^{\Gamma,\mathrm I}-S(\tilde X)_s^{\Gamma,\mathrm I}|^2\bigr]
\le
(t-s)\int_s^t \mathbb E\bigl[|S(\tilde X)_u^{\Gamma^-,\mathrm I}|^2\bigr]\,\mathrm du
\le C (t-s)^2
\le C'(t-s).
\]

If \(1\le j\le d\), then
\(
S(\tilde X)_t^{\Gamma,\mathrm I}-S(\tilde X)_s^{\Gamma,\mathrm I}
=
\int_s^t S(\tilde X)_u^{\Gamma^-,\mathrm I}\mu^j(X_u,u)\,\mathrm du
+
\sum_{\ell=1}^d \int_s^t S(\tilde X)_u^{\Gamma^-,\mathrm I}\sigma^{j\ell}(X_u,u)\,\mathrm dB_u^\ell,
\)
therefore we have,
\begin{align*}
\mathbb E\bigl[|S(\tilde X)_t^{\Gamma,\mathrm I}-S(\tilde X)_s^{\Gamma,\mathrm I}|^2\bigr]
\le
2\,\mathbb E\Big|\int_s^t S(\tilde X)_u^{\Gamma^-,\mathrm I}\mu^j(X_u,u)\,\mathrm du\Big|^2 
+
2\,\mathbb E\Big|\sum_{\ell=1}^d \int_s^t S(\tilde X)_u^{\Gamma^-,\mathrm I}\sigma^{j\ell}(X_u,u)\,\mathrm dB_u^\ell\Big|^2.
\end{align*}
For the drift term,
\[
\mathbb E\Big|\int_s^t S(\tilde X)_u^{\Gamma^-,\mathrm I}\mu^j(X_u,u)\,\mathrm du\Big|^2
\le
(t-s)\int_s^t \mathbb E\bigl[|S(\tilde X)_u^{\Gamma^-,\mathrm I}|^2|\mu^j(X_u,u)|^2\bigr]\,\mathrm du.
\]
By Hölder's inequality and Lemma \ref{lem:moment_bound_sig_main},
\(
\sup_{0\le u\le T}\mathbb E\bigl[|S(\tilde X)_u^{\Gamma^-,\mathrm I}|^2|\mu^j(X_u,u)|^2\bigr]
\le
\sup_{0\le u\le T}\bigl(\mathbb E|S(\tilde X)_u^{\Gamma^-,\mathrm I}|^4\bigr)^{1/2}
\bigl(\mathbb E|\mu^j(X_u,u)|^4\bigr)^{1/2}
<\infty,
\)
so this term is \(O((t-s)^2)\), hence \(O(t-s)\).

For the martingale term, It\^o isometry yields
\[
\mathbb E\Big|\sum_{\ell=1}^d \int_s^t S(\tilde X)_u^{\Gamma^-,\mathrm I}\sigma^{j\ell}(X_u,u)\,\mathrm dB_u^\ell\Big|^2
=
\int_s^t \mathbb E\bigl[|S(\tilde X)_u^{\Gamma^-,\mathrm I}|^2\|\sigma^{j\cdot}(X_u,u)\|^2\bigr]\,\mathrm du.
\]
Again by Hölder and Lemma \ref{lem:moment_bound_sig_main},
\(
\sup_{0\le u\le T}
\mathbb E\bigl[|S(\tilde X)_u^{\Gamma^-,\mathrm I}|^2\|\sigma^{j\cdot}(X_u,u)\|^2\bigr]
<\infty,
\)
so this term is \(O(t-s)\). This completes the induction.
\Halmos\endproof

\begin{lemma}\label{lem:stochastic_integral_continuity_main}
Let \(Z^{\pi_n}\) be predictable processes on \([0,T]\) such that
$\int_0^T \mathbb E|Z_t^{\pi_n}|^2\,\mathrm dt \rightarrow 0$ as \(\|\pi_n\|\to0,\)
and
\(
\sup_{n\ge 1}\sup_{0\le t\le T}\mathbb E|Z_t^{\pi_n}|^4<\infty.
\)
Then, for each \(j\in\{1,\dots,d+1\}\),
\[
\mathbb E\Big[\sup_{0\le t\le T}\Big|\int_0^t Z_u^{\pi_n}\,\mathrm d\tilde X_u^j\Big|^2\Big]\longrightarrow 0.
\]
\end{lemma}

\proof{Proof of Lemma \ref{lem:stochastic_integral_continuity_main}.}
We first note that the assumptions imply
\begin{equation}\label{eq:L3_from_L2_main}
\int_0^T \|Z_t^{\pi_n}\|_{L^3}^3\,\mathrm dt \longrightarrow 0.
\end{equation}
Indeed, by interpolation,
\(
\|Z_t^{\pi_n}\|_{L^3}^3 \le \|Z_t^{\pi_n}\|_{L^2}\,\|Z_t^{\pi_n}\|_{L^4}^2,
\)
hence
\[
\int_0^T \|Z_t^{\pi_n}\|_{L^3}^3\,\mathrm dt
\le
\Big(\sup_{n\ge1}\sup_{0\le t\le T}\|Z_t^{\pi_n}\|_{L^4}^2\Big)
\int_0^T \|Z_t^{\pi_n}\|_{L^2}\,\mathrm dt.
\]
By Cauchy--Schwarz,
\(
\int_0^T \|Z_t^{\pi_n}\|_{L^2}\,\mathrm dt
\le
T^{1/2}\Big(\int_0^T \mathbb E|Z_t^{\pi_n}|^2\,\mathrm dt\Big)^{1/2}\to0,
\)
which proves \eqref{eq:L3_from_L2_main}. In particular,
\begin{equation}\label{eq:L3sq_from_L2_main}
\int_0^T \|Z_t^{\pi_n}\|_{L^3}^2\,\mathrm dt \longrightarrow 0,
\end{equation}
since
\(
\int_0^T \|Z_t^{\pi_n}\|_{L^3}^2\,\mathrm dt
\le
T^{1/3}\Big(\int_0^T \|Z_t^{\pi_n}\|_{L^3}^3\,\mathrm dt\Big)^{2/3}.
\)
\paragraph{Case 1:}
If \(j=d+1\), then \(\mathrm d\tilde X_t^{d+1}=\mathrm dt\), so
\(
\sup_{0\le t\le T}\Big|\int_0^t Z_u^{\pi_n}\,\mathrm du\Big|
\le
\int_0^T |Z_u^{\pi_n}|\,\mathrm du,
\)
and therefore
\[
\Big\|\sup_{0\le t\le T}\Big|\int_0^t Z_u^{\pi_n}\,\mathrm du\Big|\Big\|_{L^2}
\le
\int_0^T \|Z_u^{\pi_n}\|_{L^2}\,\mathrm du \to 0.
\]

\paragraph{Case 2:} If \(1\le j\le d\), write
\(
\mathrm d\tilde X_t^j = \mu^j(X_t,t)\,\mathrm dt + \sum_{\ell=1}^d \sigma^{j\ell}(X_t,t)\,\mathrm dB_t^\ell.
\)

For the drift part,
\[
\Big\|\sup_{0\le t\le T}\Big|\int_0^t Z_u^{\pi_n}\mu^j(X_u,u)\,\mathrm du\Big|\Big\|_{L^2}
\le
\int_0^T \|Z_u^{\pi_n}\mu^j(X_u,u)\|_{L^2}\,\mathrm du.
\]
By Hölder's inequality with exponents \(3\) and \(6\),
\(
\|Z_u^{\pi_n}\mu^j(X_u,u)\|_{L^2}
\le
\|Z_u^{\pi_n}\|_{L^3}\,\|\mu^j(X_u,u)\|_{L^6}.
\)
Under Assumptions~\ref{ass:1}--\ref{ass:2}, standard moment bounds for \(X\) imply
\(
\sup_{0\le u\le T}\|\mu(X_u,u)\|_{L^6}<\infty,
\)
hence
\(
\Big\|\sup_{0\le t\le T}\Big|\int_0^t Z_u^{\pi_n}\mu^j(X_u,u)\,\mathrm du\Big|\Big\|_{L^2}
\le
C\int_0^T \|Z_u^{\pi_n}\|_{L^3}\,\mathrm du
\to 0
\)
by \eqref{eq:L3_from_L2_main}.
Moreover,
\(
\int_0^T \|Z_t^{\pi_n}\|_{L^3}\,dt
\le
T^{2/3}
\left(
\int_0^T \|Z_t^{\pi_n}\|_{L^3}^3\,dt
\right)^{1/3}
\to0.
\)

For the martingale part, Doob's inequality and It\^o isometry give
\[
\mathbb E\Big[\sup_{0\le t\le T}\Big|\sum_{\ell=1}^d\int_0^t Z_u^{\pi_n}\sigma^{j\ell}(X_u,u)\,\mathrm dB_u^\ell\Big|^2\Big]
\le
4\int_0^T \mathbb E\bigl[|Z_u^{\pi_n}|^2\|\sigma^{j\cdot}(X_u,u)\|^2\bigr]\,\mathrm du.
\]
Again by Hölder,
\(
\mathbb E\bigl[|Z_u^{\pi_n}|^2\|\sigma^{j\cdot}(X_u,u)\|^2\bigr]
\le
\|Z_u^{\pi_n}\|_{L^3}^2\,\|\sigma^{j\cdot}(X_u,u)\|_{L^6}^2.
\)
Since
\(
\sup_{0\le u\le T}\|\sigma(X_u,u)\|_{L^6}<\infty
\)
under Assumptions~\ref{ass:1}--\ref{ass:2}, \eqref{eq:L3sq_from_L2_main} implies that the martingale part converges to zero in \(L^2\). Combining the drift and martingale parts proves the claim.
\Halmos\endproof

\begin{lemma}[Existence of a global normalizing transformation]\label{lem:global_transform}
Under Assumptions~\ref{ass:ito_un_1}--\ref{ass:ito_un_3}, there exists a map
\(
g\in C^{2,1}(D\times[0,T];\mathbb{R}^d)
\)
such that
\[
\nabla_x g(x,t)\,\sigma(x,t)=I_d,
\qquad (x,t)\in D\times[0,T].
\]
Moreover, for each \(t\in[0,T]\), the map
\(
g_t:=g(\cdot,t):D\to g_t(D)
\)
is a global \(C^1\)-diffeomorphism.
\end{lemma}

\proof{Proof of Lemma \ref{lem:global_transform}.}
Set
\(
J(x,t):=\sigma(x,t)^{-1}.
\)
By Assumption~\ref{ass:ito_un_1}, \(J\) is continuously differentiable in \((x,t)\). By Assumption~\ref{ass:ito_un_2}, for each fixed \(t\) and each \(i=1,\dots,d\), the row vector field \(J_i(\cdot,t)\) is conservative on the convex domain \(D\). Hence there exists a scalar potential \(g_i(\cdot,t)\) such that
$
\nabla_x g_i(x,t)=J_i(x,t),
\qquad x\in D.$
Fixing a reference point \(x^\ast\in D\), we may normalize \(g_i(x^\ast,t)=0\). Define
\(
g=(g_1,\dots,g_d)^\top.
\)
Then
\(
\nabla_x g(x,t)=J(x,t)=\sigma(x,t)^{-1},
\)
and hence
\(
\nabla_x g(x,t)\,\sigma(x,t)=I_d.
\)
Because \(J\) is continuously differentiable in \((x,t)\), the map \(g\) belongs to \(C^{2,1}(D\times[0,T];\mathbb R^d)\).

We next show that, for each fixed \(t\), the map \(g_t:=g(\cdot,t)\) is injective. By Assumption~\ref{ass:ito_un_3}, there exists an invertible matrix \(A(t)\in\mathbb R^{d\times d}\) such that, for every \(x\in D\),
\(
\frac{
A(t)J(x,t)+J(x,t)^\top A(t)^\top
}{2}
\succ0.
\)
Define
\(
h_t(x):=A(t)g_t(x).
\)
Let \(x,y\in D\) with \(x\neq y\). Since \(D\) is convex, the line segment \(y+s(x-y)\) lies in \(D\) for all \(s\in[0,1]\). Note
that $
h_t(x)-h_t(y)=\int_0^1 \nabla_x h_t\bigl(y+s(x-y),t\bigr)(x-y)\,\mathrm{d}s.$
Since
\(
\nabla_x h_t(x)=A(t)\nabla_x g(x,t)=A(t)J(x,t),
\)
taking the inner product with \(x-y\) yields $
(h_t(x)-h_t(y))\cdot (x-y)
=
\int_0^1 (x-y)^\top A(t)J\bigl(y+s(x-y),t\bigr)(x-y)\,\mathrm{d}s.$
We also notice that
\(
v^\top A(t)J(x,t)v
=
v^\top \frac{A(t)J(x,t)+J(x,t)^\top A(t)^\top}{2}v
\)
for every \(v\in\mathbb{R}^d\), Assumption~\ref{ass:ito_un_3} implies that for every \(s\in[0,1]\),
\(
(x-y)^\top A(t)J\bigl(y+s(x-y),t\bigr)(x-y)>0.
\)
Therefore
\(
(h_t(x)-h_t(y))\cdot (x-y)>0.
\)
This proves that \(h_t\) is injective. Since \(A(t)\) is invertible, the injectivity of \(h_t=A(t)g_t\) implies that \(g_t\) is injective.

Since
\(
\nabla_x g(x,t)=\sigma(x,t)^{-1}
\)
and \(\sigma(x,t)\) is invertible by Assumption~\ref{ass:ito_un_1}, the Jacobian \(\nabla_x g(x,t)\) is invertible at every point of \(D\). Hence, by the inverse function theorem, \(g_t\) is a local \(C^1\)-diffeomorphism. In particular, \(g_t(D)\) is open. Because \(g_t\) is injective, these local inverses agree on overlaps, so \(g_t^{-1}:g_t(D)\to D\) is globally well defined and \(C^1\). Therefore \(g_t:D\to g_t(D)\) is a global \(C^1\)-diffeomorphism.
\Halmos\endproof

\begin{lemma}[Finite-level It\^o representation of Stratonovich coordinates]\label{lem:ito_represents_strat}
Consider the normalized constant-diffusion benchmark
\(\mathrm dX_t=b(X_t,t)\,\mathrm dt+\mathrm dW_t,\)
\(X_0=x_0\in\mathbb R^d,
\)
and let \(\tilde X_t=(X_t,t)\). For every multi-index \(\Gamma\) with \(|\Gamma|=m\), there exist coefficients \(\{c_{\Gamma,\Lambda}\}_{|\Lambda|\le m}\) such that
\[
S(\tilde X)_t^{\Gamma,\mathrm S}
=
\sum_{|\Lambda|\le m} c_{\Gamma,\Lambda}\,S(\tilde X)_t^{\Lambda,\mathrm I},
\qquad 0\le t\le T.
\]
In particular, for every \(m\ge1\),
\[
\mathrm{span}\Bigl\{S(\tilde X)_T^{\Gamma,\mathrm S}:|\Gamma|\le m\Bigr\}
\subset
\mathrm{span}\Bigl\{S(\tilde X)_T^{\Lambda,\mathrm I}:|\Lambda|\le m\Bigr\}.
\]
\end{lemma}

\proof{Proof of Lemma \ref{lem:ito_represents_strat}.}
We argue by induction on \(m=|\Gamma|\).

For \(m=1\), say \(\Gamma=(i)\), the first-level Stratonovich and It\^o signature components coincide:
\(
S(\tilde X)_t^{(i),\mathrm S}
=
\tilde X_t^i-\tilde X_0^i
=
S(\tilde X)_t^{(i),\mathrm I},  (0\le t\le T).
\)
Hence the claim.
Now suppose the claim holds for all multi-indices of length at most \(m\), and let \(\Gamma=(\Gamma^-,j)\) be a multi-index of length \(m+1\). By definition,
\(
S(\tilde X)_t^{\Gamma,\mathrm S}
=
\int_0^t S(\tilde X)_u^{\Gamma^-,\mathrm S}\circ \mathrm{d}\tilde X_u^j.
\)
By the induction hypothesis, there exist coefficients \(\{a_{\Lambda}\}_{|\Lambda|\le m}\) such that
\(
S(\tilde X)_u^{\Gamma^-,\mathrm S}
=
\sum_{|\Lambda|\le m} a_{\Lambda}\,S(\tilde X)_u^{\Lambda,\mathrm I},
\qquad 0\le u\le T.
\)
Therefore,
\(
S(\tilde X)_t^{\Gamma,\mathrm S}
=
\sum_{|\Lambda|\le m} a_{\Lambda}
\int_0^t S(\tilde X)_u^{\Lambda,\mathrm I}\circ \mathrm{d}\tilde X_u^j.
\)

Fix one such \(\Lambda\). By the It\^o--Stratonovich conversion formula,
\[
\int_0^t S(\tilde X)_u^{\Lambda,\mathrm I}\circ \mathrm{d}\tilde X_u^j
=
\int_0^t S(\tilde X)_u^{\Lambda,\mathrm I}\,\mathrm{d}\tilde X_u^j
+\frac12\bigl[S(\tilde X)^{\Lambda,\mathrm I},\tilde X^j\bigr]_t.
\]
The first term is the It\^o signature coordinate
\(
\int_0^t S(\tilde X)_u^{\Lambda,\mathrm I}\,\mathrm{d}\tilde X_u^j
=
S(\tilde X)_t^{(\Lambda,j),\mathrm I},
\)
which has order \(|\Lambda|+1\le m+1\).

It remains to compute the quadratic covariation term. If \(\Lambda=\emptyset\), then \(S(\tilde X)^{\emptyset,\mathrm I}\equiv1\), so the bracket vanishes. Suppose now that \(\Lambda=(\Lambda^-,\ell)\) is nonempty. Since the diffusion matrix of \(X\) is the identity, the local martingale part of \(X^{\ell}\) is \(W^{\ell}\), and the local martingale part of \(S(\tilde X)^{\Lambda,\mathrm I}\) is
\(
\mathbf{1}_{\{\ell\le d\}}\int_0^\cdot S(\tilde X)_u^{\Lambda^-,\mathrm I}\,\mathrm{d}W_u^\ell.
\)
Hence, if \(j=d+1\), then \(\tilde X^{d+1}=t\) has finite variation and
\(
\bigl[S(\tilde X)^{\Lambda,\mathrm I},\tilde X^{d+1}\bigr]_t=0.
\)
If \(1\le j\le d\), then
\(
\bigl[S(\tilde X)^{\Lambda,\mathrm I},\tilde X^j\bigr]_t
=
\mathbf{1}_{\{\ell=j\le d\}}
\int_0^t S(\tilde X)_u^{\Lambda^-,\mathrm I}\,\mathrm{d}u
=
\mathbf{1}_{\{\ell=j\le d\}}\,S(\tilde X)_t^{(\Lambda^-,d+1),\mathrm I}.
\)
Therefore,
\(
\int_0^t S(\tilde X)_u^{\Lambda,\mathrm I}\circ \mathrm{d}\tilde X_u^j
=
S(\tilde X)_t^{(\Lambda,j),\mathrm I}
+\frac12\,\mathbf{1}_{\{\Lambda\neq\emptyset,\ \ell=j\le d\}}\,
S(\tilde X)_t^{(\Lambda^-,d+1),\mathrm I},
\)
which is a finite linear combination of It\^o signature coordinates of order at most \(m+1\).
Substituting this identity into the expansion of \(S(\tilde X)_t^{\Gamma,\mathrm S}\) proves that \(S(\tilde X)_t^{\Gamma,\mathrm S}\) is a finite linear combination of
\(
\{S(\tilde X)_t^{\Lambda,\mathrm I}:|\Lambda|\le m+1\}.
\)
This closes the induction.
\Halmos\endproof

\subsection{Proofs}\label{appendix:subsecproofs}
\proof{Proof of Theorem \ref{thm:hedging strategy}.}
We argue by induction on \(|\Gamma|=m\).

When \(m=1\), let \(\Gamma=(i_1)\). If \(i_1\in\{1,\dots,d\}\), then by definition of the discrete It\^o signature,
\(
S(\tilde X)_{t_k}^{(i_1),\mathrm I,\pi_n}
=
\tilde X_{t_k}^{\,i_1}-\tilde X_{0}^{\,i_1}
=
X_{t_k}^{\,i_1}-X_0^{\,i_1}
=
\sum_{j=0}^{k-1}1\cdot\bigl(X_{t_{j+1}}^{\,i_1}-X_{t_j}^{\,i_1}\bigr).
\)
Hence \(S(\tilde X)_{t_k}^{(i_1),\mathrm I,\pi_n}\) is the cumulative PnL of holding one share of asset \(i_1\) over each trading interval.
If \(i_1=d+1\), then
\(
S(\tilde X)_{t_k}^{(d+1),\mathrm I,\pi_n}
=
\tilde X_{t_k}^{\,d+1}-\tilde X_0^{\,d+1}
=
t_k,
\)
which is deterministic and therefore corresponds to a cash position.

Now assume the claim holds for all multi-indices of length \(m-1\), and let
\(
\Gamma=(\Gamma^-,i_m),|\Gamma|=m.
\)
We consider three cases as follows.

\medskip
\noindent
\textbf{Case 1: \(i_m\in\{1,\dots,d\}\).}

By the recursive definition of the discrete It\^o signature,
\(
S(\tilde X)_{t_k}^{\Gamma,\mathrm I,\pi_n}
=
\sum_{j=0}^{k-1}
S(\tilde X)_{t_j}^{\Gamma^-,\mathrm I,\pi_n}
\bigl(\tilde X_{t_{j+1}}^{\,i_m}-\tilde X_{t_j}^{\,i_m}\bigr).
\)
Since \(i_m\le d\), this becomes
\(
S(\tilde X)_{t_k}^{\Gamma,\mathrm I,\pi_n}
=
\sum_{j=0}^{k-1}
S(\tilde X)_{t_j}^{\Gamma^-,\mathrm I,\pi_n}
\bigl(X_{t_{j+1}}^{\,i_m}-X_{t_j}^{\,i_m}\bigr).
\)
Moreover, \(S(\tilde X)_{t_j}^{\Gamma^-,\mathrm I,\pi_n}\) is \(\mathcal F_{t_j}\)-measurable by construction. Therefore the process
\(
\theta_j^\Gamma:=S(\tilde X)_{t_j}^{\Gamma^-,\mathrm I,\pi_n}
\)
defines an admissible trading position in asset \(i_m\), and the resulting cumulative PnL equals \(S(\tilde X)_{t_k}^{\Gamma,\mathrm I,\pi_n}\).

\medskip
\noindent
\textbf{Case 2: \(i_m=d+1\), and \(\Gamma\) contains at least one risky-asset coordinate.}

Let \(r(\Gamma)\) denote the last risky-asset coordinate in \(\Gamma\). Then \(\Gamma^-=(i_1,\dots,i_{m-1})\) still contains at least one risky-asset coordinate, and by the induction hypothesis there exist \(\mathcal F_{t_l}\)-measurable coefficients \(c_{jl}^{\Gamma,\pi_n}\) such that, for every \(j\),
\(
S(\tilde X)_{t_j}^{\Gamma^-,\mathrm I,\pi_n}
=
\sum_{l=0}^{j-1}
c_{jl}^{\Gamma,\pi_n}
\bigl(X_{t_{l+1}}^{\,i_{r(\Gamma)}}-X_{t_l}^{\,i_{r(\Gamma)}}\bigr).
\)
Since \(i_m=d+1\), the recursive definition gives
\(
S(\tilde X)_{t_k}^{\Gamma,\mathrm I,\pi_n}
=
\sum_{j=0}^{k-1}
S(\tilde X)_{t_j}^{\Gamma^-,\mathrm I,\pi_n}
\bigl(t_{j+1}-t_j\bigr):=\sum_{j=0}^{k-1}
S(\tilde X)_{t_j}^{\Gamma^-,\mathrm I,\pi_n}
\Delta t_j,
\)
here \(\Delta t_j=t_{j+1}-t_j\).
Substituting the induction representation yields,
\(
S(\tilde X)_{t_k}^{\Gamma,\mathrm I,\pi_n}
=
\sum_{j=0}^{k-1}
\Delta t_j 
\sum_{l=0}^{j-1}
c_{jl}^{\Gamma,\pi_n}
\bigl(X_{t_{l+1}}^{\,i_{r(\Gamma)}}-X_{t_l}^{\,i_{r(\Gamma)}}\bigr) 
=
\sum_{l=0}^{k-1}
\left(
\sum_{j=l+1}^{k-1} \Delta t_j c_{jl}^{\Gamma,\pi_n}
\right)
\bigl(X_{t_{l+1}}^{\,i_{r(\Gamma)}}-X_{t_l}^{\,i_{r(\Gamma)}}\bigr).
\)
Hence \(S(\tilde X)_{t_k}^{\Gamma,\mathrm I,\pi_n}\) is again the cumulative PnL of a trading strategy in asset \(i_{r(\Gamma)}\), with position
\(
\theta_l^{\Gamma,\pi_n}
:=
\sum_{j=l+1}^{k-1} \Delta t_j c_{jl}^{\Gamma,\pi_n}.
\)
Since each \(c_{jl}^{\Gamma,\pi_n}\) is \(\mathcal F_{t_l}\)-measurable, the position \(\theta_l^{\Gamma,\pi_n}\) is also \(\mathcal F_{t_l}\)-measurable.

\medskip
\noindent
\textbf{Case 3: \(\Gamma\) contains no risky-asset coordinates.}

In this case,
\(
\Gamma=(d+1,\dots,d+1),
\)
so every increment in the recursive definition is a time increment \(t_{j+1}-t_j\). Therefore \(S(\tilde X)_{t_k}^{\Gamma,\mathrm I,\pi_n}\) depends only on the partition and is deterministic. Hence it represents a cash position.
\Halmos\endproof

\vspace{\baselineskip}
\noindent\proof{Proof of Example \ref{ex:specialstrategy}.}
For notational simplicity, we write
\(
S_k^{\Lambda}:=S(\tilde X)_{t_k}^{\Lambda,\mathrm I,\pi_n}\) for \(
k=0,\dots,n.
\)
For any multi-index \(\Lambda\). Recall that
\(
S_k^{\emptyset}=1,
\)
and for \(\Lambda=(\Lambda^-,i)\),
\(
S_k^{\Lambda}
=
\sum_{j=0}^{k-1}
S_j^{\Lambda^-}\bigl(\tilde X_{t_{j+1}}^{\,i}-\tilde X_{t_j}^{\,i}\bigr).
\)
We prove the claim by induction on the number of trailing time coordinates
\(
r=m-q.
\)

\medskip
\noindent
\textbf{Base case: \(r=1\).}
In this case,
$
\Gamma=(\Gamma_0,1,2).
$
By the recursive definition of the discrete It\^o signature,
\(
S_k^{\Gamma}
=
\sum_{j=0}^{k-1} S_j^{(\Gamma_0,1)}\bigl(t_{j+1}-t_j\bigr).
\)
Since \(T=1\) and the partition is uniform, \(t_{j+1}-t_j=1/n\), so
\(
S_k^{\Gamma}
=
\frac{1}{n}\sum_{j=0}^{k-1} S_j^{(\Gamma_0,1)}.
\)
Again by the recursive definition,
\(
S_j^{(\Gamma_0,1)}
=
\sum_{l=0}^{j-1} S_l^{\Gamma_0}\bigl(X_{t_{l+1}}-X_{t_l}\bigr).
\)
Substituting this into the previous display yields
\(
S_k^{\Gamma}
=
\frac{1}{n}\sum_{j=0}^{k-1}\sum_{l=0}^{j-1}
S_l^{\Gamma_0}\bigl(X_{t_{l+1}}-X_{t_l}\bigr) 
=
\sum_{l=0}^{k-1}
\left(
\frac{1}{n}\sum_{j=l+1}^{k-1}1
\right)
S_l^{\Gamma_0}\bigl(X_{t_{l+1}}-X_{t_l}\bigr) =
\sum_{l=0}^{k-1}
\frac{k-1-l}{n}\,
S_l^{\Gamma_0}\bigl(X_{t_{l+1}}-X_{t_l}\bigr) =
\sum_{l=0}^{k-1}
\frac{\binom{k-1-l}{1}}{n}\,
S_l^{\Gamma_0}\bigl(X_{t_{l+1}}-X_{t_l}\bigr).
\)
Thus the claim holds when \(r=1\).

\medskip
\noindent
\textbf{Induction step.}
Assume the claim holds for some \(r-1\ge1\). Let
\(
\Gamma=(\Gamma_0,1,\underbrace{2,\dots,2}_{r\text{ times}})
\)
and define
\(
\Gamma':=(\Gamma_0,1,\underbrace{2,\dots,2}_{(r-1)\text{ times}}).
\)
Then, by the recursive definition,
\(
S_k^{\Gamma}
=
\sum_{j=0}^{k-1} S_j^{\Gamma'}\bigl(t_{j+1}-t_j\bigr)
=
\frac{1}{n}\sum_{j=0}^{k-1} S_j^{\Gamma'}.
\)
By the induction hypothesis, for every \(j\),
\(
S_j^{\Gamma'}
=
\sum_{l=0}^{j-1}
\frac{\binom{j-1-l}{r-1}}{n^{r-1}}\,
S_l^{\Gamma_0}\bigl(X_{t_{l+1}}-X_{t_l}\bigr).
\)
Hence
\(
S_k^{\Gamma}
=
\frac{1}{n}\sum_{j=0}^{k-1}
\sum_{l=0}^{j-1}
\frac{\binom{j-1-l}{r-1}}{n^{r-1}}\,
S_l^{\Gamma_0}\bigl(X_{t_{l+1}}-X_{t_l}\bigr) 
=
\sum_{l=0}^{k-1}
\left(
\frac{1}{n^r}\sum_{j=l+1}^{k-1}\binom{j-1-l}{r-1}
\right)
S_l^{\Gamma_0}\bigl(X_{t_{l+1}}-X_{t_l}\bigr).
\)
Using the hockey-stick identity,
\(
\sum_{j=l+1}^{k-1}\binom{j-1-l}{r-1}
=
\binom{k-1-l}{r},
\)
we obtain
\(
S_k^{\Gamma}
=
\sum_{l=0}^{k-1}
\frac{\binom{k-1-l}{r}}{n^r}\,
S_l^{\Gamma_0}\bigl(X_{t_{l+1}}-X_{t_l}\bigr).
\)
This proves the desired representation for \(r\), and the induction is complete.
\Halmos\endproof

\vspace{\baselineskip}
\noindent\proof{Proof of Theorem \ref{thm:lpconvergence}.}
We argue by induction on \(|\Gamma|\).
If \(|\Gamma|=1\), say \(\Gamma=(j)\), then by construction
\(
S(\tilde X)_{t_k}^{(j),\mathrm I,\pi_n}
=
\tilde X_{t_k}^j-\tilde X_0^j
=
S(\tilde X)_{t_k}^{(j),\mathrm I},
(k=0,1,\dots,n),
\)
so the claim is exact.

Now fix \(m\ge2\), and assume the claim holds for all multi-indices of length \(m-1\). Let \(\Gamma=(\Gamma^-,j)\) with \(|\Gamma|=m\). Define the left-step processes
\[
\bar S_t^{\pi_n}
:=
\sum_{r=0}^{n-1}
S(\tilde X)_{t_r}^{\Gamma^-,\mathrm I}\,\mathbf 1_{(t_r,t_{r+1}]}(t),
\qquad
\hat S_t^{\pi_n}
:=
\sum_{r=0}^{n-1}
S(\tilde X)_{t_r}^{\Gamma^-,\mathrm I,\pi_n}\,\mathbf 1_{(t_r,t_{r+1}]}(t).
\]
For every grid point \(t_k\),
\[
S(\tilde X)_{t_k}^{\Gamma,\mathrm I}
=
\int_0^{t_k} S(\tilde X)_u^{\Gamma^-,\mathrm I}\,\mathrm d\tilde X_u^j,
\qquad
S(\tilde X)_{t_k}^{\Gamma,\mathrm I,\pi_n}
=
\int_0^{t_k} \hat S_u^{\pi_n}\,\mathrm d\tilde X_u^j,
\]
hence
\(
S(\tilde X)_{t_k}^{\Gamma,\mathrm I}
-
S(\tilde X)_{t_k}^{\Gamma,\mathrm I,\pi_n}
=
\int_0^{t_k}
\bigl(S(\tilde X)_u^{\Gamma^-,\mathrm I}-\hat S_u^{\pi_n}\bigr)\,\mathrm d\tilde X_u^j.
\)
Therefore,
\begin{equation}\label{eq:max_error_bound_main}
\max_{0\le k\le n}
\bigl|
S(\tilde X)_{t_k}^{\Gamma,\mathrm I}
-
S(\tilde X)_{t_k}^{\Gamma,\mathrm I,\pi_n}
\bigr|
\le
\sup_{0\le t\le T}
\Big|
\int_0^t \bigl(S(\tilde X)_u^{\Gamma^-,\mathrm I}-\hat S_u^{\pi_n}\bigr)\,\mathrm d\tilde X_u^j
\Big|.
\end{equation}

We now write
\[
S(\tilde X)_u^{\Gamma^-,\mathrm I}-\hat S_u^{\pi_n}
=
\bigl(S(\tilde X)_u^{\Gamma^-,\mathrm I}-\bar S_u^{\pi_n}\bigr)
+
\bigl(\bar S_u^{\pi_n}-\hat S_u^{\pi_n}\bigr).
\]

\medskip
\noindent
\textbf{Step 1: the continuous signature is well approximated by its left-step version.}

By Lemma~\ref{lem:L2_cont_sig_main}, for every \(u\in[t_r,t_{r+1}]\),
\(
\mathbb E\bigl[|S(\tilde X)_u^{\Gamma^-,\mathrm I}-S(\tilde X)_{t_r}^{\Gamma^-,\mathrm I}|^2\bigr]
\le
C_{\Gamma^-}(u-t_r).
\)
Therefore,
\(
\int_0^T \mathbb E\bigl[|S(\tilde X)_u^{\Gamma^-,\mathrm I}-\bar S_u^{\pi_n}|^2\bigr]\,\mathrm du
=
\sum_{r=0}^{n-1}\int_{t_r}^{t_{r+1}}
\mathbb E\bigl[|S(\tilde X)_u^{\Gamma^-,\mathrm I}-S(\tilde X)_{t_r}^{\Gamma^-,\mathrm I}|^2\bigr]\,\mathrm du 
\le
C_{\Gamma^-}\sum_{r=0}^{n-1}\int_{t_r}^{t_{r+1}} (u-t_r)\,\mathrm du 
=
\frac{C_{\Gamma^-}}{2}\sum_{r=0}^{n-1}(t_{r+1}-t_r)^2=\frac{C_{\Gamma^-}}{2}\cdot T\|\pi_n\|\longrightarrow 0.
\)

Moreover, by Lemma \ref{lem:moment_bound_sig_main},
\(
\sup_{n\ge1}\sup_{0\le u\le T}
\mathbb E\bigl[|S(\tilde X)_u^{\Gamma^-,\mathrm I}-\bar S_u^{\pi_n}|^4\bigr]
<\infty.
\)
Hence Lemma~\ref{lem:stochastic_integral_continuity_main} implies
\begin{equation}\label{eq:first_term_main}
\mathbb E\Big[\sup_{0\le t\le T}\Big|
\int_0^t \bigl(S(\tilde X)_u^{\Gamma^-,\mathrm I}-\bar S_u^{\pi_n}\bigr)\,\mathrm d\tilde X_u^j
\Big|^2\Big]
\longrightarrow 0.
\end{equation}

\medskip
\noindent
\textbf{Step 2: the discrete integrand is close to the exact left-step integrand.}

By the induction hypothesis,
\(
\eta_n:=
\max_{0\le r\le n}
\mathbb E\bigl[
|S(\tilde X)_{t_r}^{\Gamma^-,\mathrm I,\pi_n}-S(\tilde X)_{t_r}^{\Gamma^-,\mathrm I}|^2
\bigr]
\longrightarrow 0.
\)
Therefore,
\(
\int_0^T \mathbb E\bigl[|\bar S_u^{\pi_n}-\hat S_u^{\pi_n}|^2\bigr]\,\mathrm du
=
\sum_{r=0}^{n-1}(t_{r+1}-t_r)\,
\mathbb E\bigl[
|S(\tilde X)_{t_r}^{\Gamma^-,\mathrm I}-S(\tilde X)_{t_r}^{\Gamma^-,\mathrm I,\pi_n}|^2
\bigr]
\le T\,\eta_n \longrightarrow 0.
\)
Again, by Lemma \ref{lem:moment_bound_sig_main},
\(
\sup_{n\ge1}\sup_{0\le u\le T}\mathbb E\bigl[|\bar S_u^{\pi_n}-\hat S_u^{\pi_n}|^4\bigr]<\infty.
\)
Hence Lemma~\ref{lem:stochastic_integral_continuity_main} yields
\begin{equation}\label{eq:second_term_main}
\mathbb E\Big[\sup_{0\le t\le T}\Big|
\int_0^t \bigl(\bar S_u^{\pi_n}-\hat S_u^{\pi_n}\bigr)\,\mathrm d\tilde X_u^j
\Big|^2\Big]
\longrightarrow 0.
\end{equation}
Finally, combining \eqref{eq:max_error_bound_main}, \eqref{eq:first_term_main}, and \eqref{eq:second_term_main}, we obtain
\(
\max_{0\le k\le n}
\mathbb E
\bigl[
|S(\tilde X)_{t_k}^{\Gamma,\mathrm I}
-
S(\tilde X)_{t_k}^{\Gamma,\mathrm I,\pi_n}|^2
\bigr]
\le
\mathbb E\max_{0\le k\le n}
\bigl|
S(\tilde X)_{t_k}^{\Gamma,\mathrm I}
-
S(\tilde X)_{t_k}^{\Gamma,\mathrm I,\pi_n}
\bigr|^2
\to0.
\)

It remains to upgrade the convergence from \(L^2\) to \(L^p\).
Let
\(
Y_{\pi_n,k}^{\Gamma}
:=
S(\tilde X)_{t_k}^{\Gamma,\mathrm I,\pi_n}
-
S(\tilde X)_{t_k}^{\Gamma,\mathrm I}.
\)
The preceding argument proves that
\(
\max_{0\le k\le n}
\|Y_{\pi_n,k}^{\Gamma}\|_{L^2(\mathbb P)}
\longrightarrow 0.
\) For the case of  \(1\le p\le2\). By Lyapunov's inequality,
\(
\|Y_{\pi_n,k}^{\Gamma}\|_{L^p(\mathbb P)}
\le
\|Y_{\pi_n,k}^{\Gamma}\|_{L^2(\mathbb P)}.
\)
Hence we have
\(
\max_{0\le k\le n}
\|Y_{\pi_n,k}^{\Gamma}\|_{L^p(\mathbb P)}
\longrightarrow 0.
\)
Now consider \(p>2\). Choose any \(q>p\). By Lemma~\ref{lem:moment_bound_sig_main} and the inequality
\(
|a-b|^q\le 2^{q-1}(|a|^q+|b|^q),
\)
we have
\(
\sup_{n\ge1}\max_{0\le k\le n}
\|Y_{\pi_n,k}^{\Gamma}\|_{L^q(\mathbb P)}
<\infty.
\)
Let \(\theta\in(0,1)\) be defined by
\(
\frac1p=\frac{\theta}{2}+\frac{1-\theta}{q}.
\)
By interpolation,
\(
\|Y_{\pi_n,k}^{\Gamma}\|_{L^p(\mathbb P)}
\le
\|Y_{\pi_n,k}^{\Gamma}\|_{L^2(\mathbb P)}^{\theta}
\|Y_{\pi_n,k}^{\Gamma}\|_{L^q(\mathbb P)}^{1-\theta}.
\)
Therefore,
\(
\max_{0\le k\le n}
\|Y_{\pi_n,k}^{\Gamma}\|_{L^p(\mathbb P)}
\longrightarrow 0.
\)
This closes the induction for the \(L^2(\mathbb P)\) convergence. The interpolation argument above then proves the stated \(L^p(\mathbb P)\) convergence for every \(p\in[1,\infty)\).
\Halmos\endproof

\vspace{\baselineskip}
\noindent\proof{Proof of Theorem \ref{thm:ito_un_general}.}
By Lemma~\ref{lem:global_transform}, there exists
\(
g\in C^{2,1}(D\times[0,T];\mathbb R^d)
\)
such that
\(
\nabla_x g(x,t)\,\sigma(x,t)=I_d,
\qquad (x,t)\in D\times[0,T],
\)
and, for each \(t\in[0,T]\), the map
\(
g_t:=g(\cdot,t):D\to g_t(D)
\)
is a global \(C^1\)-diffeomorphism.

Define the transformed process
\(
Y_t:=g(X_t,t), t\in[0,T].
\)
By It\^o's formula,
\[
\mathrm dY_t
=
\Bigl(
\partial_t g(X_t,t)
+
\nabla_x g(X_t,t)\mu(X_t,t)
+
\frac12 \sum_{a,b=1}^d
(\sigma\sigma^\top)_{ab}(X_t,t)\,\partial_{x_a x_b}g(X_t,t)
\Bigr)\mathrm dt
+
\nabla_x g(X_t,t)\sigma(X_t,t)\,\mathrm dB_t.
\]
Since \(\nabla_x g(x,t)\sigma(x,t)=I_d\), this reduces to
\(
\mathrm dY_t=\hat b(X_t,t)\,\mathrm dt+\mathrm dB_t
\)
for a suitable drift function \(\hat b\). Because \(g_t:D\to g_t(D)\) is invertible for each \(t\), we may define
\(
\tilde b(y,t):=\hat b(g_t^{-1}(y),t),
\qquad y\in g_t(D),
\)
so that
\(
\mathrm dY_t=\tilde b(Y_t,t)\,\mathrm dt+\mathrm dB_t.
\)
Thus \(Y\) is a constant-diffusion process.
Now define
\(
\Phi_g:\tilde{\mathcal K}\to C([0,T],\mathbb R^{d+1}), \Phi_g(\tilde x)_t:=(g(x_t,t),t).
\)
Since \(g\) is continuous on \(D\times[0,T]\), \(\Phi_g\) is continuous under the supremum norm. Hence
\(
\tilde{\mathcal K}_g:=\Phi_g(\tilde{\mathcal K})
\)
is compact.

We next verify that \(\tilde{\mathcal K}_g\) is a compact set of time-augmented sample paths of \(\tilde Y\). Let \(\tilde y\in\tilde{\mathcal K}_g\). By definition, there exists \(\tilde x\in\tilde{\mathcal K}\) such that
\(
\tilde y=\Phi_g(\tilde x).
\)
Since \(\tilde{\mathcal K}\) is a compact set of time-augmented sample paths of \(\tilde X\), there exists \(\omega\) such that
\(
\tilde x=\tilde X(\omega).
\)
Therefore
\(
\tilde y
=
\Phi_g(\tilde X(\omega))
=
\bigl(g(X_t(\omega),t),t\bigr)_{t\in[0,T]}
=
\tilde Y(\omega).
\)
Hence every \(\tilde y\in\tilde{\mathcal K}_g\) is a time-augmented sample path of \(\tilde Y\).

Next define
\(
\bar f:\tilde{\mathcal K}_g\to\mathbb R,
\bar f(\tilde y):=f\bigl(\Phi_g^{-1}(\tilde y)\bigr).
\)
Since \(g_t\) is injective for every \(t\), the map \(\Phi_g\) is injective on \(\tilde{\mathcal K}\). Since \(\Phi_g\) is continuous and \(\tilde{\mathcal K}\) is compact, the restriction of \(\Phi_g\) to \(\tilde{\mathcal K}\) is a homeomorphism onto \(\tilde{\mathcal K}_g\). Therefore \(\Phi_g^{-1}\) is continuous on \(\tilde{\mathcal K}_g\), and \(\bar f\) is continuous.
Apply Theorem~\ref{th:UN} to the compact set
\(\tilde{\mathcal K}_g:=\Phi_g(\tilde{\mathcal K})\) and to the continuous
function \(\bar f=f\circ \Phi_g^{-1}\). Then use
Lemma~\ref{lem:ito_represents_strat} to rewrite the resulting finite
Stratonovich-signature expansion as a finite It\^o-signature expansion of the normalized path, we obtain an integer \(m\ge1\) and coefficients \(\{\beta_{\Lambda}\}_{|\Lambda|\le m}\) such that
\(
\sup_{\tilde y\in\tilde{\mathcal K}_g}
\left|
\bar f(\tilde y)-\sum_{|\Lambda|\le m}\beta_{\Lambda}\,S(\tilde y)_T^{\Lambda,\mathrm I}
\right|
<\varepsilon.
\)
Substituting \(\tilde y=\Phi_g(\tilde x)\) yields
\(
\sup_{\tilde x\in\tilde{\mathcal K}}
\left|
f(\tilde x)-\sum_{|\Lambda|\le m}\beta_{\Lambda}\,S(\Phi_g(\tilde x))_T^{\Lambda,\mathrm I}
\right|
<\varepsilon.
\)
\Halmos\endproof

\vspace{\baselineskip}
\noindent\proof{Proof of Corollary \ref{cor:ito_un_constant}.}
Apply Theorem~\ref{th:UN} to the compact path set \(\tilde{\mathcal K}\) and the continuous function \(f\). For the given \(\varepsilon>0\), there exist a linear functional \(\ell\in T(\mathbb R^{d+1})\), an integer \(m\ge1\), and coefficients \(\{\alpha_{\Gamma}\}_{|\Gamma|\le m}\) such that
\(
\sup_{\tilde x\in\tilde{\mathcal K}}
\left|
f(\tilde x)-\sum_{|\Gamma|\le m}\alpha_{\Gamma}\,S(\tilde x)_T^{\Gamma,\mathrm S}
\right|
<\varepsilon.
\)
By Lemma~\ref{lem:ito_represents_strat}, for each multi-index \(\Gamma\) with \(|\Gamma|\le m\), there exist coefficients \(\{c_{\Gamma,\Lambda}\}_{|\Lambda|\le |\Gamma|}\) such that
\(
S(\tilde x)_T^{\Gamma,\mathrm S}
=
\sum_{|\Lambda|\le |\Gamma|} c_{\Gamma,\Lambda}\,S(\tilde x)_T^{\Lambda,\mathrm I}.
\)
Substituting these identities into the previous display and collecting coefficients yields
\(
\sum_{|\Gamma|\le m}\alpha_{\Gamma}\,S(\tilde x)_T^{\Gamma,\mathrm S}
=
\sum_{|\Lambda|\le m}\beta_{\Lambda}\,S(\tilde x)_T^{\Lambda,\mathrm I}
\)
for suitable coefficients \(\{\beta_{\Lambda}\}_{|\Lambda|\le m}\). Hence
\(
\sup_{\tilde x\in\tilde{\mathcal K}}
\left|
f(\tilde x)-\sum_{|\Lambda|\le m}\beta_{\Lambda}\,S(\tilde x)_T^{\Lambda,\mathrm I}
\right|
<\varepsilon.
\)
\Halmos\endproof

\vspace{\baselineskip}
\noindent\proof{Proof of Theorem \ref{thm:approx_hedging}.}
Fix \(\varepsilon>0\). By Corollary \ref{cor:ito_un_constant}, there exist an integer \(m\ge1\), a scalar \(\beta_0\), and coefficients \(\{\beta_{\Gamma}\}_{1\le |\Gamma|\le m}\) such that the continuous It\^o-signature model
\(
H_T^{(m)}(\tilde x)
:=
\beta_0+\sum_{1\le |\Gamma|\le m}\beta_{\Gamma}\,
S(\tilde x)_T^{\Gamma,\mathrm I}
\)
satisfies
\(
\sup_{\tilde x\in\tilde{\mathcal K}}
\bigl|F(\tilde x)-H_T^{(m)}(\tilde x)\bigr|
<
\sqrt{\varepsilon/4}.
\)
Since \(\tilde X\in\tilde{\mathcal K}\) almost surely, it follows that
\(
\mathbb E\Bigl[\bigl|F(\tilde X)-H_T^{(m)}(\tilde X)\bigr|^2\Bigr]
<
\varepsilon/4.
\)

Next define the corresponding discrete signature model
\(
H_T^{(m,\pi_n)}
:=
\beta_0+\sum_{1\le |\Gamma|\le m}\beta_{\Gamma}\,
S(\tilde X)_{T}^{\Gamma,\mathrm I,\pi_n}.
\)
By the exact replicability result for discrete It\^o-signature components established earlier, each
\(
S(\tilde X)_{T}^{\Gamma,\mathrm I,\pi_n}
\)
is the terminal wealth of a simple predictable self-financing strategy. Therefore \(H_T^{(m,\pi_n)}\) is also the terminal wealth of a simple predictable self-financing strategy. That is, there exist \(
(p_0^{m,n},\theta^{m,n})\in \mathbb R\times \mathcal L(X)
\) such that \(H_T^{(m,\pi_n)}(\beta)=\mathcal W_T(p_0^{m,n},\theta^{m,n}).\)

Moreover, for fixed \(m\) and \(\pi_n\), the finite-moment bounds for discrete It\^o-signature components in Lemma \ref{lem:moment_bound_sig_main} imply
\(\mathbb E\bigl[
|H_T^{(m,\pi_n)}(\beta)|^q
\bigr]
<\infty.\)
Since
\(H_T^{(m,\pi_n)}(\beta)=\mathcal W_T(p_0^{m,n},\theta^{m,n}),\)
we have
\(\mathcal W_T(p_0^{m,n},\theta^{m,n})\in L^q.\)
Hence
\((p_0^{m,n},\theta^{m,n})\in\mathcal H^q(X).\)

By Theorem~\ref{thm:lpconvergence} with $p=2$, for each fixed multi-index \(\Gamma\),
\(
S(\tilde X)_{T}^{\Gamma,\mathrm I,\pi_n}
\xrightarrow{L^2}
S(\tilde X)_T^{\Gamma,\mathrm I}.
\)
Since only finitely many multi-indices appear in the representation of \(H_T^{(m)}\), it follows that
\(
H_T^{(m,\pi_n)}(\beta)
\xrightarrow{L^2(\mathbb P)}
H_T^{(m)}(\tilde X)
\) as \(\|\pi_n\|\to0.
\)
Therefore we may choose \(n\) sufficiently large so that
\(
\mathbb E\Bigl[\bigl|H_T^{(m)}(\tilde X)-H_T^{(m,\pi_n)}(\beta)\bigr|^2\Bigr]
<
\varepsilon/4.
\)
Finally, using \(|a+b|^2\le 2|a|^2+2|b|^2\), we obtain
\(
\mathbb E\Bigl[\bigl|F(\tilde X)-V_T^{(m,\pi_n)}\bigr|^2\Bigr]
\le
2\mathbb E\Bigl[\bigl|F(\tilde X)-H_T^{(m)}(\tilde X)\bigr|^2\Bigr]
+
2\mathbb E\Bigl[\bigl|H_T^{(m)}(\tilde X)-H_T^{(m,\pi_n)}(\beta)\bigr|^2\Bigr]
<
\varepsilon.
\)
\Halmos\endproof

\vspace{\baselineskip}
\noindent\proof{Proof of Corollary~\ref{cor:near_optimal_signature}.}
Let \(L_P\) denote a Lipschitz constant of \(P\) such that
\(
|P(x)-P(y)|\le L_P|x-y|,
\qquad x,y\in\mathbb R.
\)
Since \(P(0)=0\) and \(P\ge0\), we have $
P(x)\le L_P|x|,
\qquad x\in\mathbb R.
$

Fix \(\eta>0\). Apply Theorem~\ref{thm:approx_hedging} with
\(
\varepsilon=\left(\frac{\eta}{L_P}\right)^2.
\), then there exists a signature-based strategy, say \((p_0^\eta,\theta^\eta)\in\mathcal H^q(X)\), whose terminal wealth
\(
\mathcal{W}_T(p_0^\eta,\theta^\eta)
\)
satisfies
\(
\mathbb E\Bigl[\bigl|F(\tilde X)-\mathcal{W}_T(p_0^\eta,\theta^\eta)\bigr|^2\Bigr]
<
\left(\frac{\eta}{L_P}\right)^2.
\)
By Cauchy--Schwarz,
\(
\mathbb E\Bigl[\bigl|F(\tilde X)-\mathcal{W}_T(p_0^\eta,\theta^\eta)\bigr|\Bigr]
\le
\Bigl(
\mathbb E\Bigl[\bigl|F(\tilde X)-\mathcal{W}_T(p_0^\eta,\theta^\eta)\bigr|^2\Bigr]
\Bigr)^{1/2}
<
\frac{\eta}{L_P}.
\)
Therefore,
\(
\mathbb E\Bigl[
P\Bigl(F(\tilde X)-\mathcal{W}_T(p_0^\eta,\theta^\eta)\Bigr)
\Bigr]
\le
L_P\,\mathbb E\Bigl[\bigl|F(\tilde X)-\mathcal{W}_T(p_0^\eta,\theta^\eta)\bigr|\Bigr]
<
\eta.
\)
\Halmos\endproof

\vspace{\baselineskip}
\noindent\proof{Proof of Corollary~\ref{cor:var_cvar_hedging_loss}.}
Fix \(\alpha\in(0,1)\) and \(\delta>0\). Choose
\(\varepsilon<(1-\alpha)^2\delta^2\). By
Theorem~\ref{thm:approx_hedging}, there exists a signature-based strategy
\((p_0,\theta)\in\mathcal H^q(X)\) such that
\(\mathbb E^{\mathbb P}[L_T(p_0,\theta)^2]<\varepsilon\). For any \(u>0\),
Markov's inequality gives
\(\mathbb P(L_T(p_0,\theta)>u)\le \varepsilon/u^2\). Taking
\(u=\sqrt{\varepsilon/(1-\alpha)}\) yields
\(\operatorname{VaR}_{\alpha}(L_T(p_0,\theta))\le \sqrt{\varepsilon/(1-\alpha)}<\delta\).

For CVaR, by the definition
\(\operatorname{CVaR}_{\alpha}(L)=\inf_{z\in\mathbb R}\{z+(1-\alpha)^{-1}\mathbb E^{\mathbb P}[(L-z)_+]\}\), taking \(z=0\) gives
\(\operatorname{CVaR}_{\alpha}(L_T(p_0,\theta))\le (1-\alpha)^{-1}\mathbb E^{\mathbb P}[L_T(p_0,\theta)]\).
By the Cauchy--Schwarz inequality,
\(\mathbb E^{\mathbb P}[L_T(p_0,\theta)]\le
\{\mathbb E^{\mathbb P}[L_T(p_0,\theta)^2]\}^{1/2}<\sqrt{\varepsilon}\).
Therefore,
\(\operatorname{CVaR}_{\alpha}(L_T(p_0,\theta))
<\sqrt{\varepsilon}/(1-\alpha)<\delta\).
This proves the claim.
\Halmos\endproof

\vspace{\baselineskip}
\noindent\proof{Proof of Theorem \ref{thm:robusterror}.}
The result follows by applying the robust high-dimensional regression theory of \citet{fan2021shrinkage} to the centered truncated discrete It\^o-signature design. The key point is that, for each fixed truncation order \(m\), the retained discrete It\^o-signature predictors have finite fourth moments by Lemma \ref{lem:moment_bound_sig_main}. Therefore the assumptions required in \citet{fan2021shrinkage} are satisfied in the present setting, and their error bounds apply directly to the truncated response-regressor pair \((\widetilde Y_i(\tau_1),\widetilde x_i(\tau_2))\).
\Halmos\endproof

\vspace{\baselineskip}
\noindent\proof{Proof of Corollary~\ref{cor:gen_hedge_error}.}
Since \(\beta_{m,\pi_n}^*\) is the population best linear predictor for the centered model, we have the exact conditional decomposition
$
\mathcal R_{m,\pi_n}(\widehat\beta_{m,\pi_n}\mid \mathcal D_N)
=
\sigma_{m,\pi_n,*}^2
+
(\widehat\beta_{m,\pi_n}-\beta_{m,\pi_n}^*)^\top
\Sigma_{m,\pi_n}
(\widehat\beta_{m,\pi_n}-\beta_{m,\pi_n}^*).
$
We have the following bound
$
\mathcal R_{m,\pi_n}(\widehat\beta_{m,\pi_n}\mid \mathcal D_N)
\le
\sigma_{m,\pi_n,*}^2
+
\lambda_{\max}(\Sigma_{m,\pi_n})\,
\|\widehat\beta_{m,\pi_n}-\beta_{m,\pi_n}^*\|_2^2.
$
The conclusion now follows directly from Theorem~\ref{thm:robusterror}.
\Halmos\endproof

\vspace{\baselineskip}
\noindent\proof{Proof of Theorem~\ref{thm:total_hedging_error}.}
For fixed \(m\) and \(\pi_n\), the centered regression problem in
\eqref{eq:centered_projection} is exactly the population linear projection of
the payoff onto the retained discrete It\^o-signature coordinates. Hence the
oracle residual variance satisfies
\(
\sigma_{m,\pi_n,*}^2=\mathcal E_{\mathrm{or}}(m,\pi_n).
\)
Applying Corollary~\ref{cor:gen_hedge_error} with the notation
\((x_{m,\pi_n},\Sigma_{m,\pi_n},p_m,K_{4,m,\pi_n})\) gives, with probability
at least \(1-3p_m^{-(\xi-2)}\),
\[
\mathbb E^{\mathbb P}
\left[
\left.
\left|
F(\tilde X_{\mathrm{new}})
-
V_{T,\mathrm{new}}^{(m,\pi_n)}(\widehat\beta_{m,\pi_n})
\right|^2
\,\right|\,\mathcal D_N
\right]
\le
\mathcal E_{\mathrm{or}}(m,\pi_n)
+
\lambda_{\max}(\Sigma_{m,\pi_n})\,
C_2\rho
\left(
\frac{K_{4,m,\pi_n}\xi\log p_m}{N}
\right)^{1-s/2}.
\]

It remains to justify the limiting statement. Let
\[
\mathcal E_{\mathrm{cont}}(m)
:=
\inf_{\beta_0,\{\beta_\Gamma:1\le |\Gamma|\le m\}}
\mathbb E^{\mathbb P}
\left[
\left|
F(\tilde X)
-
\beta_0
-
\sum_{1\le |\Gamma|\le m}
\beta_\Gamma
S(\tilde X)_T^{\Gamma,\mathrm I}
\right|^2
\right].
\]
By Corollary~\ref{cor:ito_un_constant} and the compact-support condition in
Theorem~\ref{thm:approx_hedging}, \(\mathcal E_{\mathrm{cont}}(m)\to0\) as
\(m\to\infty\). For each fixed \(m\), Theorem~\ref{thm:lpconvergence} implies
that every finite linear combination of retained discrete It\^o-signature
coordinates converges in \(L^2(\mathbb P)\) to the corresponding continuous
It\^o-signature linear combination. Therefore,
\(
\limsup_{\|\pi_n\|\to0}\mathcal E_{\mathrm{or}}(m,\pi_n)
\le
\mathcal E_{\mathrm{cont}}(m).
\)
Letting \(m\to\infty\) yields
\(
\lim_{m\to\infty}\limsup_{\|\pi_n\|\to0}
\mathcal E_{\mathrm{or}}(m,\pi_n)
=
0.
\)
\Halmos\endproof
\end{APPENDICES}
\putbib
\end{bibunit}

\end{document}